\documentclass[reprint, prd]{revtex4-2}

\usepackage[utf8]{inputenc}
\usepackage{amsmath}
\usepackage{amssymb}
\usepackage{mathrsfs}
\usepackage{times}
\usepackage{hyperref}
\usepackage{natbib}
\usepackage{graphicx}
\usepackage{hyperref}
\usepackage{natbib}

\hypersetup{
  bookmarksnumbered = true,
  bookmarksopen=false,
  pdfborder=0 0 0,         
  pdffitwindow=true,      
  pdfnewwindow=true, 
  colorlinks=true,           
  linkcolor=blue,            
  citecolor=magenta,    
  filecolor=magenta,     
  urlcolor=cyan              
}

\newcommand{\msun}{\ensuremath{M_\odot\,}}
\newcommand{\chimera}{{\sc Chimera }}
\newcommand{\gcc}{\ensuremath{{\mbox{g~cm}}^{-3}}}
\newcommand{\nn}{\nonumber}

\newcommand{\mnras}{Mon. Not. R. Astron. Soc.}

\begin{document}

\title{Core Collapse Supernova Gravitational Wave Emission for Progenitors of 9.6, 15, and 25 M$_\odot$}

\author{Anthony Mezzacappa$^{1}$, Pedro Marronetti$^{2}$, Ryan E. Landfield$^{3}$, Eric J. Lentz$^{1,4,5}$, \\
 W. Raphael Hix$^{1,4}$, J. Austin Harris$^{3}$, Stephen W. Bruenn$^{6}$, John M. Blondin$^{7}$, \\ 
O.E. Bronson Messer$^{3,4}$, Jordi Casanova$^{8}$, Luke L. Kronzer$^{4,9}$}

\affiliation{$^1$Department of Physics and Astronomy, University of Tennessee, 1408 Circle Drive, Knoxville, TN 37996-1200, USA}
\affiliation{$^2$Physics Division, National Science Foundation, Alexandria, VA 22314 USA}
\affiliation{$^3$National Center for Computational Sciences, Oak Ridge National Laboratory, P.O. Box 2008, Oak Ridge, TN 37831-6164, USA}
\affiliation{$^4$Physics Division, Oak Ridge National Laboratory, P.O. Box 2008, Oak Ridge, TN 37831-6354, USA}
\affiliation{$^5$Joint Institute for Nuclear Physics and its Applications, Oak Ridge National Laboratory, P.O. Box 2008, Oak Ridge, TN 37831-6374, USA}
\affiliation{$^6$Department of Physics, Florida Atlantic University, 777 Glades Road, Boca Raton, FL 33431-0991, USA}
\affiliation{$^7$Department of Physics, North Carolina State University,  Raleigh, NC 27695-8202, USA}
\affiliation{$^8$Physics Program, Community College of Denver, P.O. Box 173363, Denver, CO 80217-3363}
\affiliation{$^9$Department of Aerospace Engineering and Mechanics, University of Alabama, Box 870280, Tuscaloosa, AL 35487-0280}

\email{mezz@utk.edu}

\begin{abstract}
We present gravitational wave emission predictions based on three core collapse supernova simulations corresponding to
three different progenitor masses. The masses span a large range, between 9.6 and 25 \msun, are all initially non-rotating, and are of 
two metallicities: zero and Solar. We compute both the temporal evolution of the gravitational wave strains for both 
the plus and the cross polarizations, as well as their spectral decomposition and characteristic strains. The 
temporal evolution of our zero metallicity, 9.6 \msun progenitor model is distinct from the temporal evolution of our 
Solar metallicity, 15 \msun progenitor model and our zero metallicity, 25 \msun progenitor model. In the former case, 
the high-frequency gravitational wave emission is largely confined to a brief time period $\sim$75 ms after bounce, whereas 
in the latter two cases, high-frequency emission does not commence until $\sim$125 ms after bounce or later. Nonetheless,
the physical origin of the high-frequency emission in all three cases corresponds to convection in the proto-neutron star of 
both Schwarzschild and Ledoux type, and convective overshoot. The low-frequency emission in all three models exhibits very similar behavior. At 
frequencies below $\sim$250 Hz, gravitational waves are emitted by neutrino-driven convection and the SASI. This emission
extends throughout the simulations when a gain region is present. In all three models, explosion is observed, at $\sim$125,
$\sim$500, and $\sim$250 ms after bounce in the 9.6, 15, and 25 \msun progenitor models, respectively. At these times, 
the low-frequency gravitational wave emission is joined by very low frequency emission, below $\sim$10 Hz. These very 
low frequency episodes are the result of explosion and begin at the above designated explosion times in each of our 
models. Our characteristic strains tell us that in principle all three gravitational wave signals would be detectable by 
current-generation detectors for a supernova at a distance of 10 kpc. However, our 9.6 \msun progenitor model is a 
significantly weaker source of gravitational waves, with strain amplitudes approximately 5--10 times less 
than in our other two models. The characteristic strain for this model tells us that such a supernova would be detectable 
only within a much more narrow frequency range around the maximum sensitivity of today's detectors. Finally, in our 9.6 \msun 
progenitor model, we see very high frequency gravitational radiation, extending up to $\sim 2000$ Hz. 
This feature results from the interaction of shock- and deleptonization-induced convection with perturbations introduced 
in the progenitor by nuclear burning during core collapse. While unique to the 9.6 \msun progenitor model analyzed 
here, this very high frequency emission may in fact be a generic feature of the predictions for the gravitational wave emission 
from all core collapse supernova models when simulations are performed with three-dimensional progenitors.
\end{abstract}



\maketitle

\section{Introduction}

The anticipation of a Galactic core collapse supernova and its multi-messenger detection in photons, neutrinos, and gravitational 
waves is driving in part the development of three-dimensional supernova models and methods to analyze supernova gravitational wave data. 
Results from three-dimensional supernova models have now been published by a number of groups 
\cite{HaMuWo13,LeBrHi15,MeJaBo15,MeJaMa15,KuKoTa16,KuTaKo16,RoOtHa16,KuKoHa17,MuMeHe17,OcCo18,MuGaHe18,SuJaMe18,BuRaVa19,GlJuJa19,MuTaHe19,Powell2019,Radice2019,VaBuRa19,BuRaVa20,KuArTa20,MuVa20,PoMu20,ShKuKo20,StJaKr20,WaTaJa20,PaLiCo21,PaWaCo21,ShKuKo21,VaLaRe21,BuCo22,VaCoBu22,VaMuSc22}.
The prediction and analysis of gravitational wave emission in these models has followed
\cite{KuKoTa16,Andresen2017,KuKoHa17,OcCo18,Powell2019,Radice2019,SrBaBr19,VaBuRa19,MeMaLa20,PoMu20,ShKuKo20,VaBu20,AnGlJa21,PaLiCo21,PaWaCo21,ShKuKo21,PoMu22,RiZaAn22,VaMuSc22}. 
The latter has opened a window onto a rich and complex gravitational wave phenomenology. 

Core collapse supernova gravitational wave emission prediction has two main thrusts. First among these is detection. Second 
among these is the extraction of information about the explosion and its remnant proto-neutron star, given a detection. The full taxonomy 
of core collapse supernova explosions presents a challenge for both thrusts. There are at least two possible explosion mechanisms: neutrino-driven 
explosions and magnetohydrodynamically-driven explosions, which give rise to different explosion dynamics on different time scales. The 
two categories of explosions stem from the two categories of stellar progenitors in Nature, slowly and rapidly rotating massive stars, with 
rapidly rotating progenitors providing the rotational energy necessary to ramp up and organize the magnetic fields such that the outflows 
may be generated largely by magnetohydrodynamic forces, not neutrino heating. And within each category of progenitor, the further distinguishing 
characteristics of mass and metallicity can also lead to different explosion dynamics and time scales. We can expect an imprint of these differences 
on the gravitational wave emission. The variety of emission characteristics is arguably not ``templatable,'' though within a class of core collapse 
supernova explosions -- e.g., those arising from more massive, slowly rotating progenitors that are neutrino-driven -- there are certainly common 
characteristics. Given we will begin the studies presented here with a broad range of progenitor masses, at least within the context of one of the 
explosion mechanisms listed above --  neutrino-driven explosions -- we will see examples of the different gravitational wave emission characteristics
we can anticipate in a future detection.

Of course, the use of a core collapse supernova gravitational wave detection to cull information about the explosion and remnant proto-neutron
star will be possible only if we derive a clear understanding of the physical origins of the gravitational wave emission in such an event. As we will 
hope to demonstrate, this, in turn, will require that we develop an ability to compute the gravitational wave emission from separate regions in the post-bounce 
stellar core -- specifically, from regions below the supernova shock wave, including regions between the proto-neutron star surface and the shock and
within the proto-neutron star itself. We hope to show that indirect determinations of the origins of low- and high-frequency gravitational wave emission
may lead to the wrong, or at the very least incomplete, conclusions.

Kuroda, Kotake, and Takiwaki \cite{KuKoTa16} conclude that the high-frequency emission in their model is the result of $g$-mode oscillations of 
the proto-neutron star ``surface,'' though they do not discuss how they define this surface. They reach this conclusion based on the fact their peak 
emission frequency tracks the fit proposed by M\"{u}ller, Janka, and Marek \cite{MuJaMa13}, which is based on the assumption that the emission 
corresponds to excitations of $g$-modes in the Ledoux stable region within the proto-neutron star, though this is not necessarily a ``surface'' region 
(see Andresen et al. \cite{Andresen2017} and Mezzacappa et al. \cite{MeMaLa20}). The models we show here, which exhibit similar fits yet also 
exhibit significant high-frequency emission from both Ledoux stable and Ledoux unstable regions, will demonstrate that the increase in the peak 
frequency of the high-frequency emission in accord with the M\"{u}ller et al.-proposed fit is not alone sufficient to delineate all of the major sources 
of high-frequency gravitational wave emission in the models. It should also be noted that the definition of the proto-neutron star ``surface'' varies 
across groups, some using $\rho = 1\times 10^{11}$ \gcc to define it, others using $\rho = 1\times 10^{10}$ \gcc, still others using some other criterion. 
This is an important issue that requires further discussion, particularly as it pertains to efforts to ascertain information about the ``proto-neutron star'' 
through gravitational wave detection. Finally, Kuroda, Kotake, and Takiwaki use spatial decomposition, as we do here, in tracking the origin of the 
low-frequency emission in their model. They track the origin to a deep layer in the proto-neutron star, between 10 and 20 km, and attribute the cause 
to convection- and SASI-modulated accretion flows penetrating deep within the star -- i.e., they conclude that the low-frequency emission stems from 
the proto-neutron star itself, not from the post-shock region where these modulated accretion flows originate. In the models we show here, the latter 
region dominates the low-frequency emission, although there is also evidence of some low-frequency emission emanating from the proto-neutron star 
due to excitation from accretion, as discussed above.

Andresen et al. \cite{Andresen2017} conclude that the high-frequency gravitational wave emission in their models results from a combination of 
Ledoux convection in the convective unstable layer within the proto-neutron star and convective overshoot into the convectively stable layer above it, 
with the emission from the latter being primary. They reached this conclusion by spatially dividing the region below the shock in their models into 
three layers. The deepest layer includes the convective and convective overshoot layer. Above it is the proto-neutron star surface layer, with the surface 
in their analysis defined by $\rho = 1\times 10^{10}$ \gcc, above which is the layer between the proto-neutron star surface and the shock. Given this
spatial decomposition, and taking care regarding issues arising at the boundaries between these regions, which was emphasized again more recently 
by Eggenberger Andersen et al. \cite{EggenbergerAndersen21} and which we will discuss in detail in our analysis later, they computed the square of the Fourier 
amplitudes of the gravitational wave amplitudes in each of the three regions and, from this, determined that the deepest layer dominates. They found this 
surprising -- i.e., such a conclusion was reached only through an attempt to spatially decompose the gravitational wave emission. 
Note further, the conclusion that the convective overshoot sublayer of the deepest layer in the 
Andresen et al. analysis dominates the high-frequency gravitational wave emission was based on how well the peak frequency in their models tracked the 
M\"{u}ller et al. fit, which assumes g-mode emission. That is, Andresen et al. did not compute the gravitational wave emission from the convective sublayer 
and the convective overshoot sublayer, both contained within their deepest layer, individually. In the models we present here, we compute the gravitational 
wave emission from these sublayers individually. In our models, the peak frequency evolution is described well by the M\"{u}ller et al. fit, at densities consistent 
with the deep gravitational wave emission, yet we find significant gravitational wave emission from both the convective and convective overshoot regions. 
Discerning this was possible only through a sufficiently refined spatial decomposition of the gravitational wave emission in the proto-neutron star. Moving 
now to the low-frequency signal, using the same spatial analysis Andresen et al. conclude that the low-frequency gravitational wave emission emanates 
from all three layers but predominantly from their two deepest layers within it, excited by convection- and SASI-modulated accretion onto it. The accretion 
flows in their models penetrate deeply into the proto-neutron star. Moreover, their conclusions are the same for their exploding and non-exploding models. It is 
here where the greatest differences lie between what we find (see Mezzacappa et al. \cite{MeMaLa20} and the results presented here) and what Andresen 
et al. find in their first study. In all of our models, all of which explode, the low-frequency emission 
stems largely from the gain region itself. Having said that, the later study by Andresen et al. \cite{Andresen2019}, focused on the impact of rotation on core collapse 
supernova gravitational wave emission, concluded the opposite for their fastest rotating model, due to the presence of a strong spiral SASI and a quicker drop in 
the mass accretion rate onto the proto-neutron star. This suggests that the predominant origin of the low-frequency gravitational wave emission depends on 
the details of the post-bounce explosion dynamics and that one conclusion cannot be drawn for all models.

Powell and M\"{u}ller \cite{Powell2019} attribute the high-frequency emission in their models to excitations of PNS $g$-modes by matter accreting onto 
the PNS surface, tracked in their study by density contours at $\rho = 1\times 10^{11}$ \gcc and $\rho = 1\times 10^{12}$ \gcc, from the region between 
the PNS surface and the shock. Their conclusion is based on the fact that the peak in the high-frequency emission in their models occurs just prior to 
shock revival and declines afterward (e.g., see their Figures 7 and 8). Yet, the high-frequency emission persists for the duration of their runs, especially 
for model He3.5. 
A spatial decomposition of the gravitational wave emission can shed 
some light here. High-frequency emission can be excited simultaneously from within the PNS, through, for example, sustained Ledoux convection due to 
continued core deleptonization after bounce, and from above, due to convection- and SASI-modulated, penetrating accretion flows onto the PNS surface. 
We will discuss a model here where clearly both excitation mechanisms occur simultaneously but where there is a transition to excitation from Ledoux convection 
after shock revival, with a corresponding decline, though not as pronounced of a decline as that reported by Powell and M\"{u}ller, in the high-frequency 
gravitational wave emission. In this model, we also see a transition from a broader-band emission prior to explosion, due to the stochastic nature of the 
accretion onto the proto-neutron star surface, to a narrower-band emission after explosion, characteristic of excitation by Ledoux convection. Thus, a decline 
in the high-frequency emission is not alone sufficient to rule out excitation of this emission from deep within the proto-neutron star by Ledoux convection.

Radice et al. \cite{Radice2019} observe initial high-frequency emission due to prompt convection in the proto-neutron star in all of their models. However, 
they attribute the high-frequency emission after this initial convective period to the excitation of $g$-modes in the proto-neutron star by convection- and 
SASI-modulated accretion onto it from the post-shock region. They conclude this based on two considerations: (1) In one of their models (their 9 M$_\odot$ 
model) proto-neutron star convection remains ``vigorous'' but the gravitational wave luminosity drops off precipitously after 
$\sim$300 ms (the model is run for 800 ms). (2) They find a close correlation between the total amount of turbulent energy accreted onto the proto-neutron 
star and the total energy radiated in gravitational waves, across the broad range of models considered: 9 to 60 M$_\odot$. In this paper, we present an 
example of continued high-frequency gravitational wave emission from the proto-neutron star after the onset of explosion and a decline in the gravitational 
wave emission from the proto-neutron star surface regions, as well as deeper regions, resulting from accretion. The evolution of the high-frequency emission in our 
example, whose start is well correlated with the onset of Ledoux convection in the proto-neutron star, and whose continuance long after explosion is initiated 
and accretion is reduced and whose amplitude is not significantly affected by either, points instead to excitation by Ledoux convection.

Powell and M\"{u}ller \cite{PoMu20} observe prompt convection in all of their models, including one model (their 18 M$_\odot$ model) with rotation.
This is followed by high-frequency emission attributed to the excitation of $f$- and $g$-modes in the proto-neutron star. For the non-rotating models,
the peak frequency evolution is in good agreement with the predictions of the M\"{u}ller et al. fit. The authors do not discuss the excitation mechanisms
responsible for the high-frequency emissions in their models, though, at least in their non-rotating models, the evolution of the peak frequencies 
suggests that one mechanism is the excitation of the convectively stable region in the proto-neutron stars in these models. Whether excitation results
from convective overshoot deep within the proto-neutron star or from convection- and SASI-modulated accretion onto it is not considered. In two of 
the three models we present here, we will show that the convectively unstable as well as the convective overshoot regions both contribute 
significantly to the high-frequency gravitational wave emission while at the same time our peak frequencies evolve according to the M\"{u}ller 
et al. fit, albeit at a density $\sim 10^{12}$ \gcc, not at the density we use to define the proto-neutron star surface ($10^{11}$ \gcc ). As discussed 
above, the evolution of the peak frequency alone is insufficient to identify all significant contributions to the high-frequency emission.

Pajkos et al. \cite{PaWaCo21} report on gravitational emission from prompt convection in the proto-neutron star, as well as on high-frequency emission 
through the first several hundred milliseconds after bounce, which they attribute to emission excited by stochastic accretion onto the proto-neutron star 
from the postshock region above it. In their study, the focus is on the impact of rotation. They did not perform a study of the origin of the high-frequency 
emission in their models. The analyses we present here point to the necessity of attempts to further explore the origins of the high-frequency emission
due to the redundancy in the ways it can be excited and, most important, to the fact that excitation mechanisms internal to the proto-neutron star may 
dominate over excitations external to it in producing such emissions.

Pan et al. \cite{PaLiCo21} report on high-frequency gravitational wave emission in their model, with rising peak frequency, which is seen by all groups, but do not discuss 
what they attribute to be its origins. They also report on low-frequency frequency gravitational wave emission in their model and, citing the conclusions drawn by others 
(discussed above), attribute it to convection- and SASI-modulated-accretion--induced, low-frequency excitations of the proto-neutron star.

Focusing on their non-rotating model, Shibagaki et al. \cite{ShKuKo21} report on gravitational wave emission due to prompt convection and then go on 
to focus on the low-frequency SASI-modulated emission in their models -- particularly due to the spiral SASI modes.

To summarize: (1) Gravitational wave emission due to prompt convection has been reported in multiple studies, which, to begin with, demonstrates 
that the excitation mechanisms for gravitational wave emission from the proto-neutron star are not all external to it. (2) Indirect evidence of the excitation of 
g-modes within the proto-neutron star as the primary driver of high-frequency emission from it, the most compelling of which is the evolution of the peak 
frequency (at least in non-rotating models), is not sufficient to rule out other significant drivers -- e.g., Ledoux convection. An accurate assessment of the origins 
of the gravitational wave emission from the proto-neutron star -- specifically, the high-frequency emission -- requires a spatial decomposition of the signal 
sufficiently fine to capture all of the potential excitation mechanisms. (3) Low-frequency gravitational wave emission from neutrino-driven convection and 
the SASI emanates from both the proto-neutron star, as a result of convection- and SASI-modulated accretion onto it, and the gain region itself, as a result 
of convective- and SASI-modulated flows there. The dominance of one mode of excitation over the other depends on the details of the explosion dynamics. 
Thus, in this case too we can expect both internal and external excitations of gravitational waves in this part of the spectrum, as we go from model to model.

It is with all of this in mind we now endeavor to understand the gravitational wave emission and its origins in the three-dimensional models presented here. 

\section{Models and Methods}

\subsection{Core Collapse Supernova Models}

The three-dimensional core collapse supernova simulations used for this study are part of the D-series of simulations performed with the \chimera code \cite{BrBlHi20}, which will be presented in detail elsewhere \cite{LeHiHa21,LeXX21}.
\chimera utilizes multi-group flux-limited diffusion neutrino transport in the ray-by-ray approximation, Newtonian self-gravity with a monopole correction to account for the effects of general relativity, Newtonian hydrodynamics, and a nuclear reaction network. 
Neutrino--matter interactions in \chimera include electron capture on protons and nuclei, the latter using the LMSH capture rates, electron--positron annihilation, and nucleon--nucleon bremsstrahlung, along with their inverse weak interactions.
The included scattering processes are coherent isoenergetic scattering on nuclei, as well as neutrino--electron (large-energy transfer) and neutrino--nucleon (small-energy transfer) scattering.
The equation of state for densities above $10^{11}$ \gcc, is that of \citet{LaSw91} with a bulk incompressibility of K = 220 MeV. At lower densities the equation of state is an enhanced version of that of \citet{Cooperstein85}, used in conjunction with an active nuclear network \cite{HiTh99} at lower temperatures.
Of particular note here, all three simulations analyzed were performed using a Yin--Yang grid with angular resolution equivalent to one degree in $\theta$ and $\phi$.
Complete details of the \chimera code and improvements that were made to the code to perform the D-series simulations, relative to the C-series simulations \cite{LeBrHi15}, are given in \cite{BrBlHi20}.

The simulations used for this study were all initiated from non-rotating progenitors covering a range of mass and metallicity.
The lightest progenitor used was a 9.6 \msun model of zero metallicity (A. Heger, private communication) evolved as a low-mass extension of the \citet{HeWo10} set.
\chimera simulation D9.6-sn160-3D was computed with a 160-species nuclear network and 540 radial zones \cite{LeHiHa21}.
The other two progenitors are the 15 \msun Solar-metallicity progenitor from \citet{WoHe07} used in our previous study \cite{MeMaLa20} and a 25 \msun zero-metallicity progenitor from \citet{HeWo10}.
The corresponding \chimera simulations, D15-3D and D25-3D, were computed with an alpha network and 720 radial zones \cite{LeXX21}.
Given that all of the simulations considered in this study were computed in 3D, we will shorten the \chimera designations to D9.6, D15, and D25.

\subsection{Gravitational Wave Extraction Methods}

We employ the quadrupole approximation for extracting the gravitational wave signals from the mass motions.
We begin with the lowest multipole (quadrupole) moment of the Transverse-Traceless gravitational wave strain \cite{Kotake2006}
\begin{equation}
h_{ij}^{\rm TT}=\frac{G}{c^4}\frac{1}{r}\sum_{m=-2}^{+2}\frac{d^{2}I_{2m}}{dt^{2}}(t-\frac{r}{c})f_{ij}^{2m},
\label{eq:quadrupoleexpansion}
\end{equation}

\noindent where $i$ and $j$ run over $r$, $\theta$, and $\phi$ and where $f_{ij}^{2m}$ are the tensor spherical harmonics, given by

\begin{equation}
f_{ij}^{2m} = \alpha r^{2}
\begin{pmatrix}
0 & 0            & 0                                  \\
0 & W_{2m} & X_{2m}                         \\
0 & X_{2m}  & -W_{2m}\sin^{2}\theta  \\
\end{pmatrix},
\label{eq:tensorsphericalharmonics}
\end{equation}

\noindent with

\begin{equation}
X_{2m}=2\frac{\partial}{\partial\phi}\left(\frac{\partial}{\partial\theta}-\cot\theta\right) Y_{2m}(\theta,\phi)
\label{eq:Xlm}
\end{equation}

\noindent and 

\begin{equation}
W_{2m}=\left(\frac{\partial ^2}{\partial \theta^2}-\cot\theta\frac{\partial}{\partial\theta}-\frac{1}{\sin^2\theta}\frac{\partial^2}{\partial\phi^2}\right)Y_{2m}(\theta, \phi).
\label{eq:Wlm}
\end{equation}

\noindent The normalization, $\alpha$, is determined by

\begin{equation}
 \int d\Omega \left(f_{l m}\right)_{ab}
\left(f_{l' m'}^*\right)_{cd}
\gamma^{ac}\gamma^{bd}
= r^4 \delta_{l l'}\delta_{m m'},
\end{equation}

\noindent where $a,b,c,d=\theta,\phi$, and $\gamma_{ab}$ is the 2-sphere metric

\begin{equation}
 \gamma_{ab} = 
  \begin{bmatrix}
   1 & 0 \\
   0 & \sin^2\theta
  \end{bmatrix}.
\end{equation}

\noindent For $l=2$,  $\alpha = \frac{1}{4\sqrt{3}}$.

The mass quadrupole is

\begin{equation}
I_{2m}=\frac{16\sqrt{3}\pi}{15} \int \tau_{00}Y_{2m}^{*}r^2dV.
\label{eq:massquadrupole}
\end{equation}

\noindent In equation (\ref{eq:massquadrupole}), $dV=r^{2}\sin\theta dr d\theta d\phi$, and $\tau_{00}$ is simply the rest-mass density, $\rho$, for the weak fields assumed here. We define the gravitational wave amplitude

\begin{equation}
A_{2m}\equiv\frac{G}{c^4}\frac{d^{2}I_{2m}}{dt^{2}}.
\label{eq:gravwaveamplitude}
\end{equation}

\noindent or

\begin{equation}
A_{2m}\equiv\frac{dN_{2m}}{dt},
\label{eq:N2mdot}
\end{equation}

\noindent where

\begin{equation}
N_{2m}=\frac{G}{c^4}\frac{dI_{2m}}{dt}.
\label{eq:N2mdot}
\end{equation}

\noindent Combining equations (\ref{eq:massquadrupole}) and (\ref{eq:N2mdot}), we obtain

\begin{eqnarray}
\label{eq:N2mdot2}
N_{2m} & = & \frac{16\sqrt{3}\pi G}{15c^4}\frac{d}{dt} \int \rho Y_{2m}^{*}r^2dV \\ \nn
& = & \frac{16\sqrt{3}\pi G}{15c^4}\int \frac{\partial \rho}{\partial t}Y_{2m}^{*}r^2dV. \nn
\end{eqnarray}

\noindent The continuity equation can be used to eliminate the time derivative in equation (\ref{eq:N2mdot2}) in an effort to 
minimize the numerical noise associated with computing the second time derivative of $I_{2m}$ numerically, by reducing the 
calculation in advance to computing the first time derivative of $N_{2m}$ numerically, which gives \cite{Finn90}

\begin{eqnarray}
\label{eq:n2m-integration}
& &  N_{2m} = \frac{16\sqrt{3}\pi G}{15c^4}
 \int_0^{2\pi}d\varphi' \int_0^\pi d\vartheta' \int_a^b dr'~r'^3 \\ \nn
  && \left[
2\rho v^{r}Y^*_{2m}\sin\vartheta'
+ \rho v^{\vartheta}\sin\vartheta'
\frac{\partial}{\partial\vartheta'}Y^*_{2m} \right. 
\left.+ \rho v^{\varphi}
\frac{\partial}{\partial\varphi'}Y^*_{2m}
\right] \\ \nn
&& -\frac{16\sqrt{3}\pi G}{15c^4}\int_0^{2\pi}d\varphi' \int_0^\pi d\vartheta' Y^*_{2m}\sin\vartheta' 
(r_{b}^4\rho_{b}v^{r}_{b}-r_{a}^4\rho_{a}v^{r}_{a}) \\ \nn
&& \equiv \bar{N}_{2m} + \Delta N_{2m},
\end{eqnarray}
where 
\begin{eqnarray}
\label{eq:deltan2m}
& &  \Delta N_{2m} \equiv -\frac{16\sqrt{3}\pi G}{15c^4} \\ \nn 
& & \times \int_0^{2\pi}d\varphi' \int_0^\pi d\vartheta' Y^*_{2m}\sin\vartheta' 
(r_{b}^4\rho_{b}v^{r}_{b}-r_{a}^4\rho_{a}v^{r}_{a}),
\end{eqnarray}
and where we integrate over the source coordinates $r'$, $\vartheta'$ and $\varphi'$. $r_a$ and $r_b$ are the inner
and outer radial boundaries, respectively, of the region for which we are calculating the strain. For a whole-star calculation, $r_a=0$ and $r_b=\infty$. In this case, the surface term,
which is the last term in equation (\ref{eq:n2m-integration}), vanishes because $r_a=0$ and $\rho_b=0$. Eggenberger Andersen et al. \cite{EggenbergerAndersen21} emphasized 
the fact that the surface term cannot be neglected for $r_a$ and $r_b$ nonzero and finite. 
In this analysis, we will compute $A_{2m}$ in three ways, by evaluating numerically (1) the {\em second} time derivative of $I_{2m}$, 
(2) the {\em first} time derivative of $N_{2m}$, and (3) the first time derivative of $\bar{N}_{2m}$. We will compare results obtained 
with all three methods in Section \ref{sec:results} and discuss the limitations of all of them. Given these limitations, we 
use method (3) throughout this paper to {\em guide} our analysis, but we do so in light of its limitations, and use it accordingly.

Finally, we compute the gravitational wave strains for both polarizations, which are related to $h^{TT}_{ij}$ by:

\begin{eqnarray}
\label{eq:rhpus}
h_+ & = & \frac{h^{TT}_{\theta\theta}}{r^2}, \\ 
h_{\times} & = & \frac{h^{TT}_{\theta\phi}}{r^2 \sin\theta}. \\ \nonumber
\end{eqnarray}

The total luminosity emitted in gravitational waves is given by \cite{Thorne80}

\begin{equation}
\frac{dE}{dt}=\frac{c^3}{G}\frac{1}{32\pi}\sum_{m=-2}^{+2}\langle\left|\frac{dA_{2m}}{dt}\right|^2\rangle,
\label{eq:gravwaveluminosity}
\end{equation}

\noindent where the $\langle\rangle$ indicate averaging over several wave cycles. To compute the spectral signatures, we must relate the gravitational wave  luminosity to its spectrum, using Parseval's Theorem:

\begin{equation}
\int_{-\infty}^{+\infty}|x(t)|^2dt=\int_{-\infty}^{+\infty}|\tilde{x}(2\pi f)|^2df.
\label{eq:ParsevalsTheorem}
\end{equation}

\noindent Here, $\tilde{x}(2\pi f)$ is the Fourier transform of $x(t)$. The total energy emitted in gravitational waves is 

\begin{eqnarray}
\label{eq:totalGWenergyf}
E=\int_{-\infty}^{+\infty}\frac{dE}{dt}dt & = & \frac{c^3}{32\pi G}\sum_{m=-2}^{+2}\int_{-\infty}^{+\infty}|\dot{A}_{2m}|^2dt \\ \nonumber
& = &\frac{c^3}{32\pi G}\sum_{m=-2}^{+2}\int_{-\infty}^{+\infty}|\tilde{\dot{A}}_{2m}(2\pi f)|^2df \\ \nonumber
& = &\frac{c^3}{16\pi G}\sum_{m=-2}^{+2}\int_{0}^{+\infty}|\tilde{\dot{A}}_{2m}(2\pi f)|^2df. \\ \nonumber
\end{eqnarray}

\noindent where the over-dot now represents the time derivative. The time derivative of $\tilde{A}_{2m}$ in equation (\ref{eq:totalGWenergyf}) can be eliminated using the standard property of Fourier transforms -- i.e.,

\begin{equation}
|\tilde{\dot{A}}_{2m}(2\pi f)|^2=(2\pi f)^2|\tilde{A}_{2m}(2\pi f)|^2.
\label{eq:FTproperty}
\end{equation}

\noindent Inserting equation (\ref{eq:FTproperty}) in equation (\ref{eq:totalGWenergyf}) and taking the derivative with respect to frequency gives

\begin{equation}
\frac{dE}{df}=\frac{c^3}{16 \pi G}(2\pi f)^2\sum_{m=-2}^{+2}|\tilde{A}_{2m}|^2.
\label{eq:dedf}
\end{equation}

\noindent The stochastic nature of gravitational wave signals from core collapse supernovae prompts the use of short-time Fourier transform (STFT)
techniques to determine $\tilde{A}_{2m}$ \cite{Murphy09}:
\begin{equation}
  \textrm{STFT}\{A_{2m}(t)\}\left(\tau, f\right) =  \int\limits_{-\infty}^{\infty}
            A_{2m}(t)\,H(t - \tau)e^{-i\,2\pi ft}dt
\end{equation}
where $H(t - \tau)$ is the Hann window function.
In our analysis, we set the window width to $\sim$15 ms. The sampling interval of our data is 0.2 ms and, within the 
resolution of the individual time steps in our run, which are $\sim$0.1 $\mu$s, is uniform throughout the evolution 
of all three models considered here. Finally, we relate $dE/df$ to the characteristic gravitational wave strain, defined 
by \cite{Flanagan98}

\begin{equation}
h^{2}_{\rm char}(f)=\frac{2G(1+z)^2}{\pi^2 c^3 D^2(z)}\frac{dE}{df}[(1+z)f],
\label{eq:hchar}
\end{equation}

\noindent where $z$ is the source's redshift. Here we assume $z=0$, as for a Galactic supernova, and $D(0)\equiv D=10$ kpc. Then equation (\ref{eq:hchar}) becomes

\begin{equation}
h_{\rm char}(f)=\sqrt{\frac{2G}{\pi^2 c^3 D^2}\frac{dE}{df}}.
\label{eq:hchar3}
\end{equation}

\section{Results}
\label{sec:results}

\begin{figure}
\includegraphics[width=0.98\columnwidth,clip]{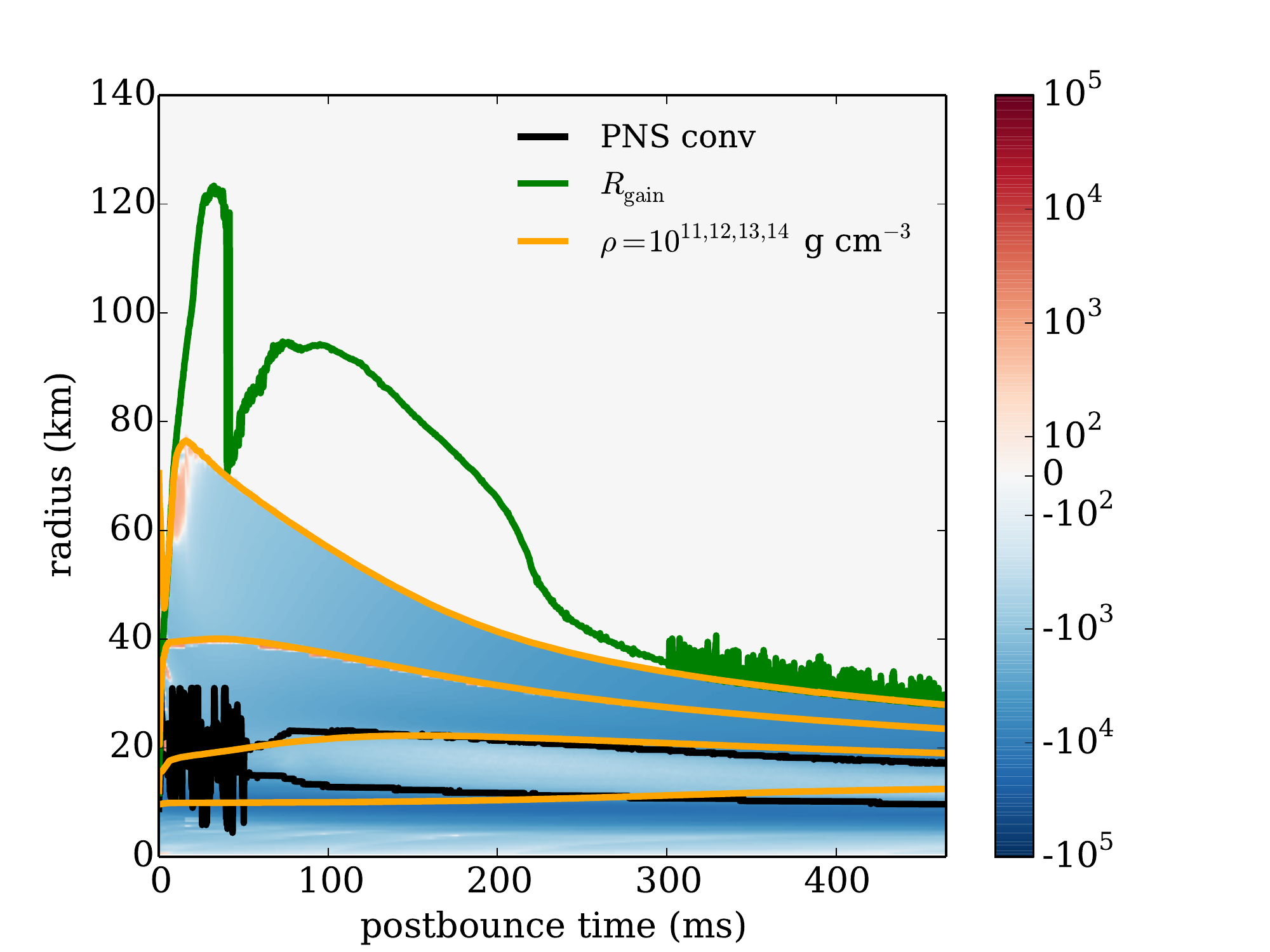}
\includegraphics[width=0.98\columnwidth,clip]{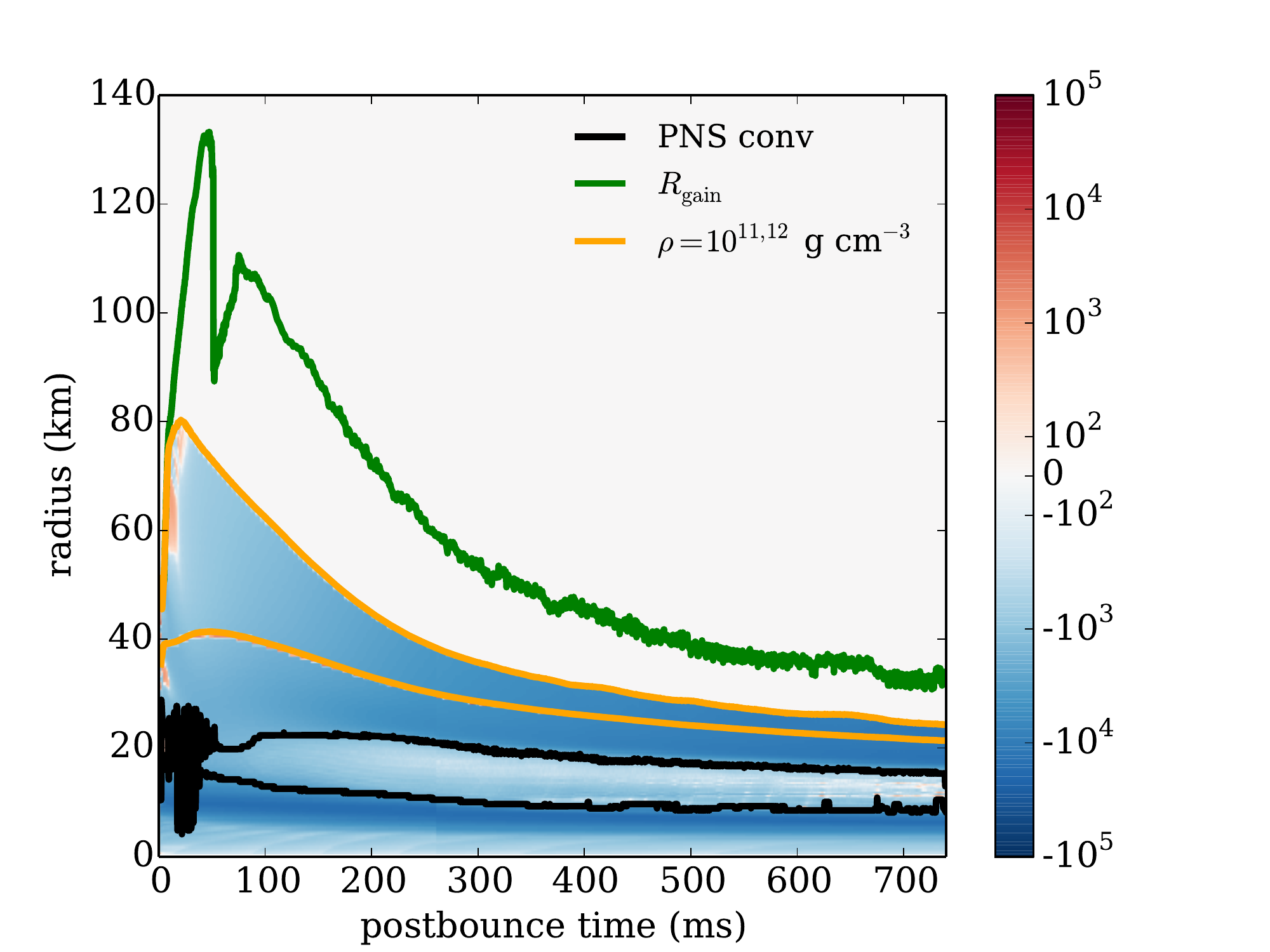}
\includegraphics[width=0.98\columnwidth,clip]{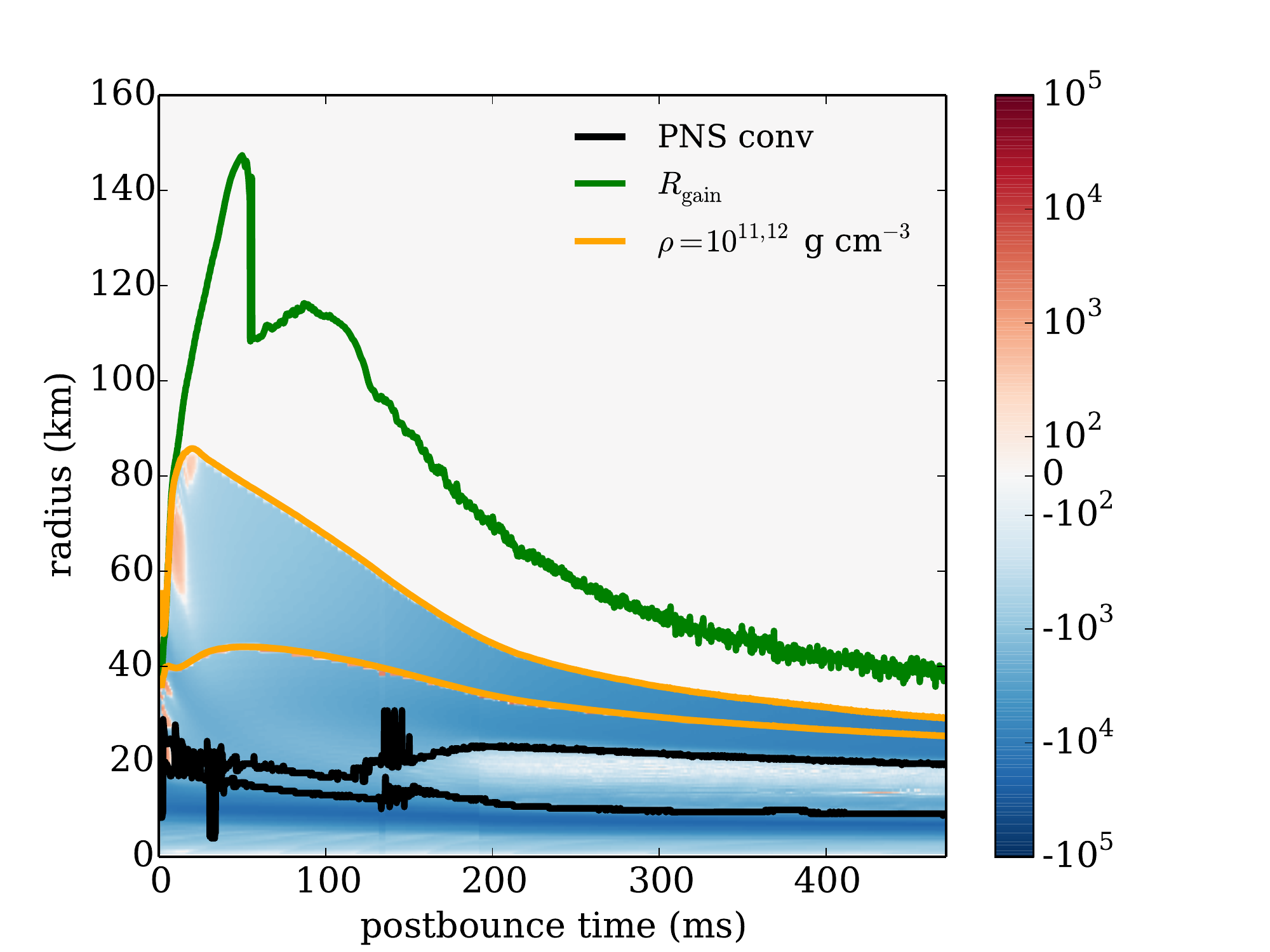}
\caption{Region boundaries for D9.6, D15, and D25, respectively, together with the Brunt--V\"{a}is\"{a}l\"{a} frequency within the proto-neutron star (superimposed).}
\label{fig:regionboundaries}
\vspace{-10pt}
\end{figure}

\subsection{Defining Gravitational Wave Source Boundaries}

To begin our analysis, we first map out the region below the supernova shock wave in each of our models. Plots of the evolution of these regions as a function
of time are shown in Figure \ref{fig:regionboundaries}. The postshock region in each model divides into five regions from which gravitational waves are emitted. It 
is necessary to separate the discussion of D9.6 from D15 and D25 for reasons that will become clear. Beginning 
with D15 and D25, Region 1, the deepest region of the five, corresponds to the region of the core that is Ledoux convective. 
Its boundaries are defined by the contours at which the convective mass flux is 5\% of its peak value. (This was also true in Mezzacappa et al. 
\cite{MeMaLa20}, though incorrectly stated there as where the convective velocity, not the convective mass flux, is 5\% of its peak value.) Region 2 
is bounded from above by the $10^{12}$ g/cm$^{3}$ density contour. Region 3 is bounded from below and from above by the $10^{12}$ g/cm${^3}$ 
and $10^{11}$ g/cm$^{3}$ density contours, respectively, the latter of which we take to define the proto-neutron star surface in our models. Region 4 
corresponds to the net-neutrino-cooling layer between the proto-neutron star surface and the gain radius, and Region 5 corresponds to the gain 
layer. Superimposed on all three regional plots is the Brunt--V\"{a}is\"{a}l\"{a} frequency below the proto-neutron star surface, with values indicated 
by the color bars.

For all three of our models, Region 1 is not well defined until $\sim$50 ms after bounce. Given that the gravitational wave emission for 
D9.6 occurs largely within an $\sim$75 ms window of bounce, this presents a challenge and requires that we instead, in this case, 
define our innermost Regions 1 and 2 by the constant-density contours at $10^{14}$ \gcc, and $10^{13}$ \gcc, with the inner (outer) 
boundary of Region 1 defined by the $10^{14}$ \gcc ($10^{13}$ \gcc) contour. For D9.6, Region 4 is 
also not well defined for post-bounce times greater than $\sim$300 ms, nor is the gain radius. Outside of these exceptions, all five regions are well 
characterized during the time periods in which the dominant gravitational wave emission occurs.

\begin{figure*}
\includegraphics[width=\columnwidth,clip]{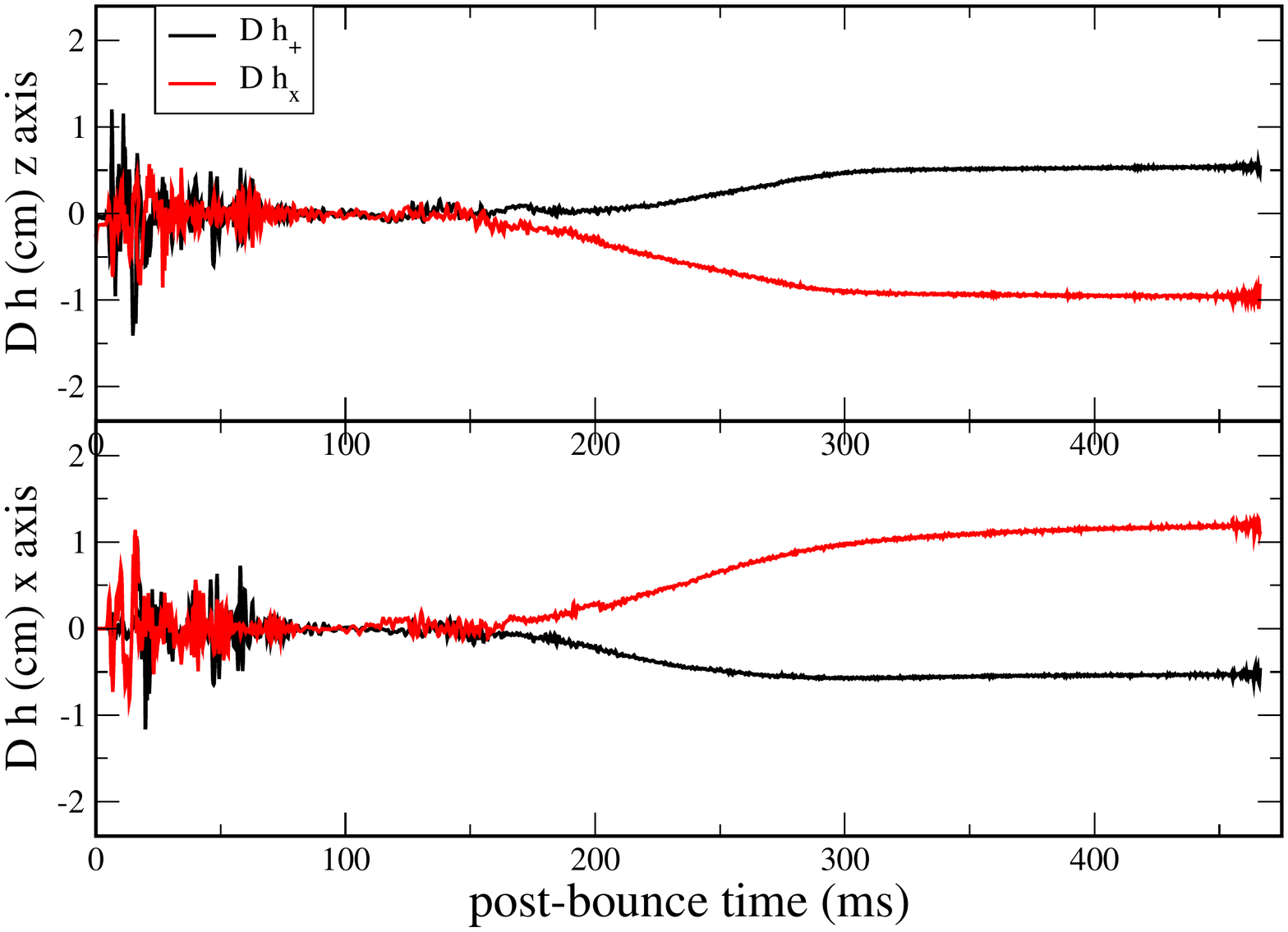}
\includegraphics[width=\columnwidth,clip]{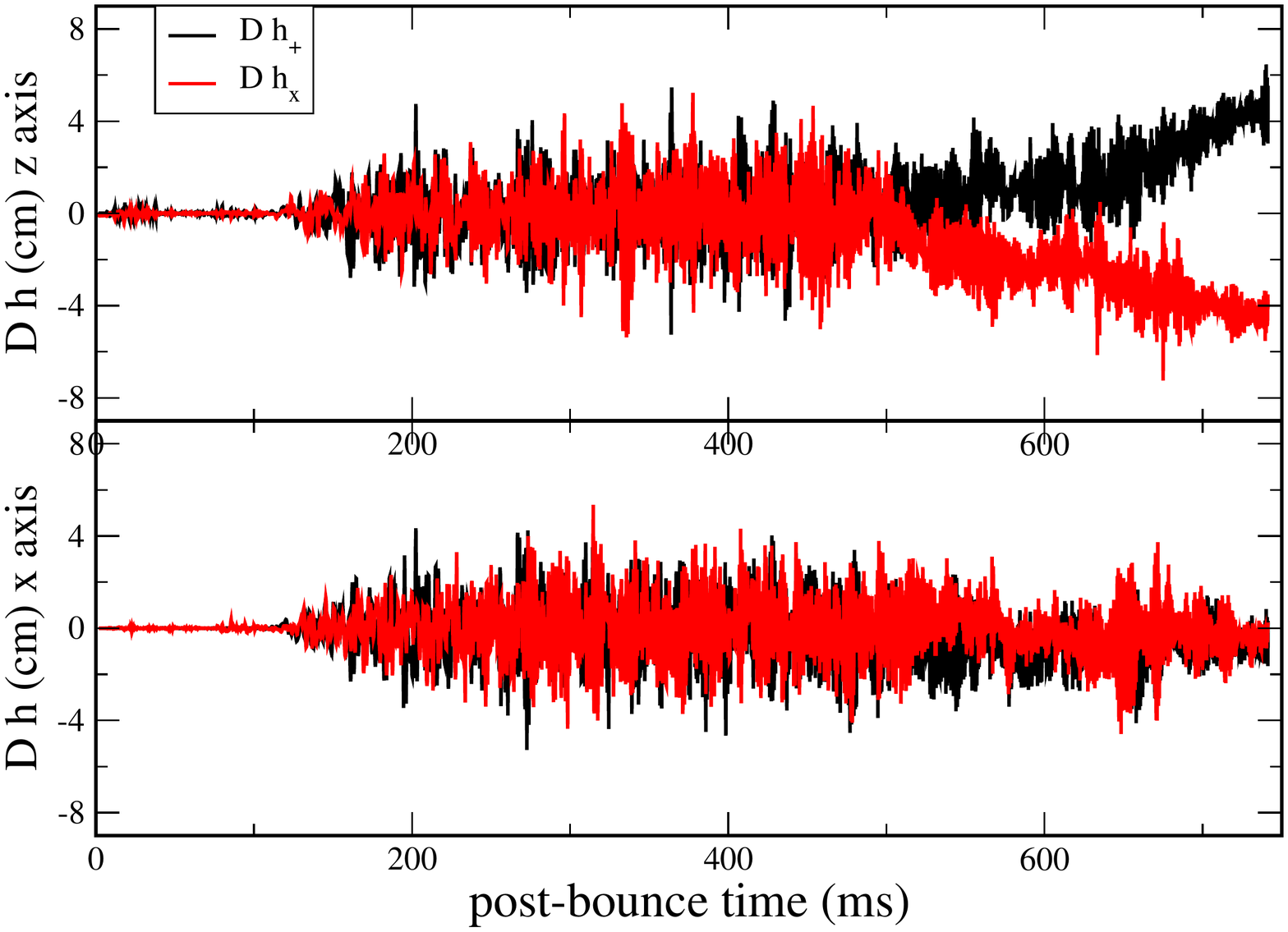}
\includegraphics[width=\columnwidth,clip]{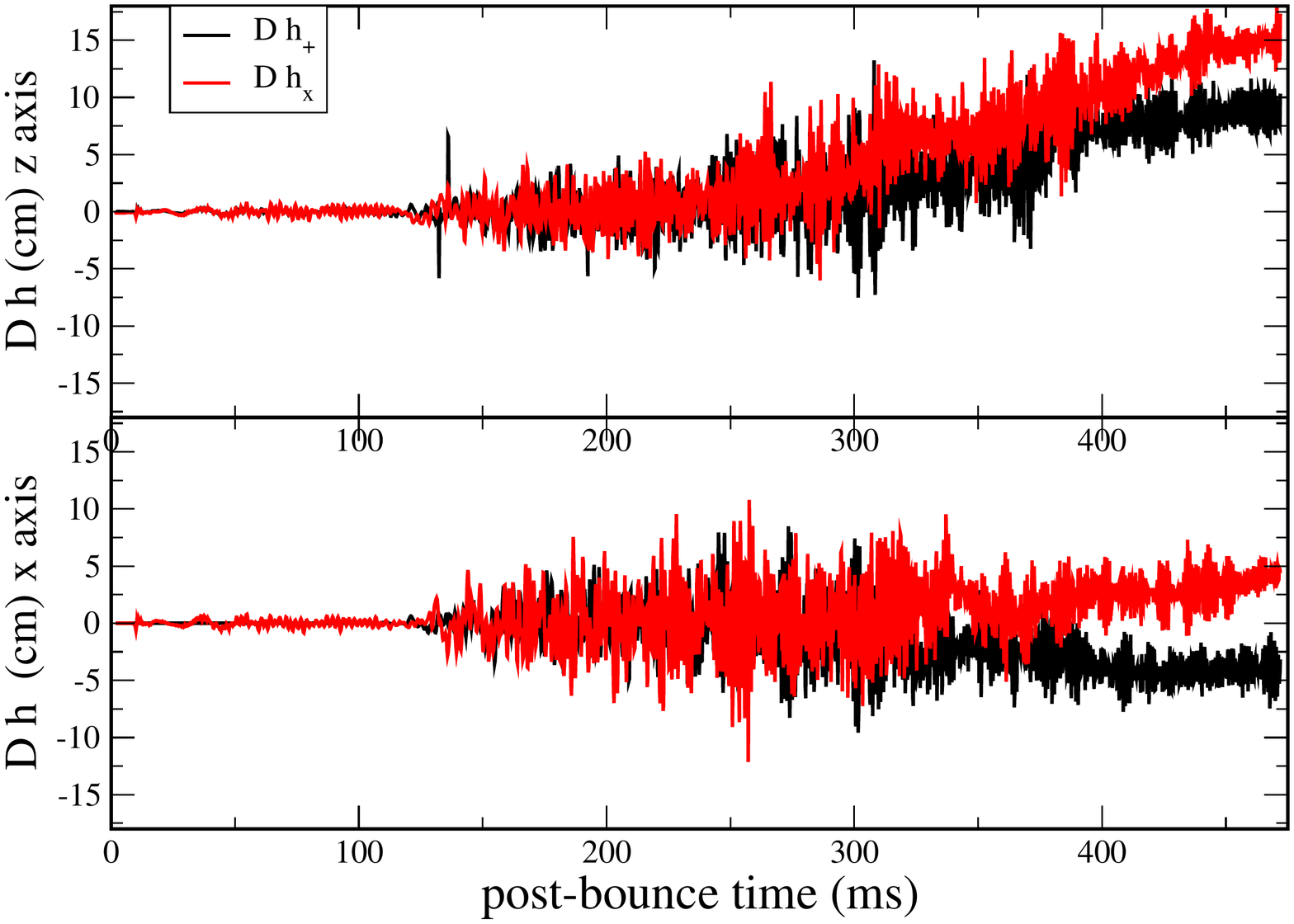}
\caption{Gravitational wave strains for D9.6, D15, and D25, respectively.}
\label{fig:strains}
\end{figure*}

\subsection{Gravitational Wave Strains}

In Figure \ref{fig:strains}, we plot the gravitational wave strains, for both polarizations, as a function of post-bounce time and for all three 
progenitors. Comparing the strain evolution across the three models, D9.6 is obviously distinct from the other two, which 
correspond to more massive progenitors. In the former case, with the exception of an offset that occurs late in the simulation, the gravitational 
wave signal is largely confined to a very brief period of time after bounce of $\sim$75 ms, whereas the gravitational wave emission in D15 and 
D25 is largely emitted after $\sim$125 ms.

\begin{figure}
\includegraphics[width=\columnwidth,clip]{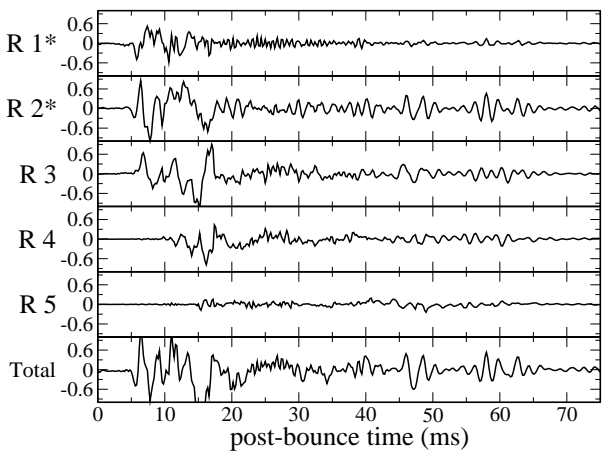}
\caption{Gravitational wave strains, $Dh_{+}$ (cm), by region for D9.6, during the first 75 ms post bounce.}
\label{fig:strainsbyregion96}
\end{figure}

The very early post-bounce gravitational wave emission for D9.6 arises largely from two distinct episodes of convection within the proto-neutron star, both short-lived.  To highlight this convection, Figure 4 shows the entropy per baryon in two-dimensional slices along the xy-plane, at 4 different times.  The first convective episode, an episode of prompt Ledoux convection, begins at 4 ms after bounce and remains confined to the inner 40 km of the star -- i.e., to Regions 1 and 2 (see Figure 4, upper left). This episode of convection dissipates quickly and is largely complete after 20 ms (see Figure 4, upper right). At 9 ms after bounce, the first convective episode is joined by a second -- in this case, an episode of Schwarzschild convection -- confined to a region above 40 km -- i.e., to Region 3 (also see Figure 4, upper right). These episodes correlate well with the evolution of the strain amplitude for the plus polarization, detailed in Figure 3 over the first 75 ms after bounce. For example, one can see the slight delay of the rise of the strain amplitude in Region 3, relative to the rise in the amplitudes in Regions 1 and 2, resulting from the slight delay of the onset of the second convective episode relative to the first. From Figure 3, we can also see that the largest strains in this model are produced within the first 20 ms post bounce.  Neutrino-driven convection, which develops later, when a sustained gain region is established, is shown in the lower panels of Figure 4, at two different times, early and late in its development.  At these later times, the central region seems quiescent in these entropy plots. However, Figure 5, a two-dimensional slice along the xy-plane of the electron fraction, reveals that proto-neutron star convection continues, driven by the continued deleptonization.

\begin{figure*}
\includegraphics[width=0.5\textwidth]{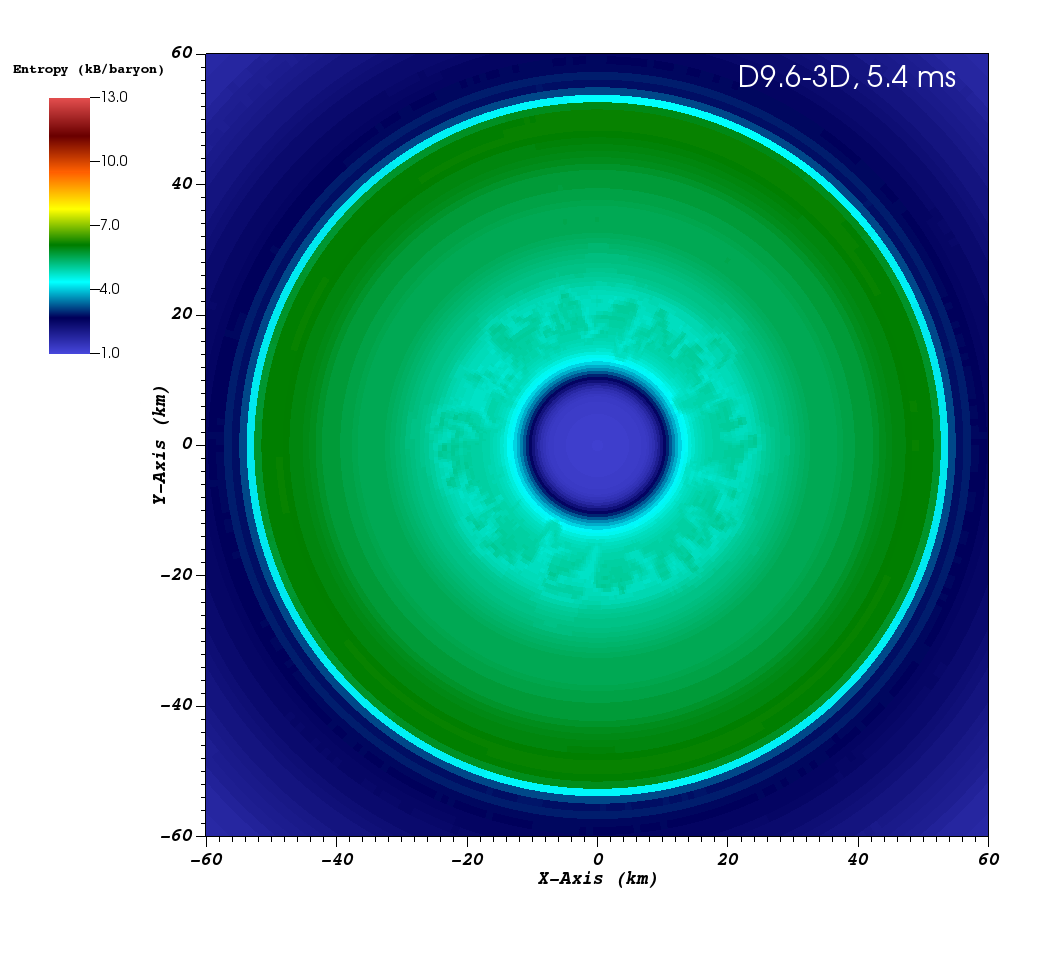}\hfill
\includegraphics[width=0.5\textwidth]{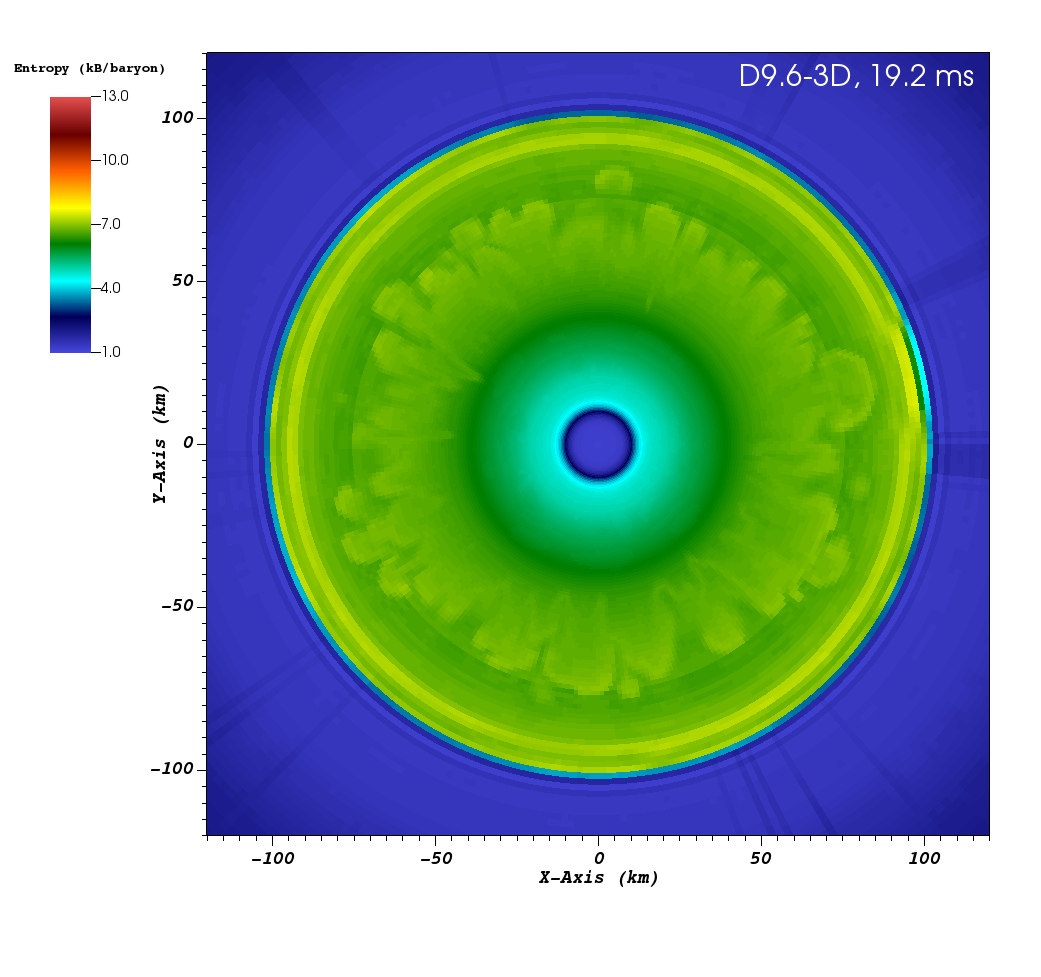}\\
\includegraphics[width=0.5\textwidth]{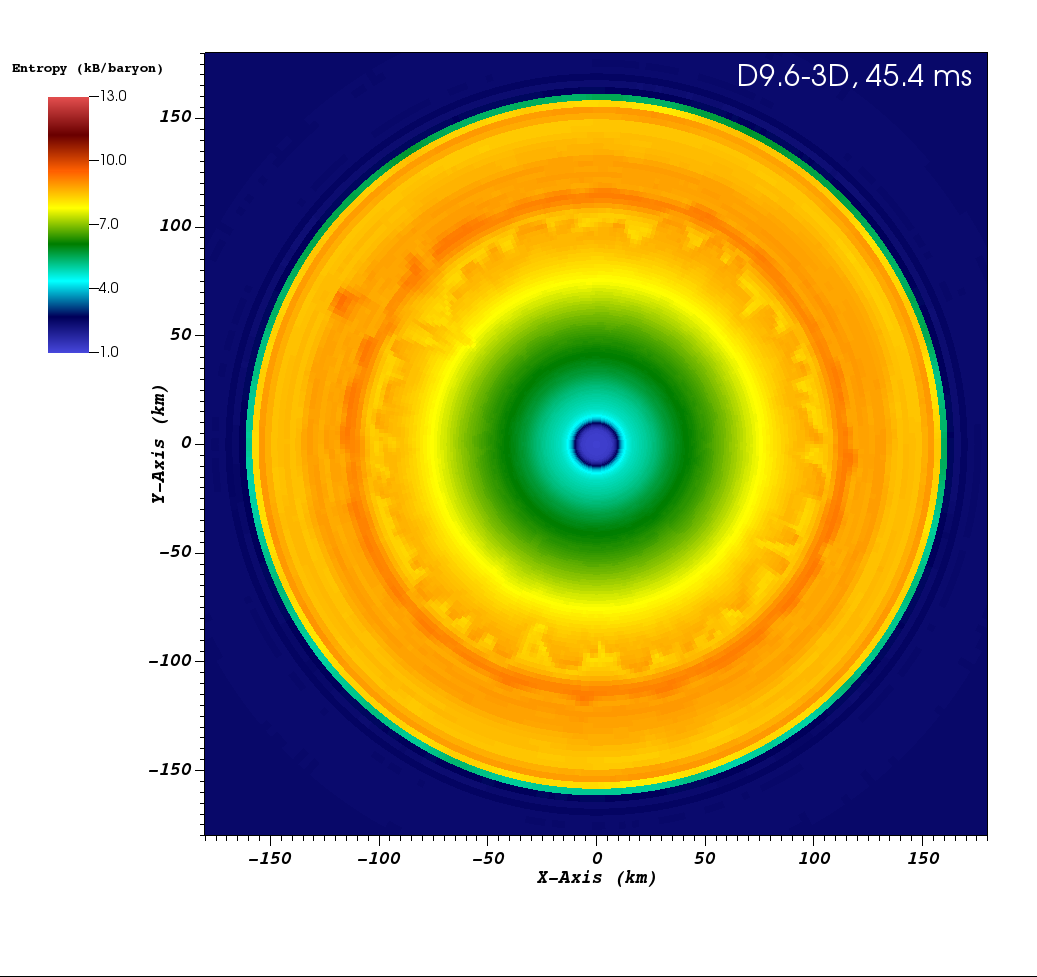}\hfill
\includegraphics[width=0.5\textwidth]{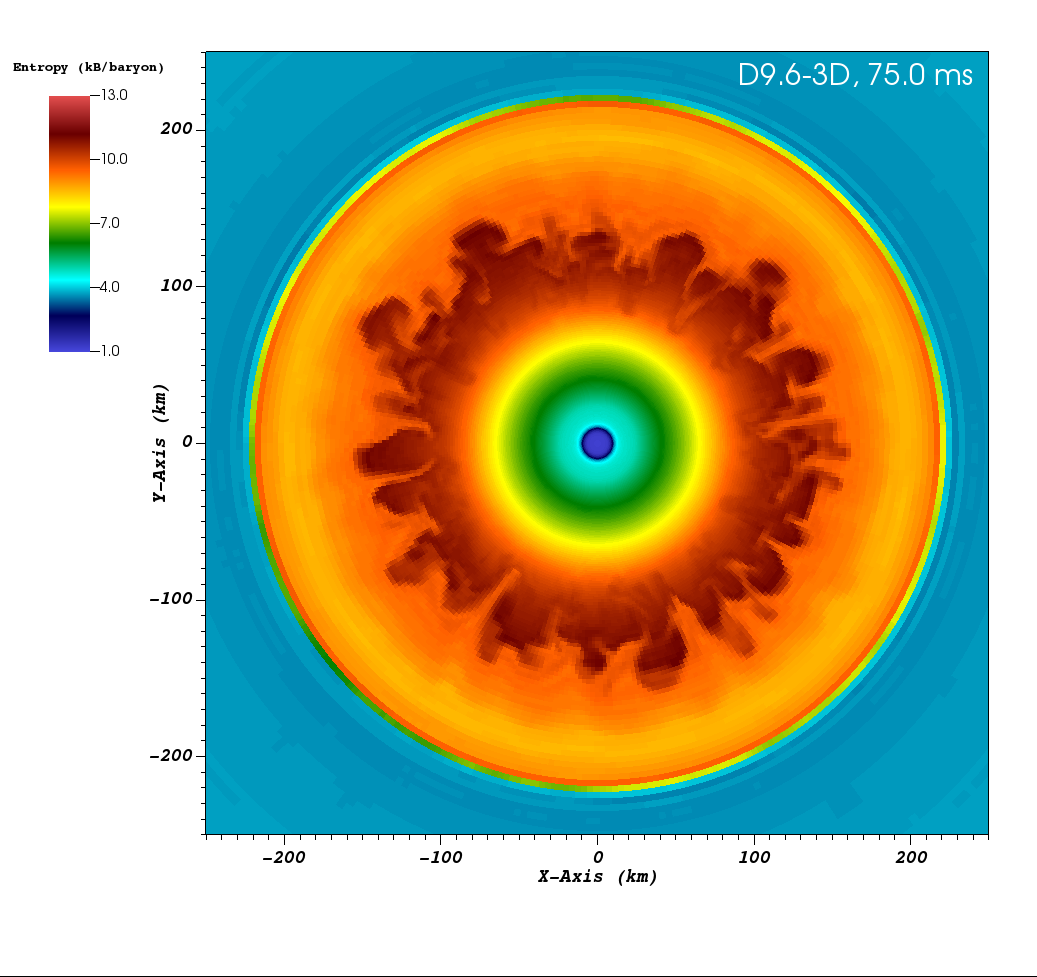}
\caption{Entropy per baryon cross sections in the $xy$-plane at select times post bounce in D9.6.}
\label{fig:slices}
\end{figure*}

\begin{figure}
\includegraphics[width=0.5\textwidth]{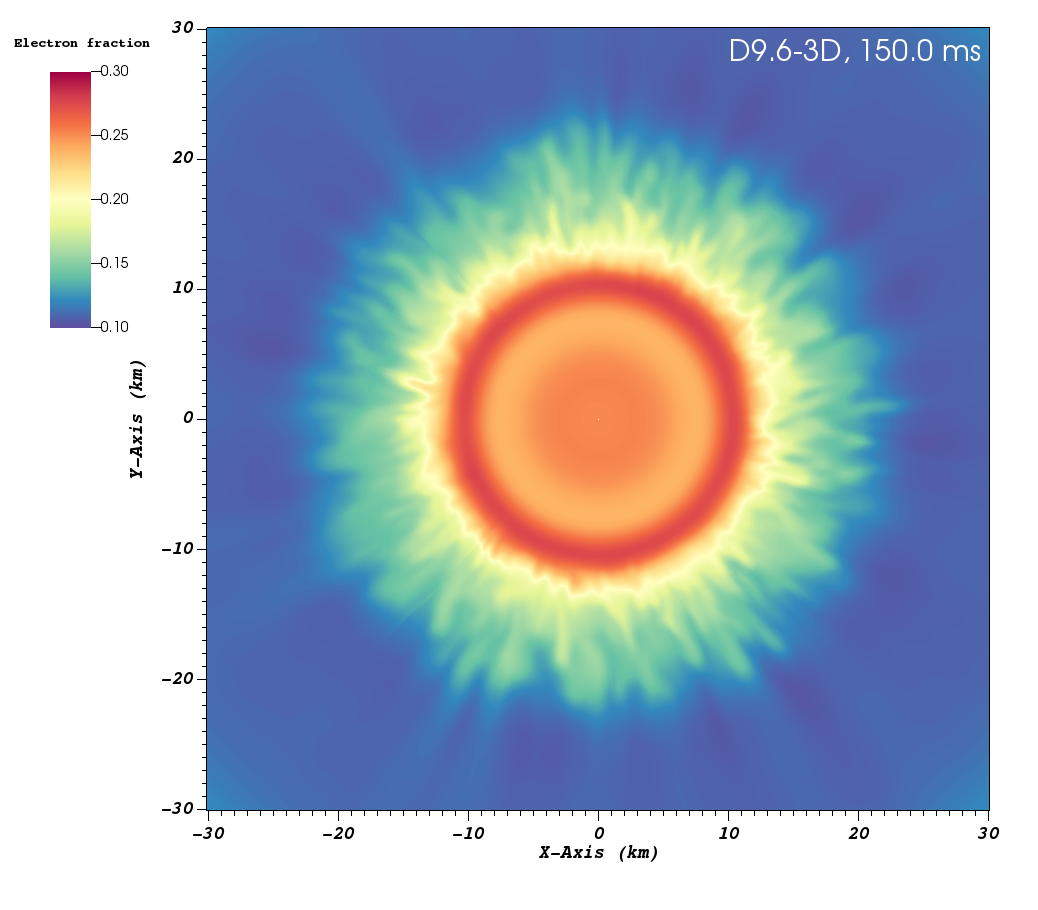}\hfill
\caption{Electron fraction cross section in the $xy$-plane at 150 ms post bounce in D9.6.}
\label{fig:electronfractionslice}
\end{figure}

For D15 and D25, the early evolution is relatively quiet with regard to gravitational wave emission. In these models, emission begins 
in earnest at $\sim$150 ms after bounce. They exhibit the behavior delineated in \cite{MeMaLa20}: High-frequency gravitational wave emission begins 
after the formation of the Ledoux convective region (Region 1) deep within the proto-neutron star and the subsequent generation of gravitational waves 
from Ledoux convection and convective overshoot into the Ledoux stable above (Region 2). This is joined by low-frequency gravitational wave 
emission below $\sim$250 Hz resulting from neutrino-driven convection in the gain region (Region 5) and the SASI.

In all three models, explosion is exhibited by the offset of the strains above and below zero (depending on the viewing angle) at late times in the models. 
It is accompanied by very low frequency gravitational wave emission below $\sim$10 Hz. This occurs at $\sim$150 ms, $\sim$500 ms, and $\sim$250 ms 
for D9.6, D15, and D25, respectively. Note, the time to the offset is not a monotonic function of the progenitor mass, reflecting the fact that 
the time to explosion is a function of the detailed structure of the progenitor, which cannot be captured by a single parameter. When viewed along 
the $z$-axis, in all three cases the positive offset in the amplitude of the plus polarization indicates that the explosion takes on a prolate shape. For an 
$x$-axis view and for the same polarization, the sign of the offset reverses in all three cases, indicative of the fact that the explosion takes on an oblate 
shape when viewed along this direction. The two views together provide a three-dimensional picture of the explosion's morphology, for all three models.

\begin{figure*}
\includegraphics[width=\columnwidth,clip]{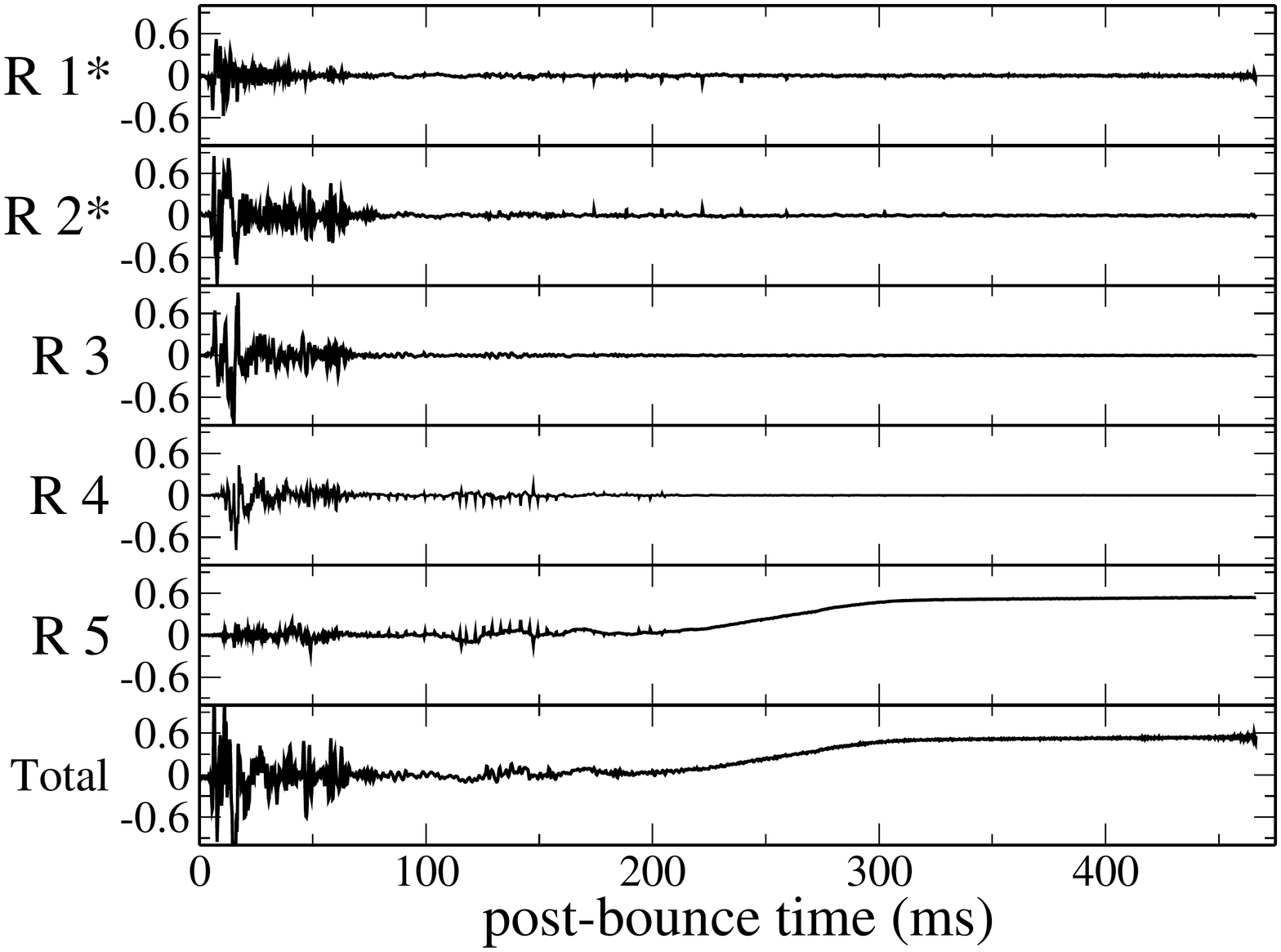}
\includegraphics[width=\columnwidth,clip]{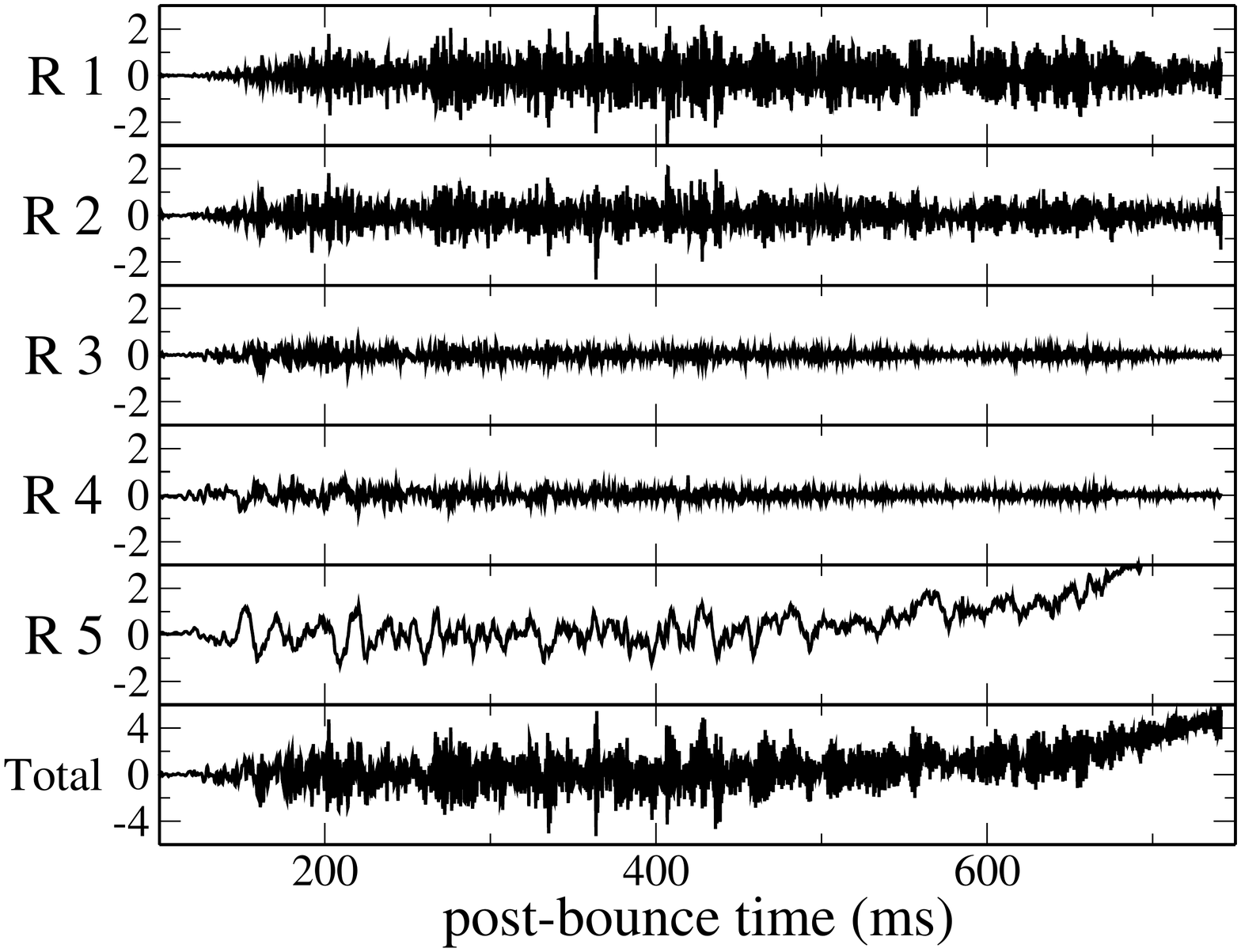}
\includegraphics[width=\columnwidth,clip]{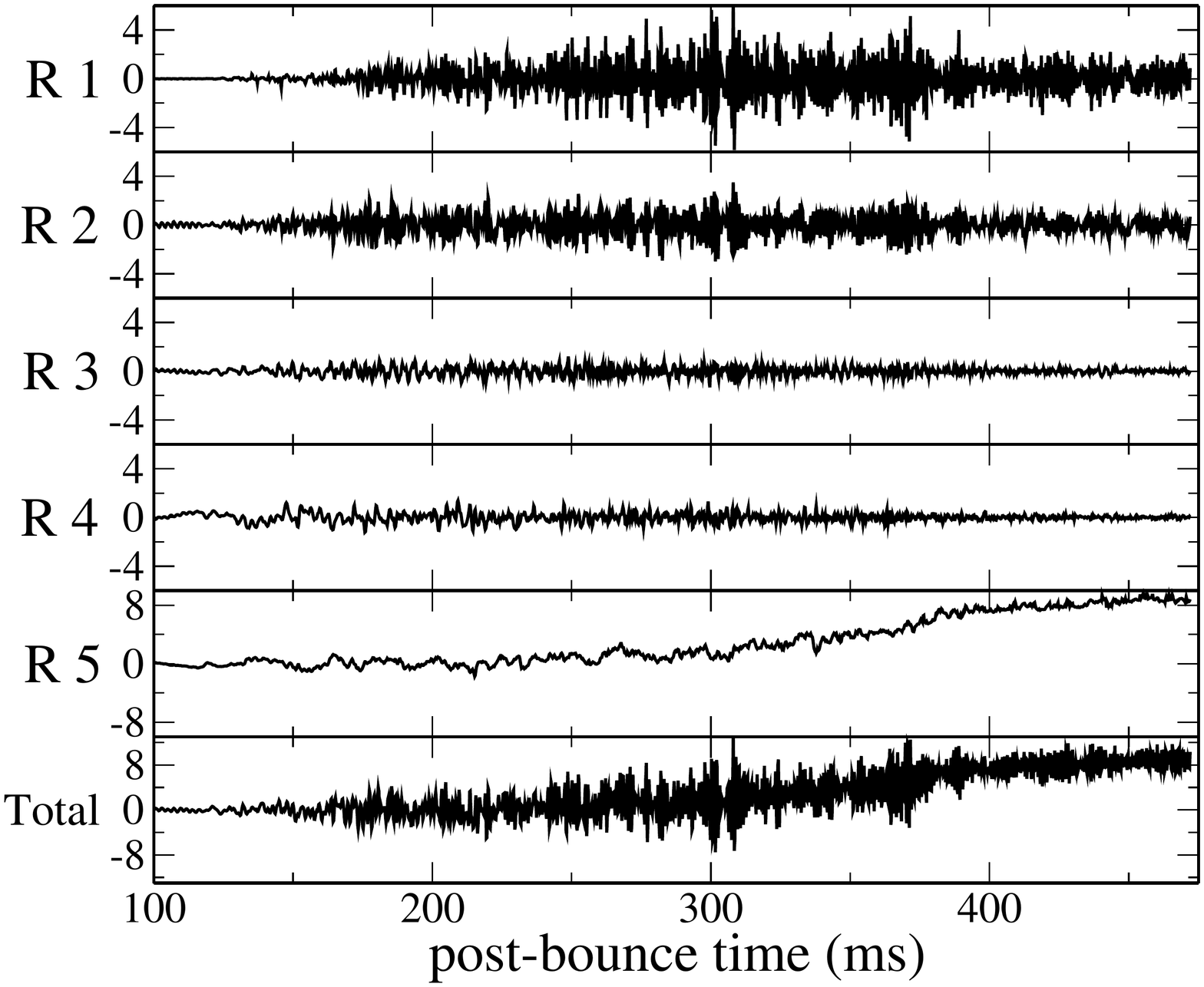}
\caption{Gravitational wave strains, $Dh_{+}$ (cm), by region for D9.6, D15, and D25, respectively.}
\label{fig:strainsbyregion}
\end{figure*}

Figure \ref{fig:strainsbyregion} gives the breakdown by region of the total strain 
for the plus-polarization and all three models. For D9.6, 
significant (relative to the total amplitude) gravitational wave emission emanates from multiple regions all the way up to the gain radius, dominated by emission 
from the innermost three regions. (Regions 1 and 2 here are delineated with an asterisk to indicate they have been defined by density contours, not convective activity.) 
Instead, for D15 and D25, gravitational wave emission emanates largely from a more restricted radial region deep 
within the proto-neutron star, Regions 1 and 2. (This remains true when the other two methods for the regional strain decomposition 
are used.) High-frequency gravitational wave emission is the result of Ledoux 
convection in Region 1 and convective overshoot into the Ledoux convectively stable region above it, Region 2. This is consistent with what was found in the analysis of the 
gravitational wave emission in our first three-dimensional model \cite{MeMaLa20}. 

In all three cases, low-frequency gravitational wave emission, including the gravitational wave strain offset, emanates from Region 5, just below the shock. 
This part of the gravitational wave emission spectrum is the result of neutrino-driven convection, the SASI, and explosion.

\subsection{Heat Maps}

\begin{figure*}
\includegraphics[width=0.5\textwidth]{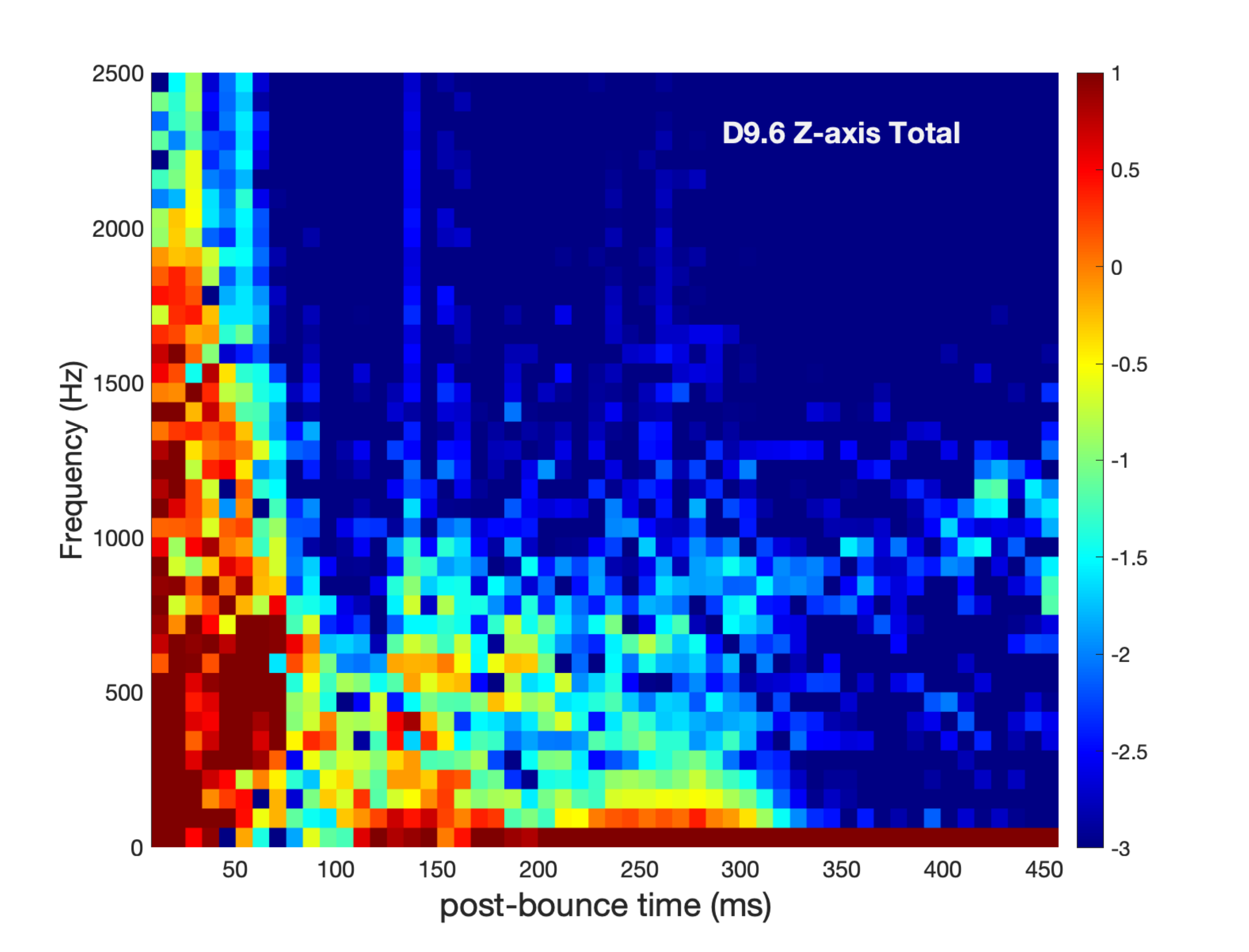}\hfill
\includegraphics[width=0.5\textwidth]{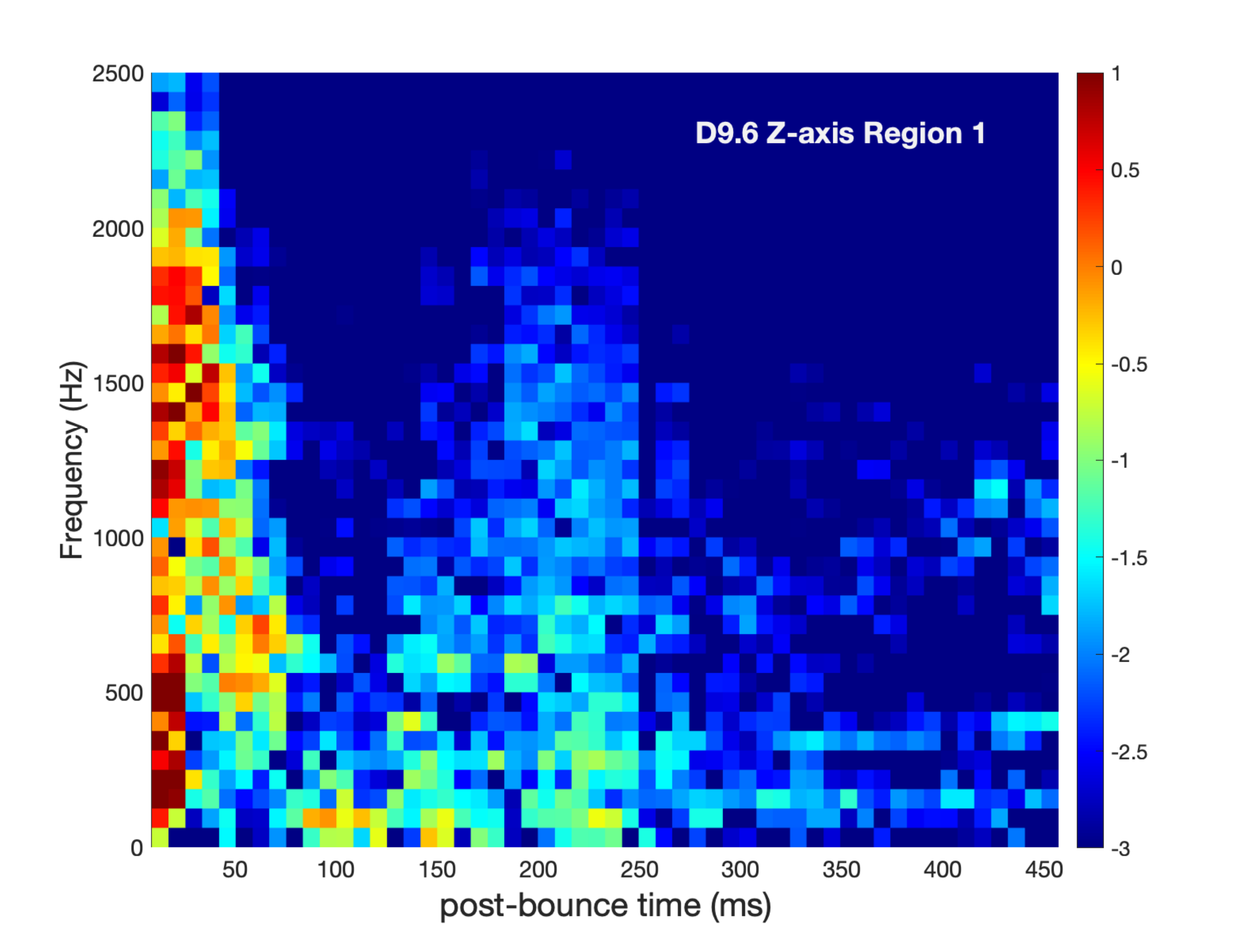}\\
\includegraphics[width=0.5\textwidth]{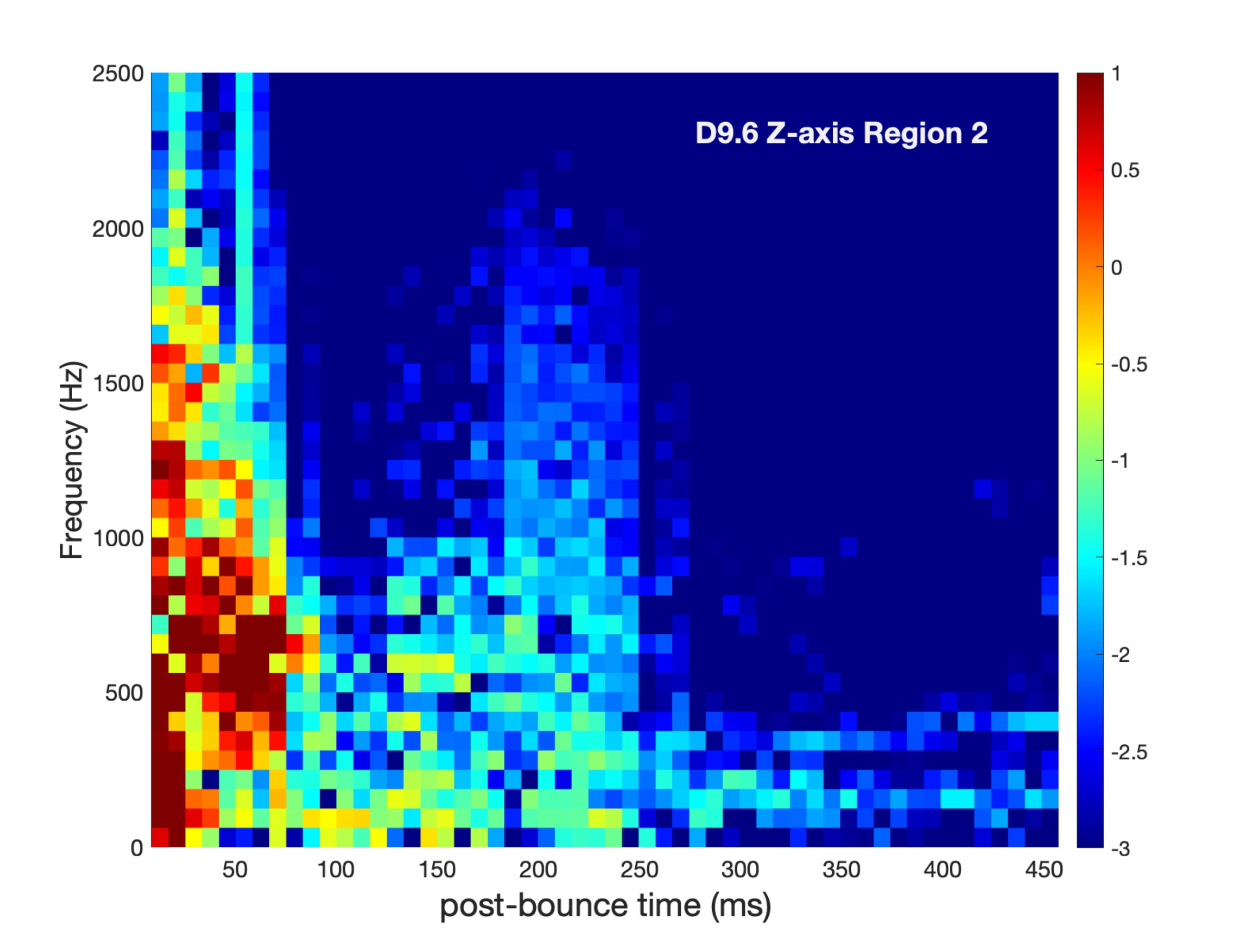}\hfill
\includegraphics[width=0.5\textwidth]{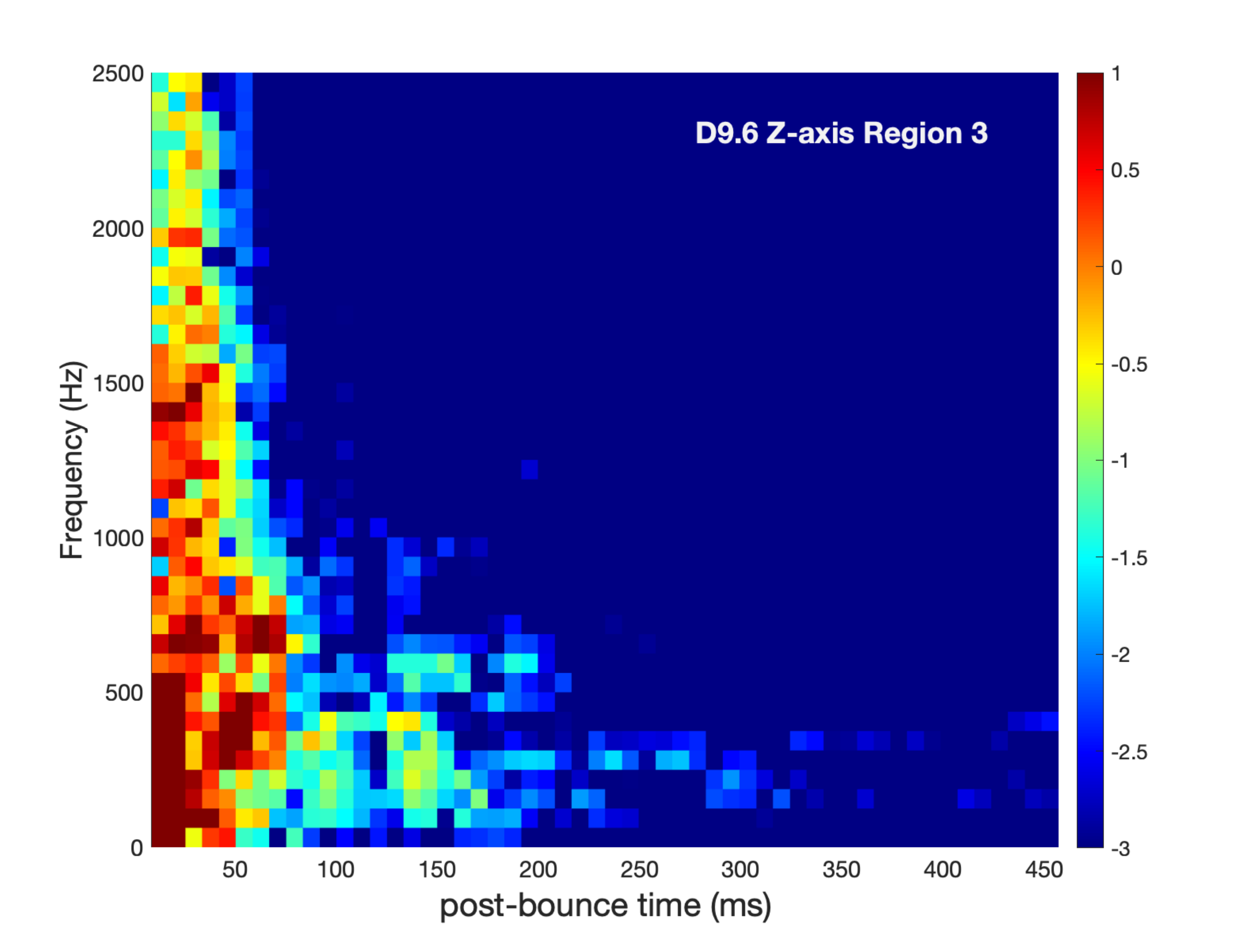}\\
\includegraphics[width=0.5\textwidth]{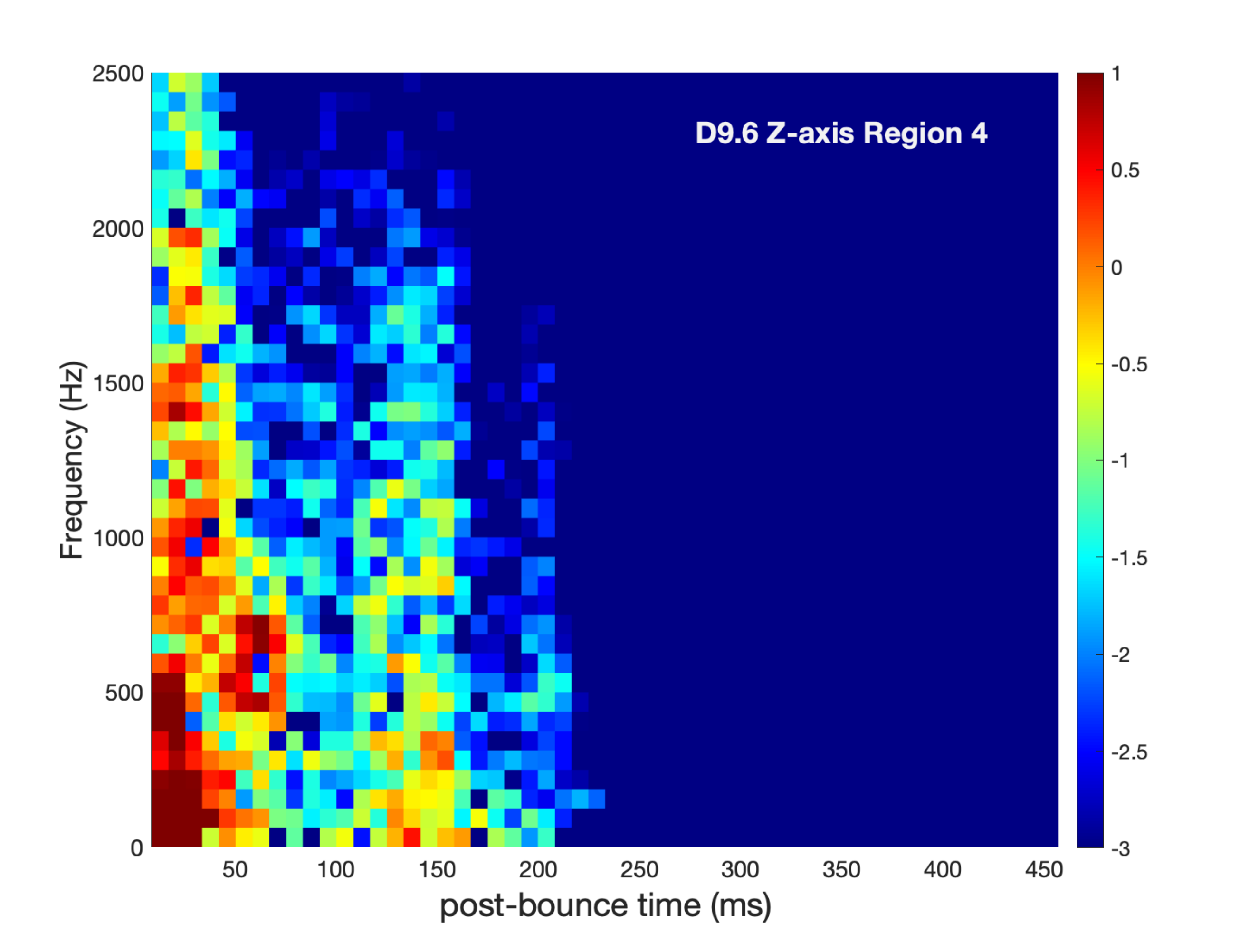}\hfill
\includegraphics[width=0.5\textwidth]{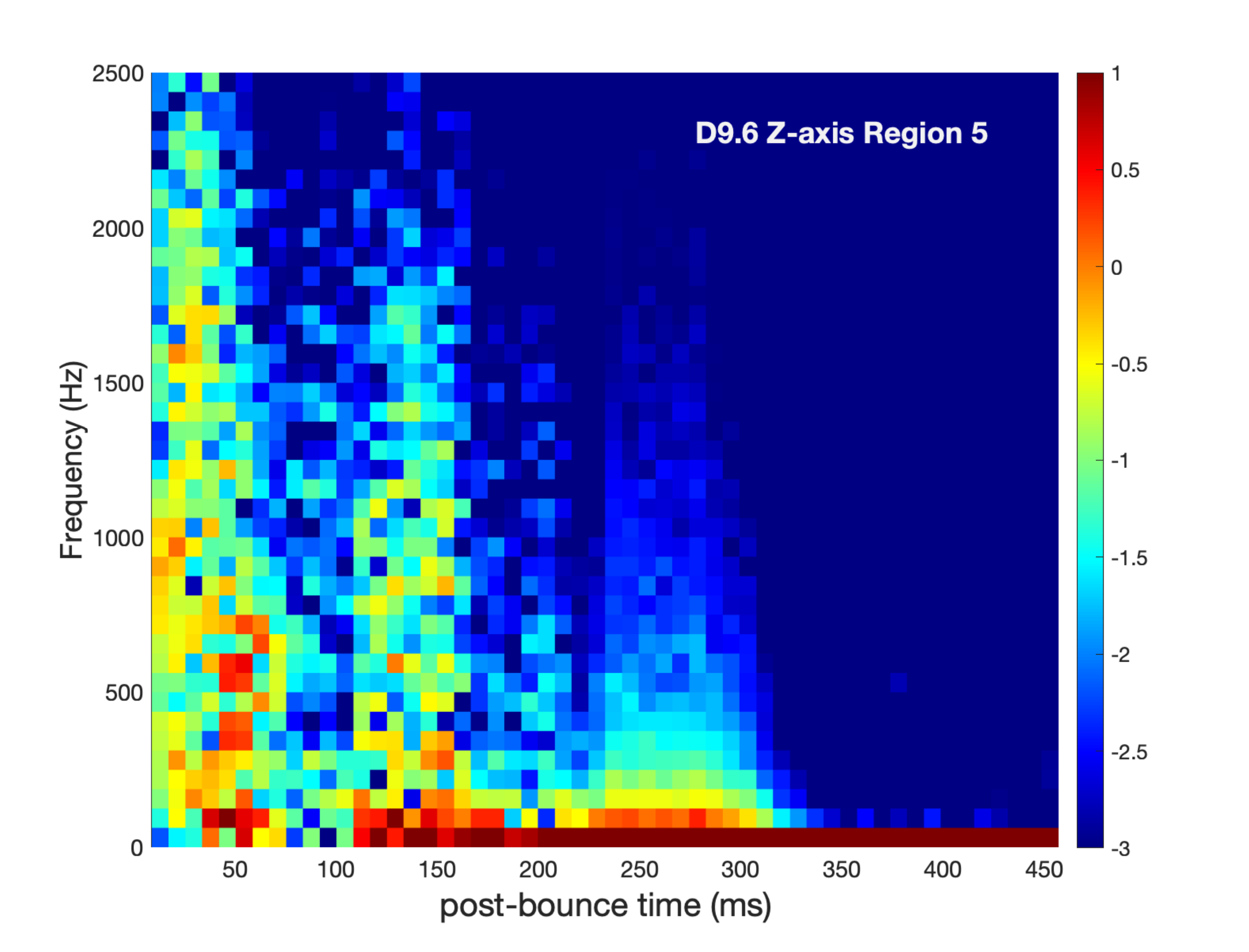}
\caption{Heat maps for Regions 1 through 5 for D9.6.}
\label{fig:heatmaps96}
\end{figure*}

\begin{figure*}
\includegraphics[width=0.5\textwidth]{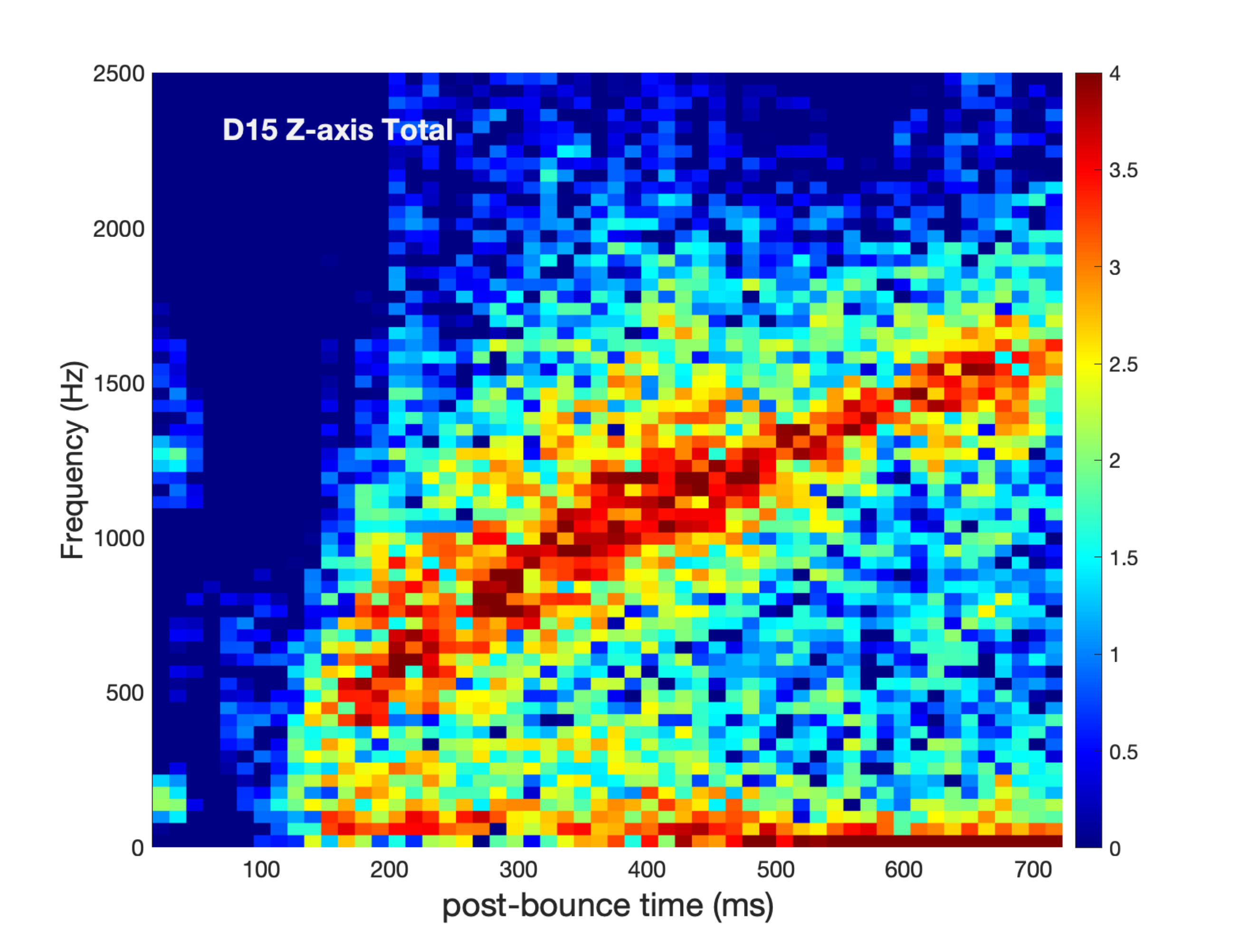}\hfill
\includegraphics[width=0.5\textwidth]{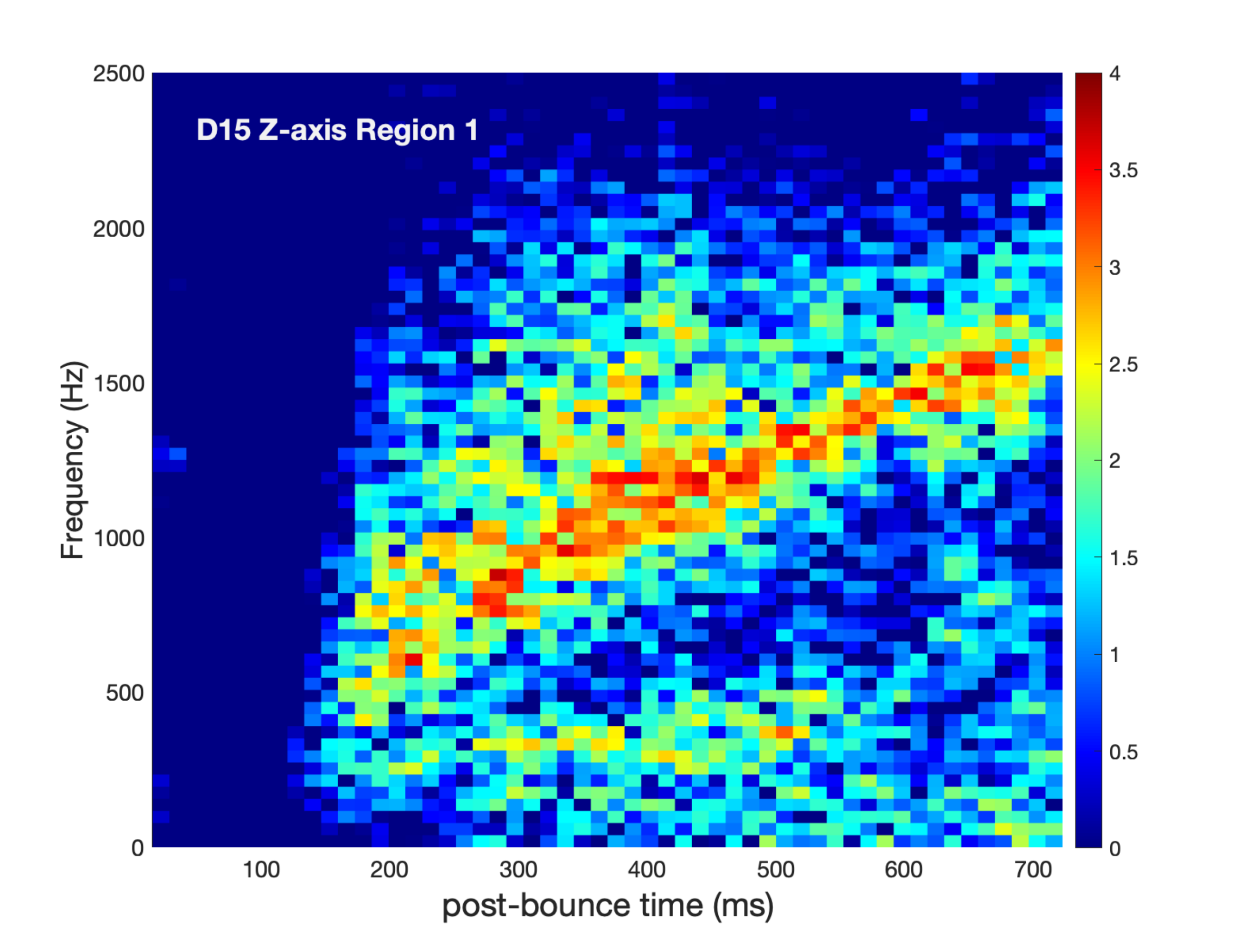}\\
\includegraphics[width=0.5\textwidth]{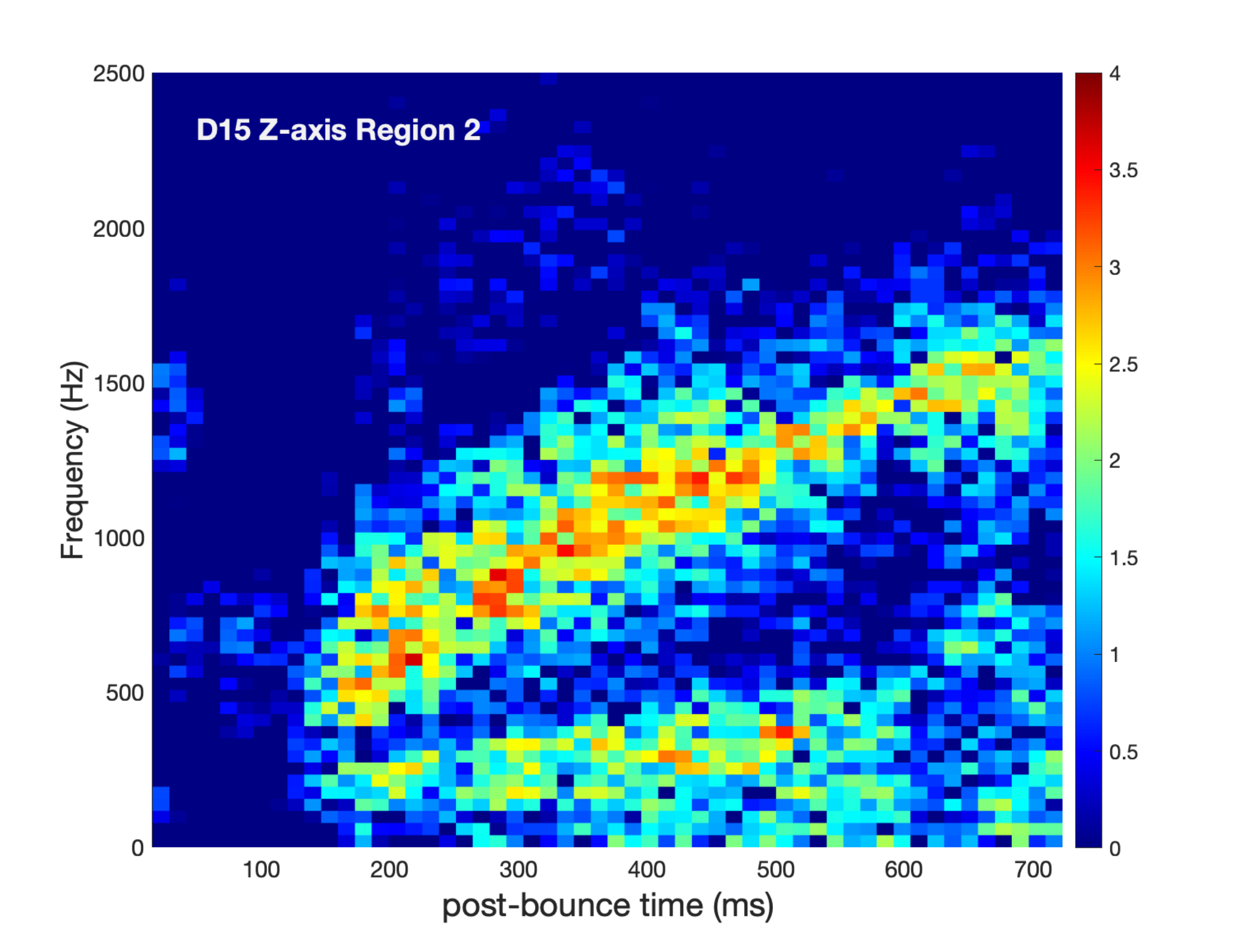}\hfill
\includegraphics[width=0.5\textwidth]{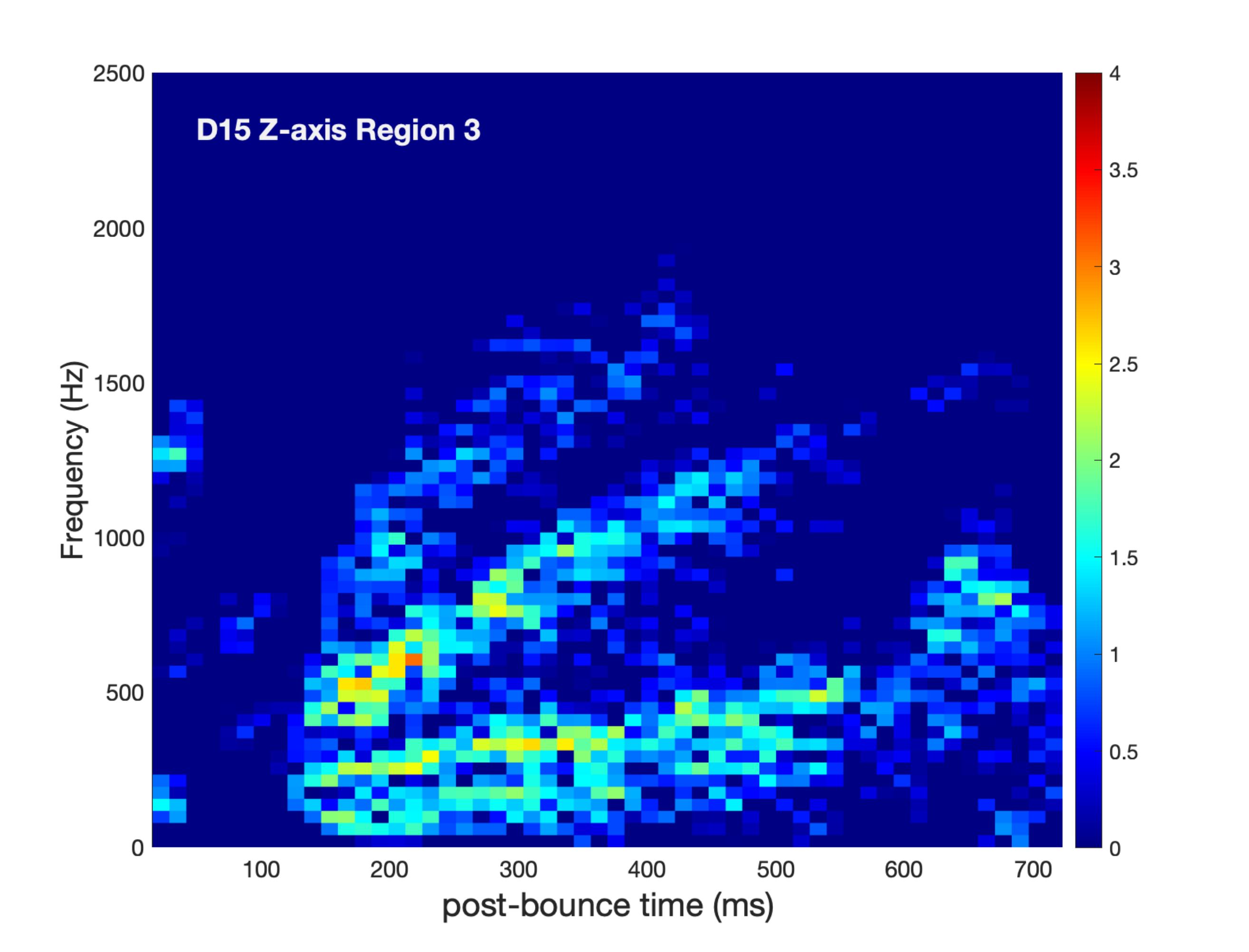}\\
\includegraphics[width=0.5\textwidth]{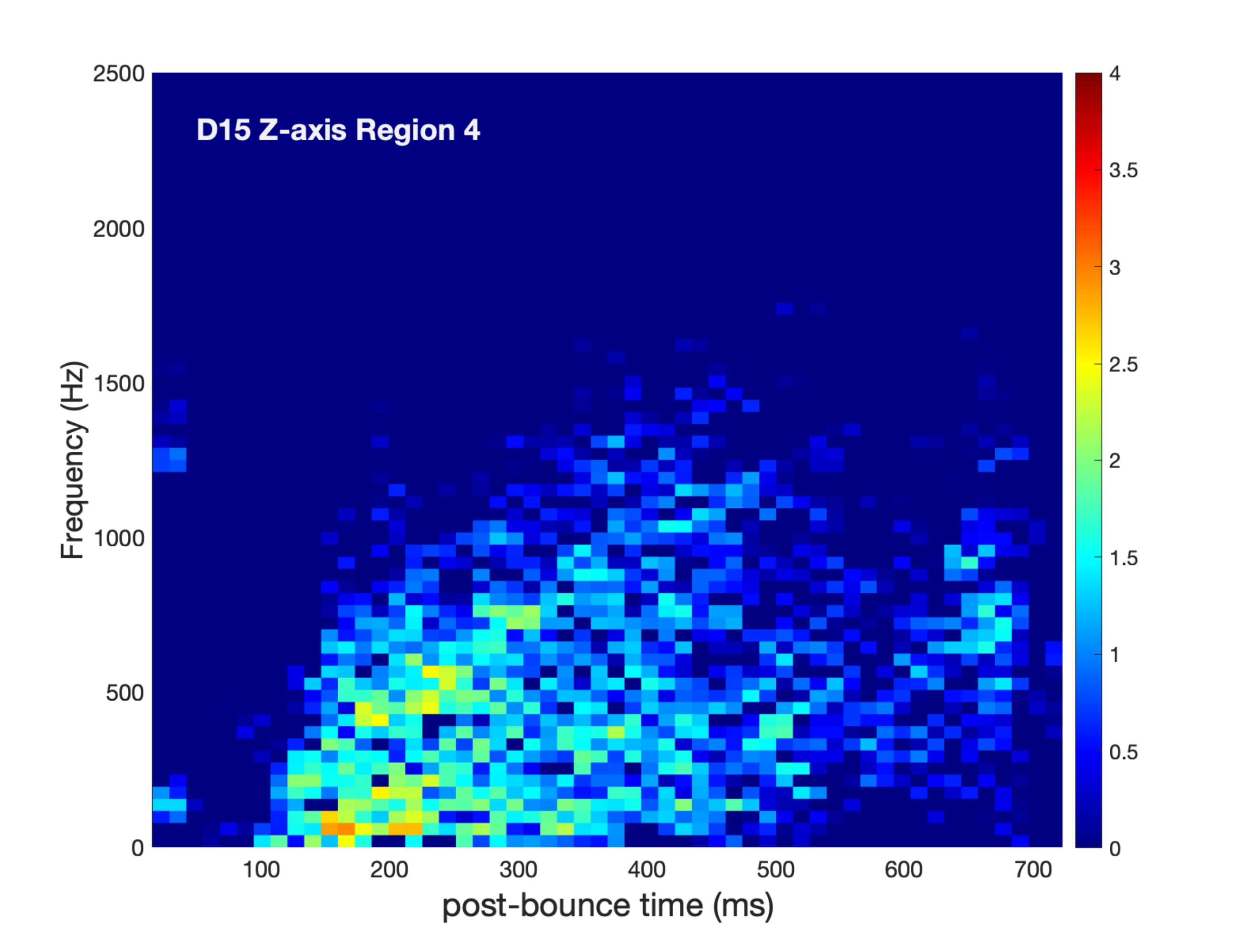}\hfill
\includegraphics[width=0.5\textwidth]{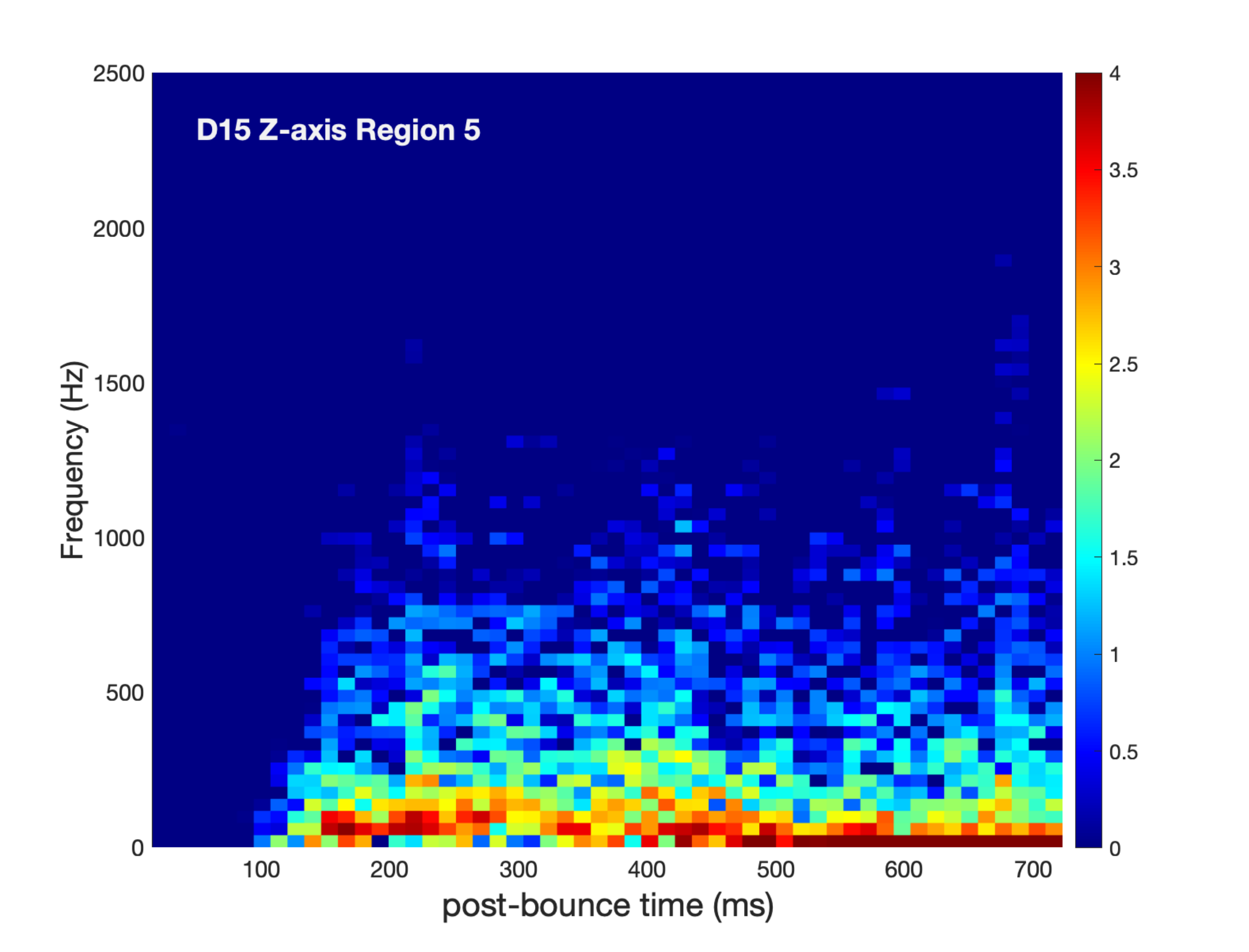}
\caption{Heat maps for Regions 1 through 5 for D15.}
\label{fig:heatmaps15}
\end{figure*}

\begin{figure*}
\includegraphics[width=0.5\textwidth]{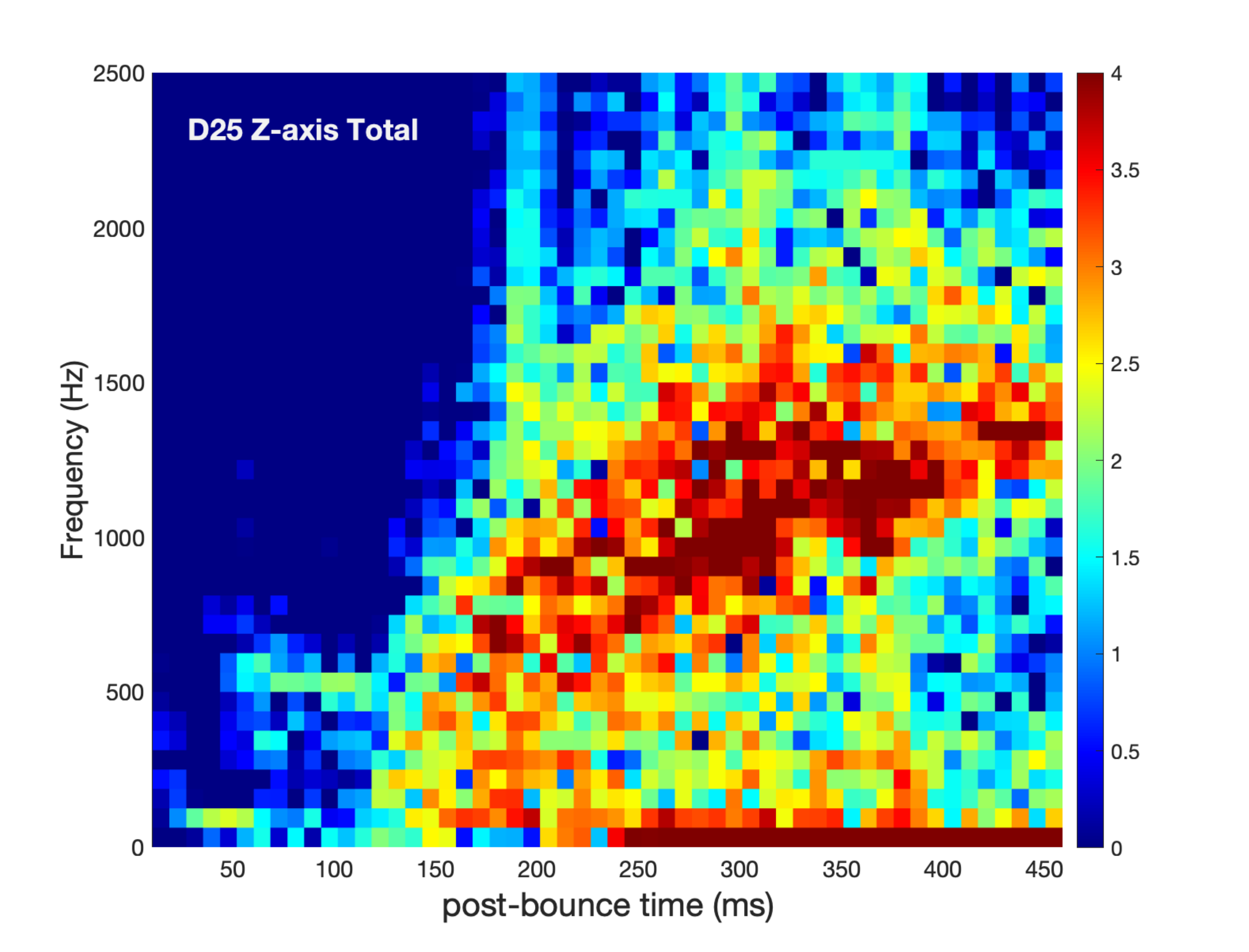}\hfill
\includegraphics[width=0.5\textwidth]{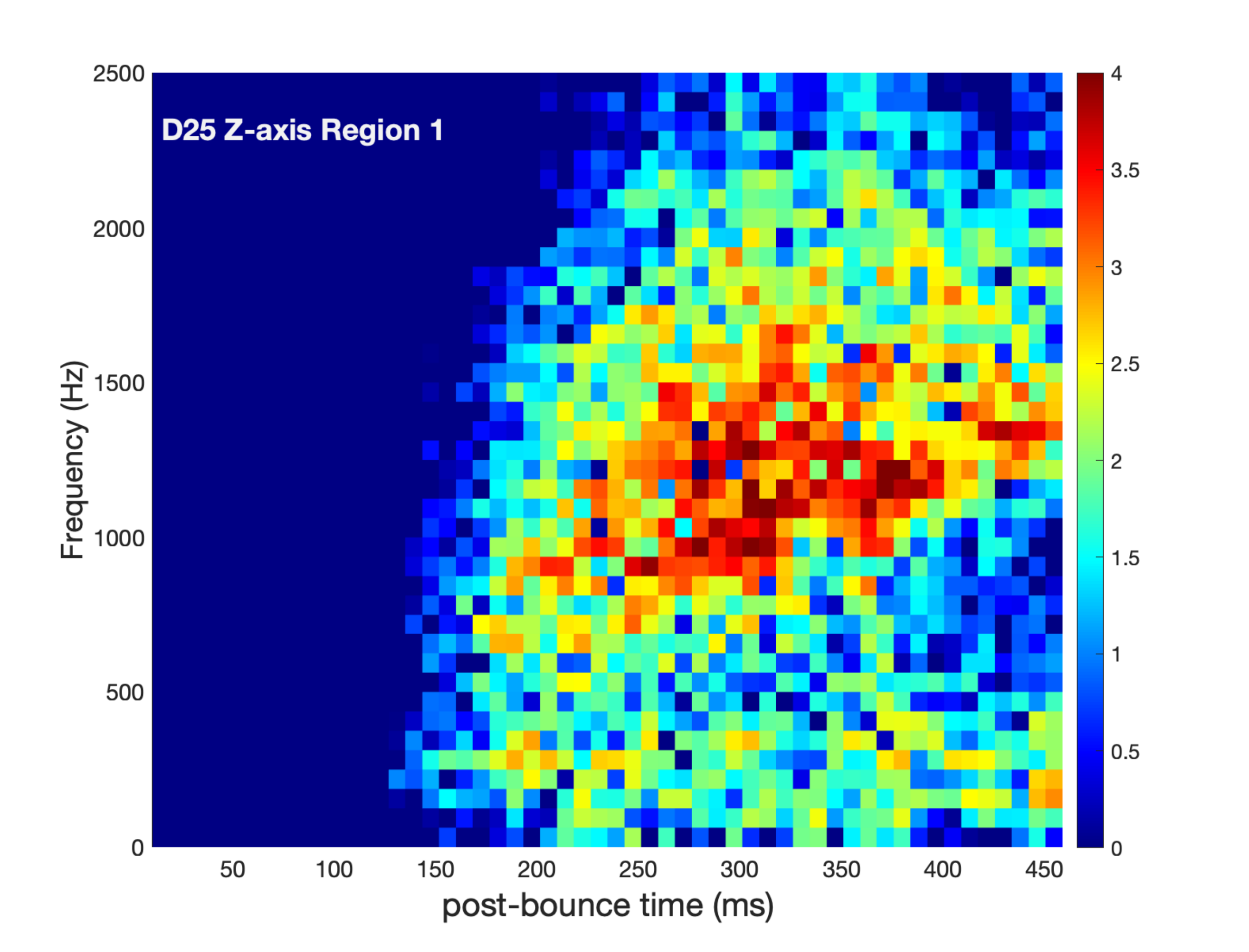}\\
\includegraphics[width=0.5\textwidth]{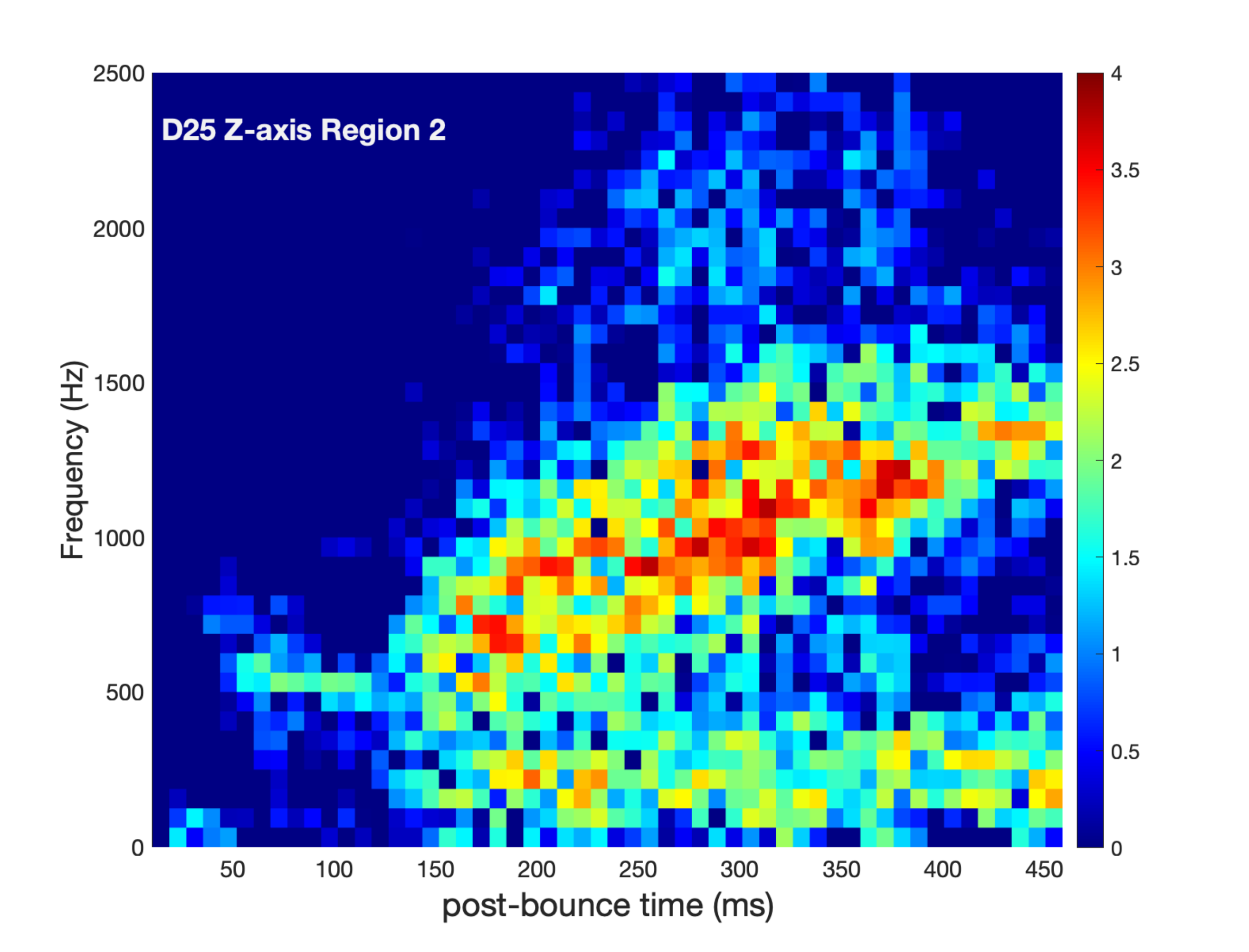}\hfill
\includegraphics[width=0.5\textwidth]{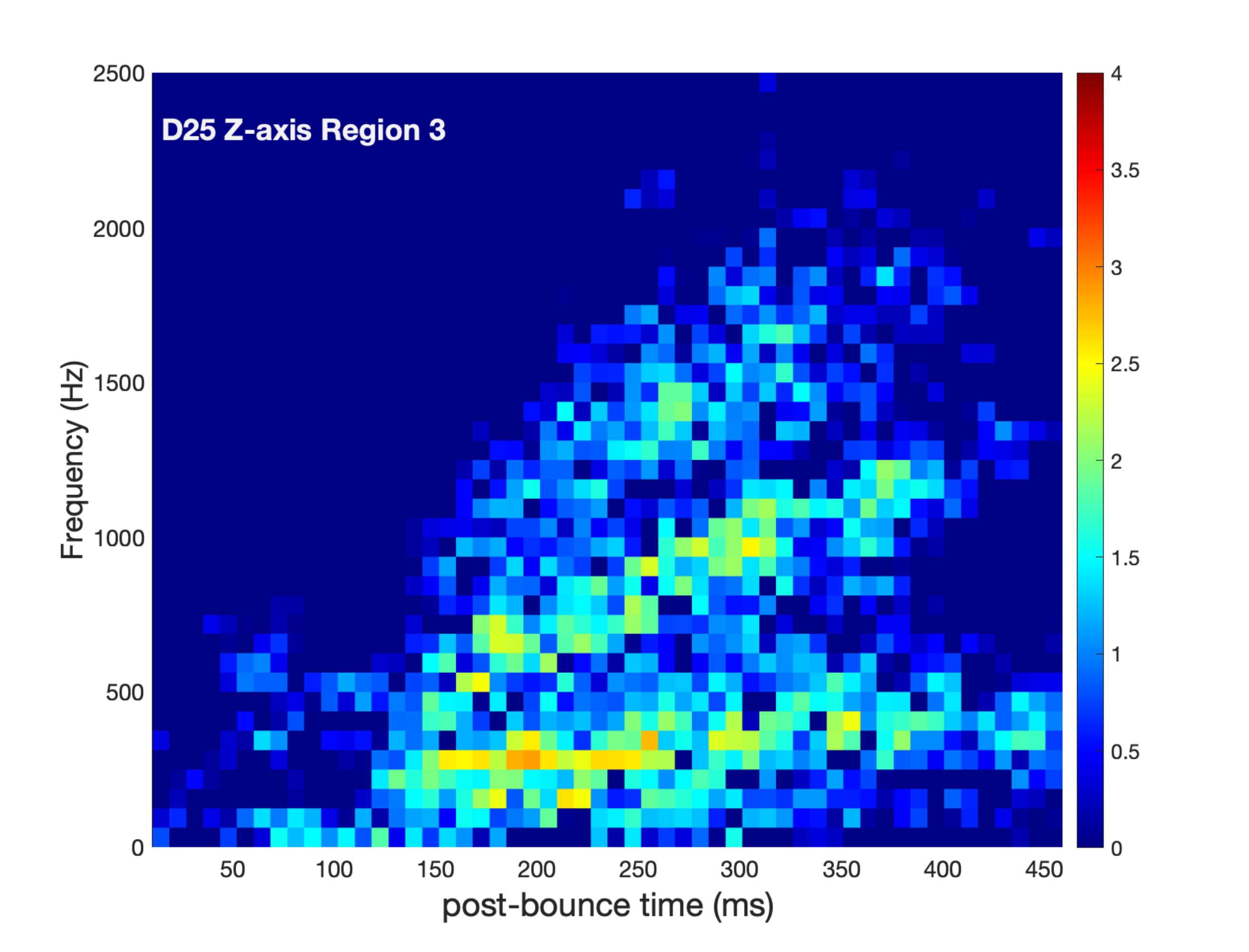}\\
\includegraphics[width=0.5\textwidth]{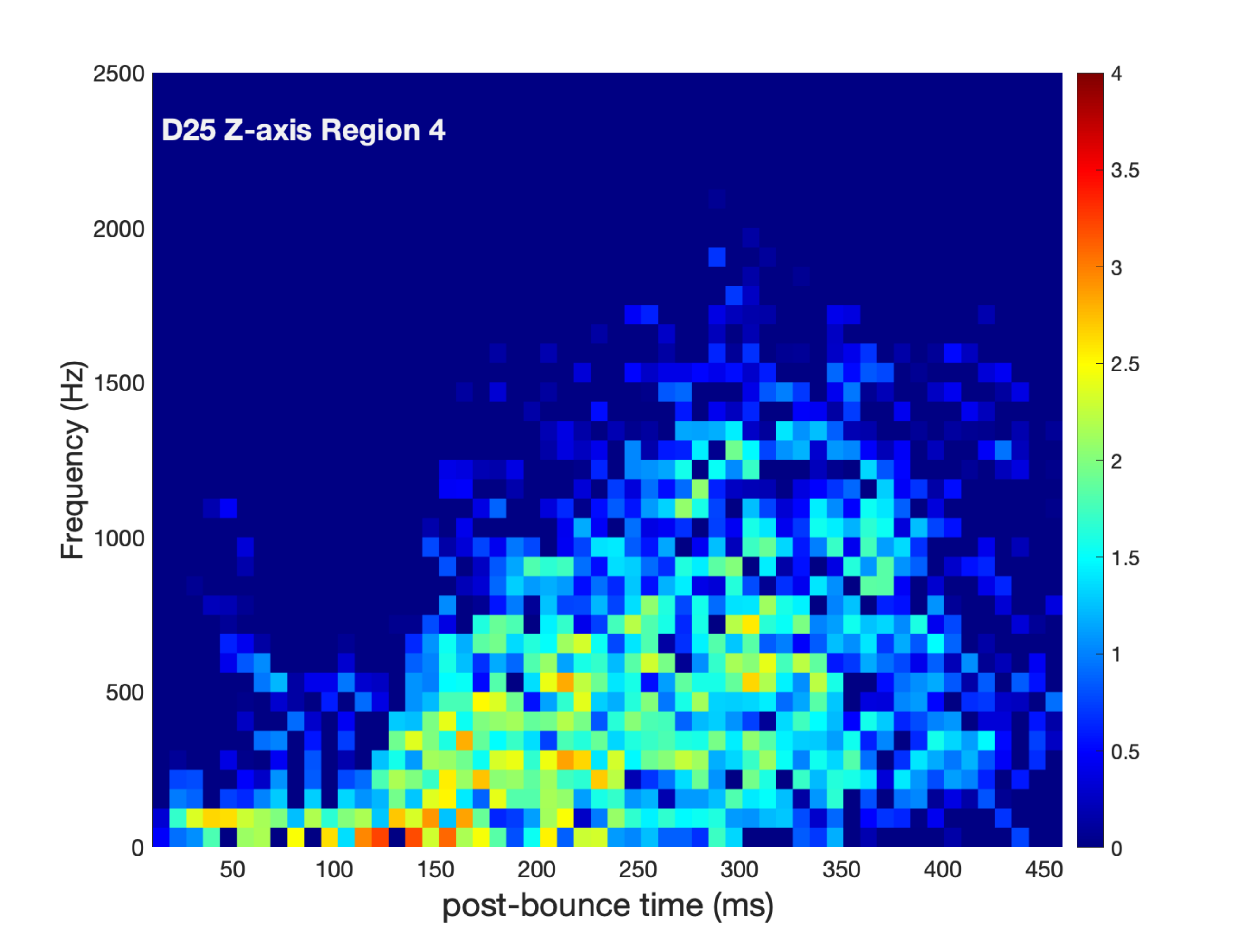}\hfill
\includegraphics[width=0.5\textwidth]{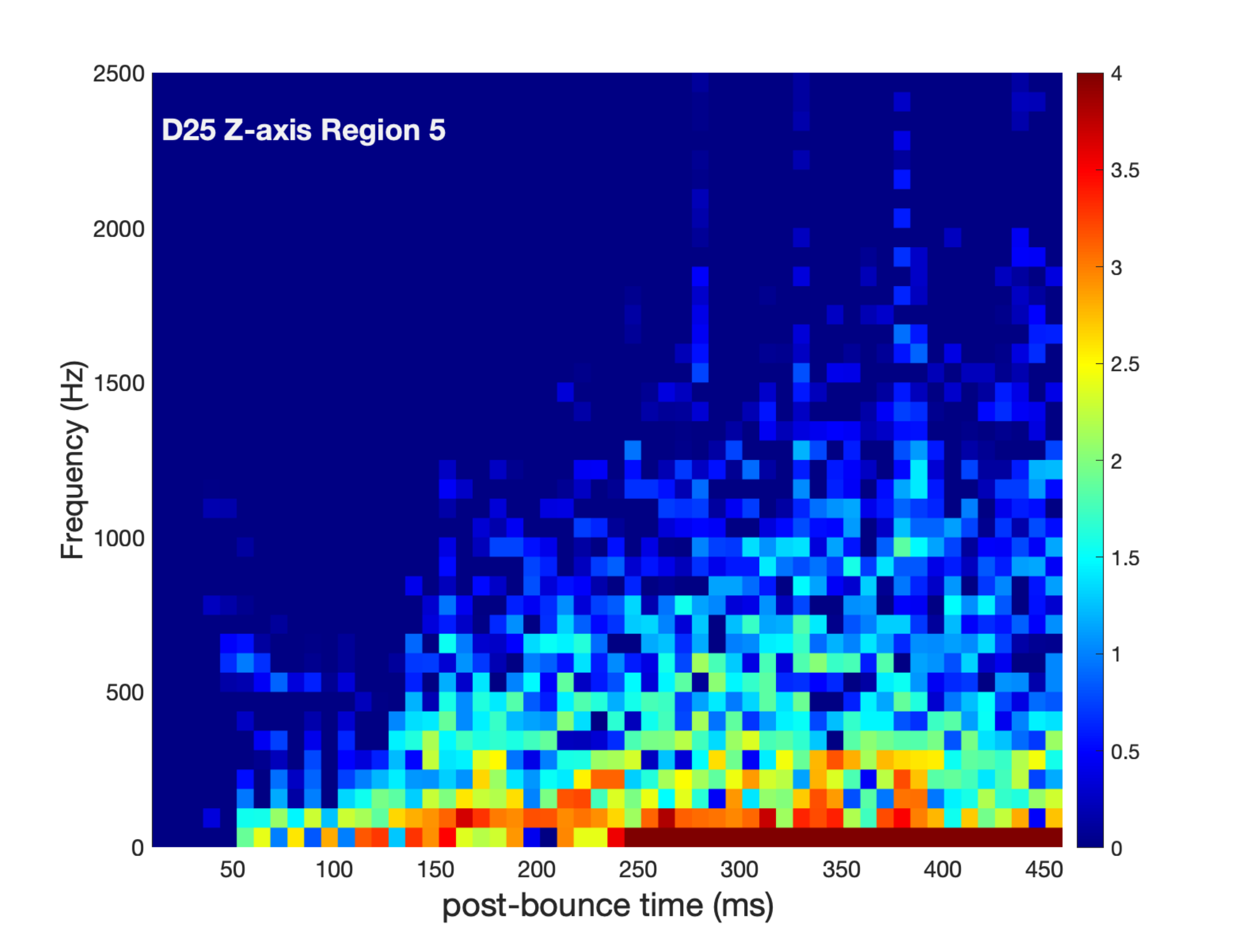}
\caption{Heat maps for Regions 1 through 5 for D25.}
\label{fig:heatmaps25}
\end{figure*}

The ``heat maps,'' Figures \ref{fig:heatmaps96}, \ref{fig:heatmaps15}, and \ref{fig:heatmaps25}, plot the binned gravitational wave emission 
spectrum as a function of time for each of the five regions discussed above, across all three of our models. As in the case of the evolution of the 
gravitational wave strains, the distinct nature of the gravitational wave emission from D9.6 relative to D15 and D25 is also evident here. 

In the case of D9.6, the heat map is particularly informative, especially after the first 20 ms post bounce, when the gravitational wave strain 
amplitudes are greatly reduced:

(1) The early stochastic gravitational wave emission, from Regions 1 through 5 and extending up to very high frequencies of $\sim$2000 Hz, is seeded initially 
by two episodes of convection in the proto-neutron star, one in Regions 1 and 2 and the other in Region 3, in conjunction with perturbations introduced throughout 
the core by nuclear burning during collapse. These perturbations lead to far more stochastic gravitational wave emission than would be caused by convection alone 
in an otherwise initially spherical medium (as in D15 and D25). While the gravitational wave strains are greatly reduced after only 20 ms post bounce, 
low-amplitude, high-frequency emission persists, visible here until $\sim$75 ms post bounce, and continuing for the duration of the run as the proto-neutron star continues 
to convect -- the result of continued deleptonization by neutrino diffusion (see Figure \ref{fig:electronfractionslice}).

(2) At $\sim$125 ms after bounce, Regions 4 and 5 give rise to stochastic gravitational wave emission up to frequencies $\sim$1200 Hz, as the explosion
powers up and neutrino-driven convection and SASI activity increase.

(3) There is low-frequency gravitational wave emission at frequencies below $\sim$250 Hz, which fundamentally changes character at $\sim$300 ms after 
bounce. At this point, the gain region is replaced by a neutrino-driven wind (see Figure \ref{fig:regionboundaries}), indicative of the fact we are in a post-explosion 
epoch, and emission from neutrino-driven convection and the SASI ceases. After 300 ms, the only (very) low-frequency emission is the result of explosion itself.
Evidence of the SASI presents itself after 100 ms post bounce, at frequencies of $\sim$100 Hz, twice the frequency of the $\ell=1$ SASI modes (see the 
discussion of Figure \ref{fig:com}). Other SASI modes -- e.g., the $\ell=2$ mode -- also contribute to the emission during this time. Overall, 
the SASI emission exhibits some intermittency prior to $\sim$300 ms after bounce, with two periods lasting of $\sim$150 ms separated by a period of 
little gravitational wave emission of $\sim$50 ms.

D15 and D25 exhibit the features documented in \cite{MeMaLa20}. For D15, the heat map 
reveals there is very little gravitational wave emission, at any frequency, until $\sim$150 ms, after which the emission is characterized by two 
primary features: (1) high-frequency emission above $\sim$500 Hz, with peak frequency increasing with time and (2) low-frequency 
emission below $\sim$250 Hz with multiple peak frequencies, each corresponding to an important time period in the model. The high-frequency 
feature corresponds to gravitational wave emission from Ledoux convection and convective overshoot, in Regions 1 and 2 of the proto-neutron 
star, respectively. 
As the proto-neutron star contracts with post-bounce time, the peak frequency of the gravitational wave emission from these regions 
increases. The low-frequency emission leads off with peak emission between $\sim$40 Hz and $\sim$80 Hz. This is consistent with strong SASI activity 
from the $\ell=1$ SASI modes. (See the plots of $\tilde{h}_{+}(f)$ shown in Figure \ref{fig:R5SpectraPreE} and the associated 
discussion.) SASI activity associated with multiple of its modes persists throughout the majority of our run. Contributions to the low-frequency 
emission from neutrino-driven convection are also present throughout. As the explosion powers up, the low-frequency emission becomes increasingly 
stochastic, as can be seen at a post-bounce time of $\sim$400 ms.
The heat map for D25 is similar in most respects to the heat map for D15, with the exception that 
the high-frequency emission remains more stochastic throughout our run as the result of the different and more significant mass accretion history in this 
model. 
At low frequencies, the emission is initiated between 100 and 150 ms post bounce and is consistent with emission associated with convective and 
$\ell=2$ SASI activity, with peak frequency below $\sim$60 Hz. (Again, see the plots of $\tilde{h}_{+}(f)$ shown in Figure \ref{fig:R5SpectraPreE} and 
the associated discussion.) This initial phase of low-frequency emission is followed by emission between 150 and 200 ms post bounce, at 
frequencies between $\sim$60 Hz and $\sim$125 Hz, consistent with SASI activity from its $\ell=1$ modes, which becomes more stochastic as the 
explosion powers up prior to the explosion time of $\sim$250 ms. SASI and convective activity then continue post explosion, becoming 
increasingly stochastic, until $\sim$400 ms.

\begin{figure*}
\includegraphics[width=\columnwidth,clip]{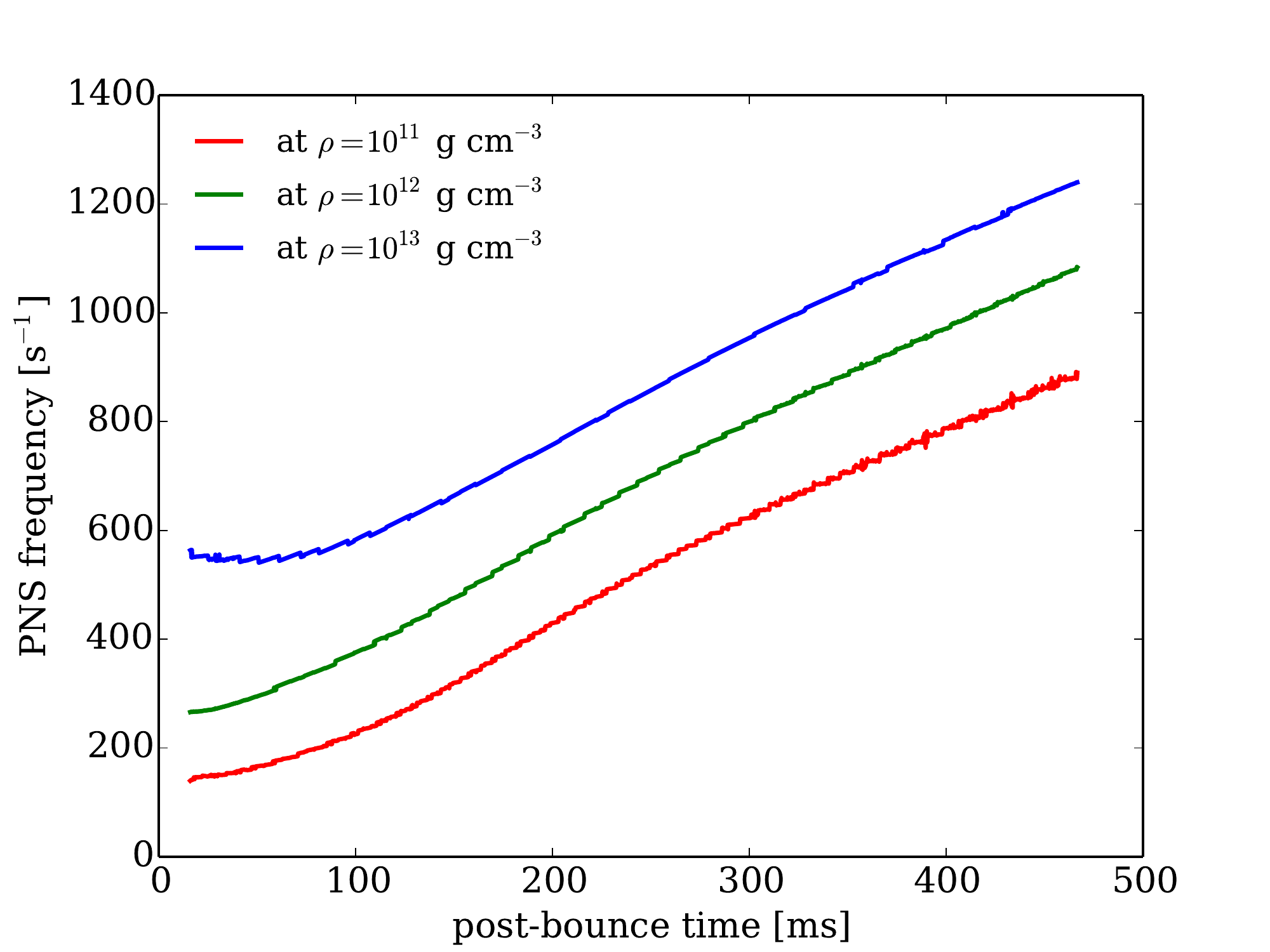}
\includegraphics[width=\columnwidth,clip]{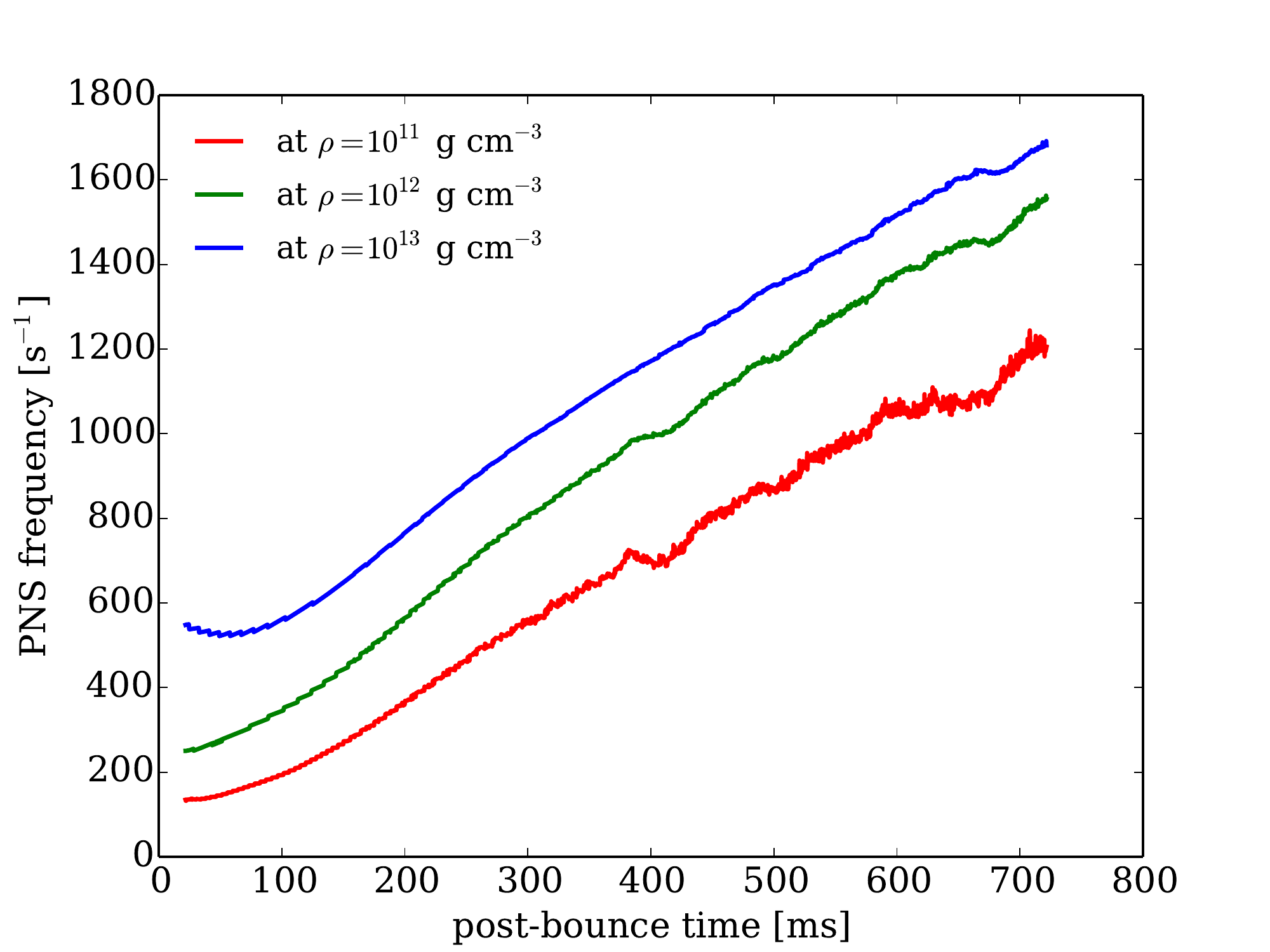}
\includegraphics[width=\columnwidth,clip]{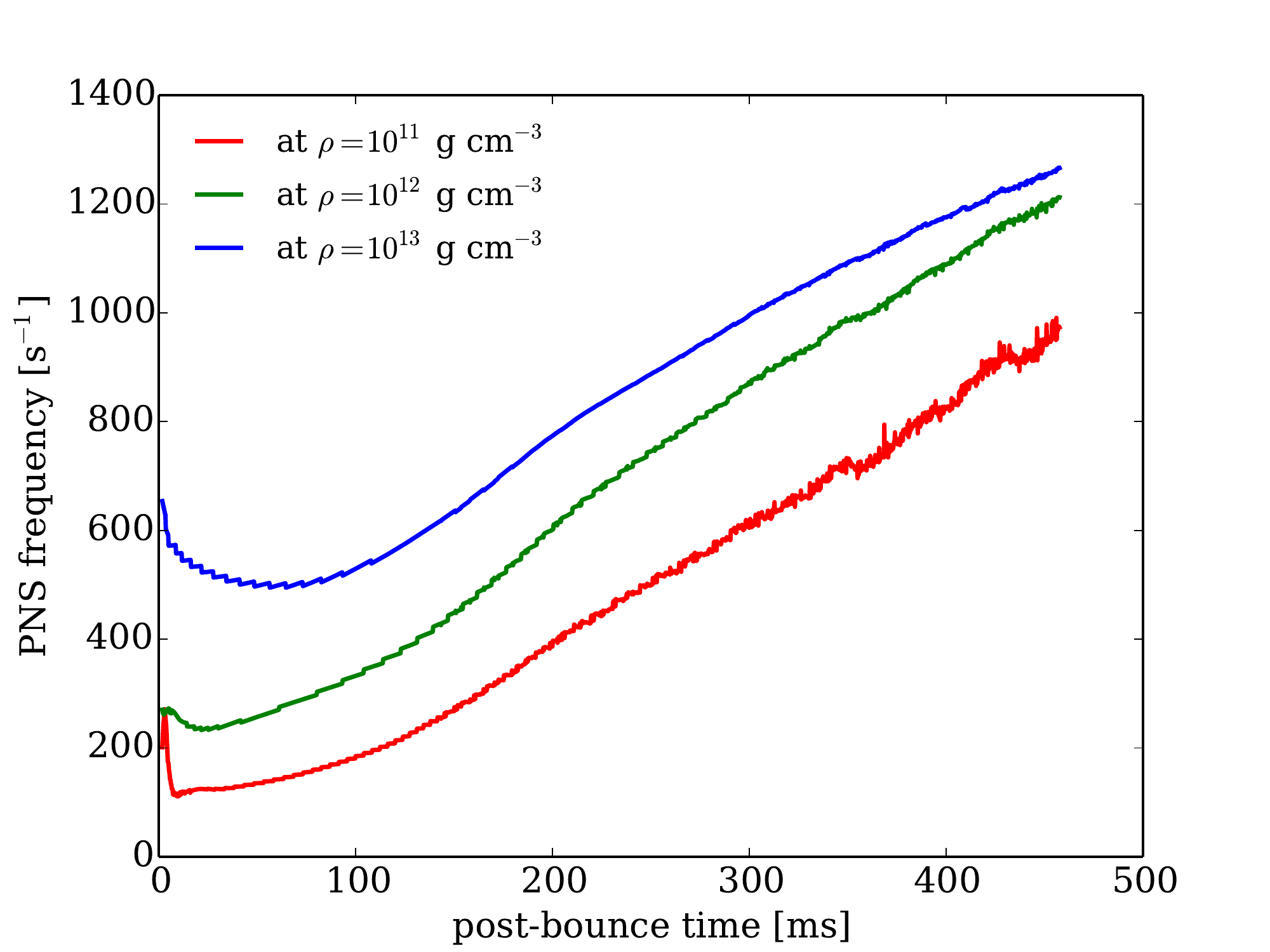}
\caption{Peak frequency, determined by Equation (\ref{eq:peakfrequency2}), as a function of post-bounce time for D9.6, D15, and D25, respectively.}
\label{fig:peakfreqfits}
\end{figure*}

\begin{figure*}
\includegraphics[width=\columnwidth,clip]{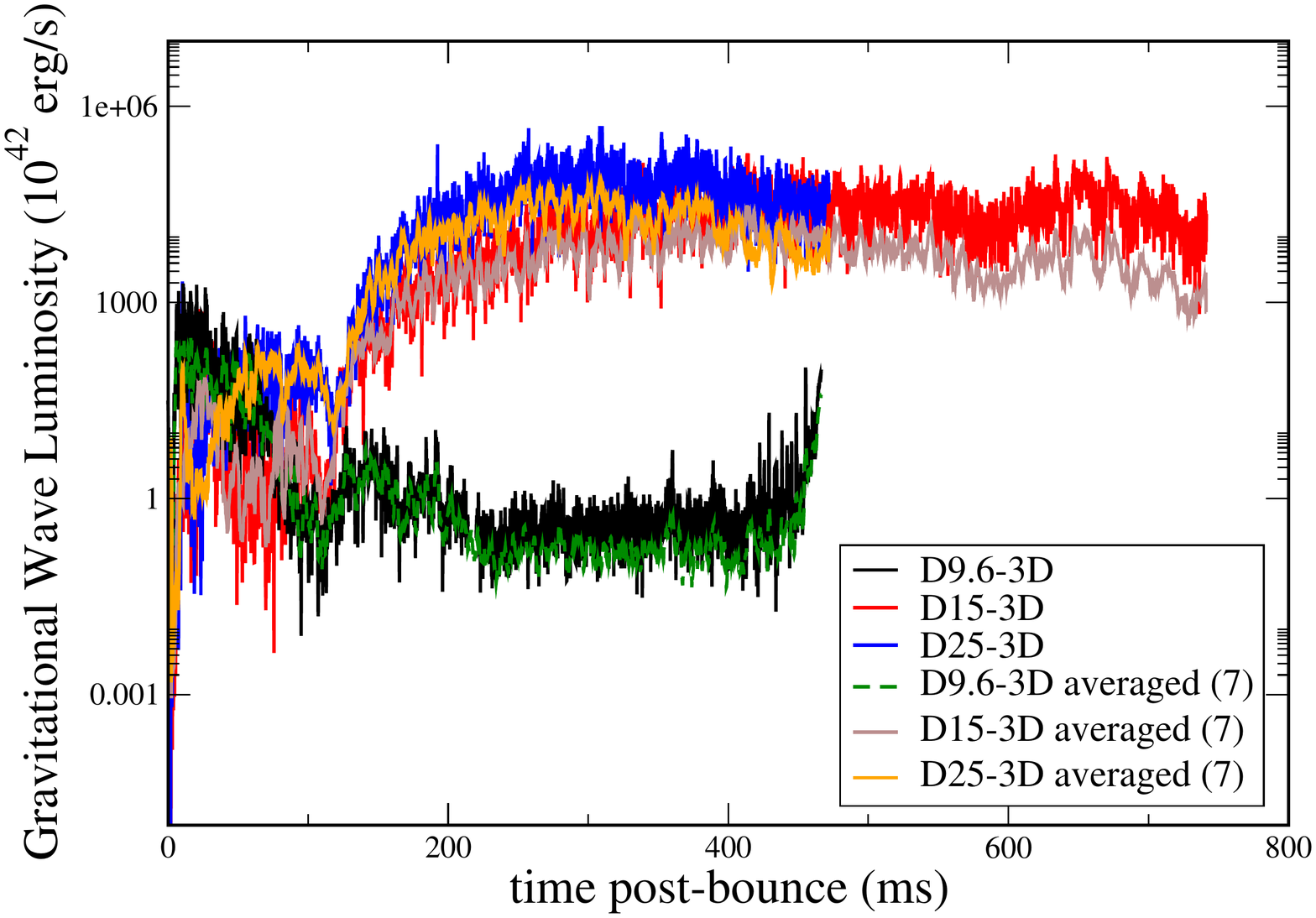}
\includegraphics[width=\columnwidth,clip]{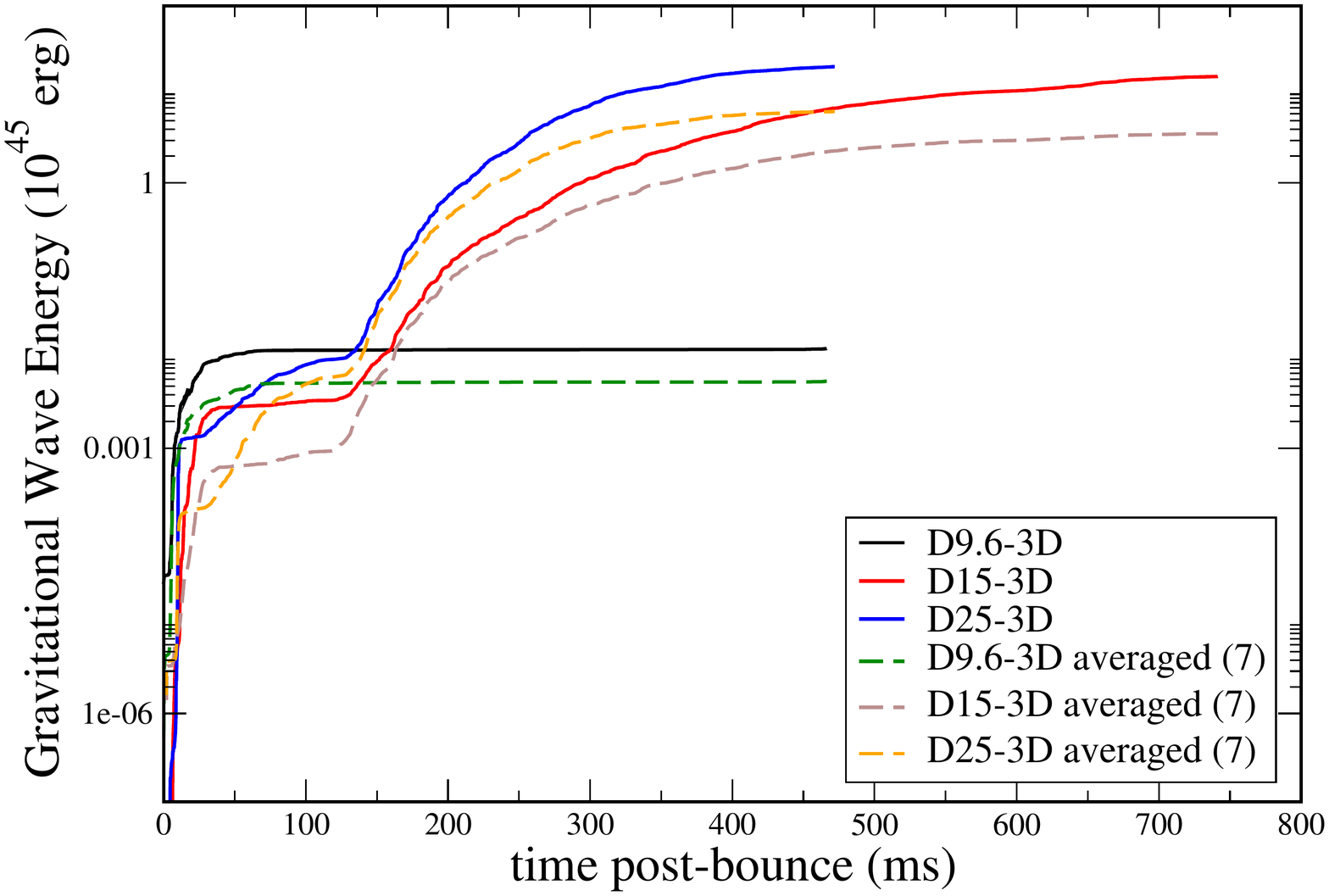}
\includegraphics[width=\columnwidth,clip]{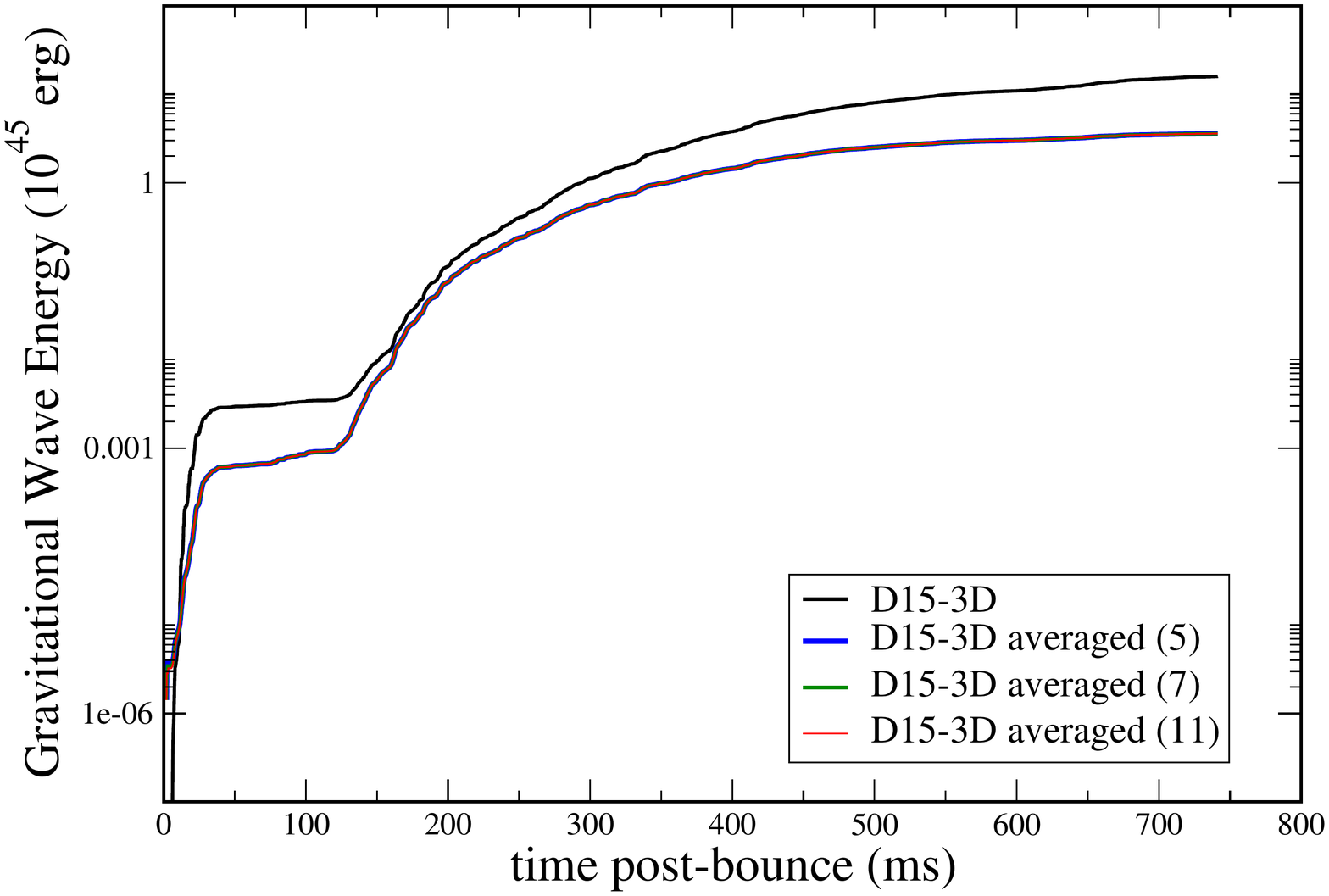}
\caption{Gravitational wave luminosity and total energy emitted for all three models.}
\label{fig:gwenergyluminosity}
\end{figure*}

While the above discussion captures the primary features of the gravitational wave emission, there is further information contained within 
the heat maps that should be mined. For example, looking at the heat map for Region 2 of D15, we see 
a notable decrease in the high-frequency emission after the onset of explosion at $\sim$500 ms, whereas Region 1 is less impacted. 
This demonstrates that, while the high-frequency emission in this model is dominated by convection and convective overshoot in 
Regions 1 and 2, some of the high-frequency emission prior to explosion is excited from above, by accretion funnels impinging 
on the surface regions of the proto-neutron star. This also explains why the high-frequency emission is more stochastic in the case
of D25, given its different and more significant accretion history.

In all three models, explosion is represented by the very low frequency emission below $\sim$10 Hz, beginning at $\sim$150 ms, $\sim$500 ms, and $\sim$250 ms 
after bounce for D9.6, D15, and D25, respectively. This correlates well with the development of the offsets in the strain amplitudes themselves, 
shown in Figures \ref{fig:strains} and \ref{fig:strainsbyregion}, another marker of explosion.

\subsection{Peak Frequency Evolution}

M\"{u}ller et al. \cite{MuJaMa13} derived a formula relating the peak frequency of (high-frequency) gravitational wave emission to, among other things,
the mass and radius of the proto-neutron star:

\begin{equation}
f_{p}=\frac{1}{2\pi}\frac{GM}{R^{2}}\frac{1}{c_s}\sqrt{\Gamma -1}(1-\frac{GM}{Rc^{2}})^{3/2}
\label{eq:peakfrequency}
\end{equation}
Here, $M$, $R$, $c_s$, and $\Gamma$ are the proto-neutron star mass, proto-neutron star radius, sound speed, and adiabatic index, respectively.
In the weak-field limit, we can write the spacetime metric as

\begin{equation}
ds^2=-(1-\frac{2GM}{Rc^2})dt^2+(1+\frac{2GM}{Rc^2})dr^2+r^2(d\theta^2+\sin^2\theta d\phi^2)
\label{eq:metric}
\end{equation}
With this in mind, we can express $f_p$ in terms of $M$, $R$, and the lapse function, $\alpha$, defined as

\begin{equation}
\alpha = 1-\frac{GM}{Rc^2}
\label{eq:lapse}
\end{equation}
to get

\begin{equation}
f_{p}=\frac{1}{2\pi}\frac{GM}{R^{2}}\frac{1}{c_s}\sqrt{\Gamma -1}\alpha^{3/2}.
\label{eq:peakfrequency2}
\end{equation}
$\Gamma$ can be determined by inverting the definition of the sound speed:

\begin{equation}
\Gamma=\frac{\rho c_s^2}{P}.
\label{eq:soundspeed}
\end{equation}
To find $f_p$ at any instant of time during the course of our simulation, we evaluate the right-hand side of equation (\ref{eq:peakfrequency2}) 
along three different constant-density contours at $\rho=10^{11, 12, 13}$ \gcc. The input to equation (\ref{eq:peakfrequency2}) is obtained 
by computing an angular average of the relevant quantities over a radial shell.

Comparing Figure \ref{fig:peakfreqfits} with Figures \ref{fig:heatmaps96}, \ref{fig:heatmaps15}, and \ref{fig:heatmaps25}, it is obvious  
the evolution of the peak frequency of the high-frequency component of our heat maps does not agree well with the evolution of the peak 
frequency as given by Equation (\ref{eq:peakfrequency2}) evaluated along the $\rho=10^{11}$ \gcc, contour, which we have
defined to be our proto-neutron star surface. For example, given the narrow-band character of the high-frequency emission for D15, 
it is easily seen that the agreement is better along the $\rho=10^{12}$ \gcc, contour. This should come as no surprise given we 
have demonstrated that Regions 1 and 2, both at densities above $\rho=10^{12}$ \gcc, provide the strongest source of high-frequency 
gravitational wave emission in this case. 
The detection of the gravitational wave signal from a Galactic core collapse supernova will present an opportunity to attempt to extract 
information about the progenitor, proto-neutron star, and high-density equation of state, but the analysis we present here suggests that 
what we can glean about the proto-neutron star will depend on further consideration of how we define it. 

\subsection{Gravitational Wave Luminosity and Total Energy}

In Figure \ref{fig:gwenergyluminosity} we plot the gravitational wave luminosity and total energy emitted in gravitational
waves as a function of time after bounce for all three of our models. The trends in the gravitational wave emission that were discussed 
above are reflected here as well. For D9.6, the gravitational wave luminosity peaks early and drops
off quickly within the first $\sim$100 ms. Concommitantly, the energy emitted in gravitational waves in this model rises rapidly 
and levels off by $\sim$50 ms after bounce. However, neither the luminosity nor the energy emitted are particularly large in this case. 
For D15 and D25, gravitational wave emission remains low until $\sim$125 ms after bounce, at 
which point the gravitational wave luminosity in both cases rises rapidly and to significant levels, orders of magnitude above
the level seen in D9.6. This sharp rise occurs in sync with the development of Ledoux instability and 
subsequent convection in the proto-neutron star. A concommitant rapid increase in the total energy emitted in gravitational 
waves in these models also occurs at this time. To gauge the maturity of the models with regard to gravitational wave emission,
we can see that the total energy emitted continues to increase at the end of both runs, but the rate of increase has slowed 
considerably, indicating we have captured the bulk of the gravitational wave emission in both models, at $\sim$500 and $\sim$750 
ms for D15 and D25, respectively. In both plots, two sets of curves are shown. In one of the two cases, 
our simulation data are averaged over 7 cycles. This was done to eliminate high-frequency noise in the emission, which in turn 
lowers the luminosities and emitted energies. The averaging was done over 5, 7, and 11 cycles. As shown in Figure 
\ref{fig:gwenergyluminosity}, the results in all three cases are insensitive to the number of cycles used (at least above 5 cycles).

\begin{figure*}
\includegraphics[width=\columnwidth,clip]{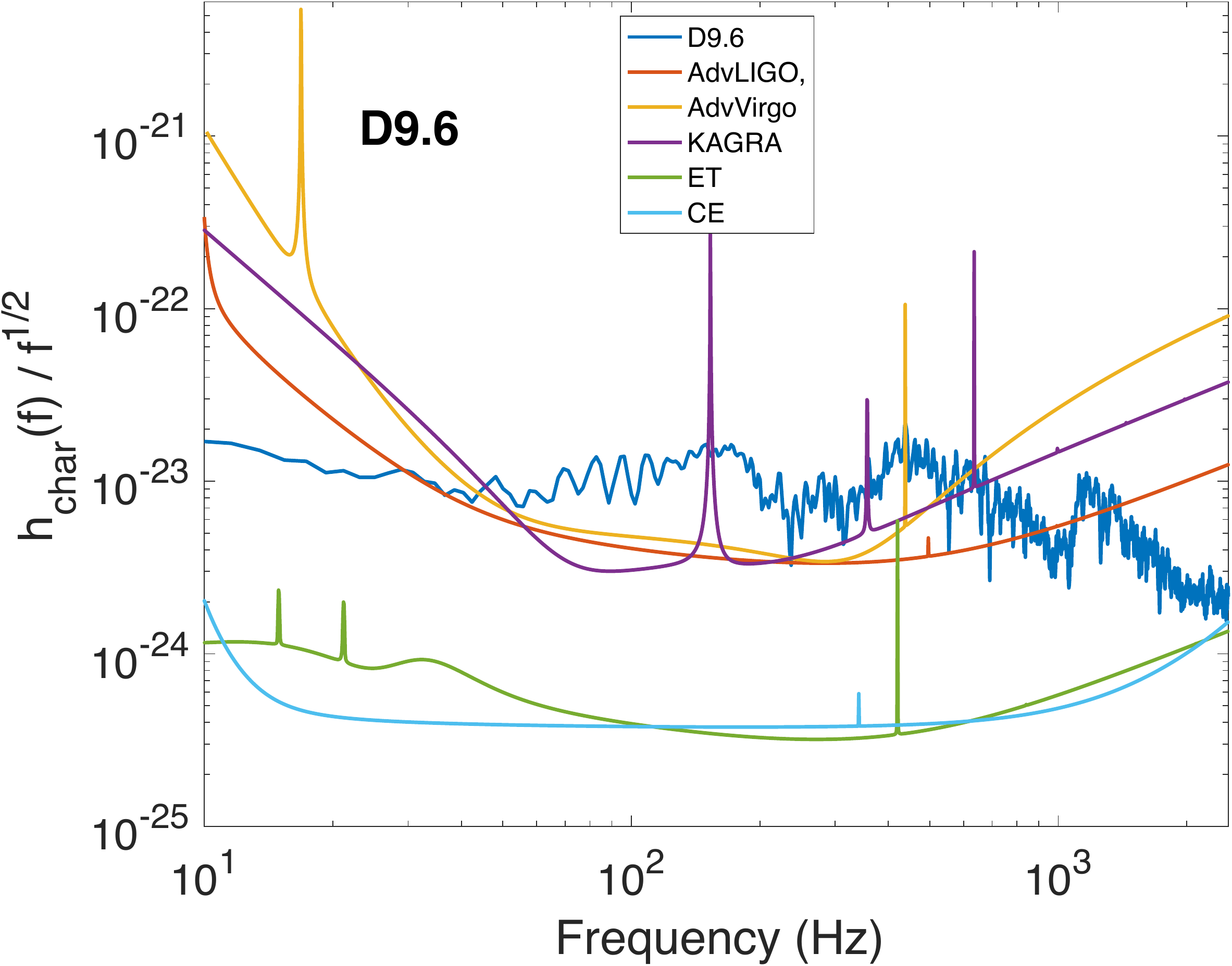}
\includegraphics[width=\columnwidth,clip]{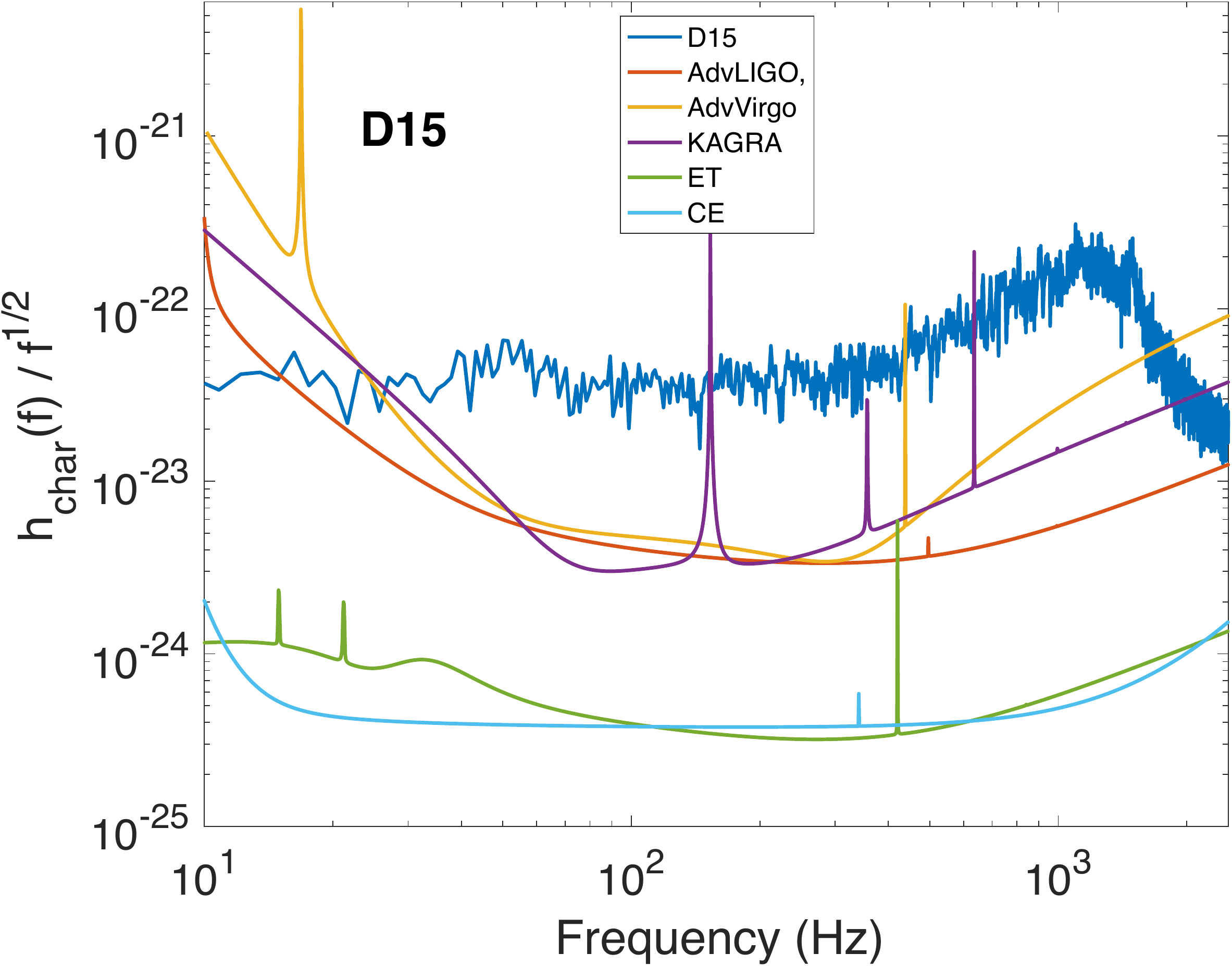}
\includegraphics[width=\columnwidth,clip]{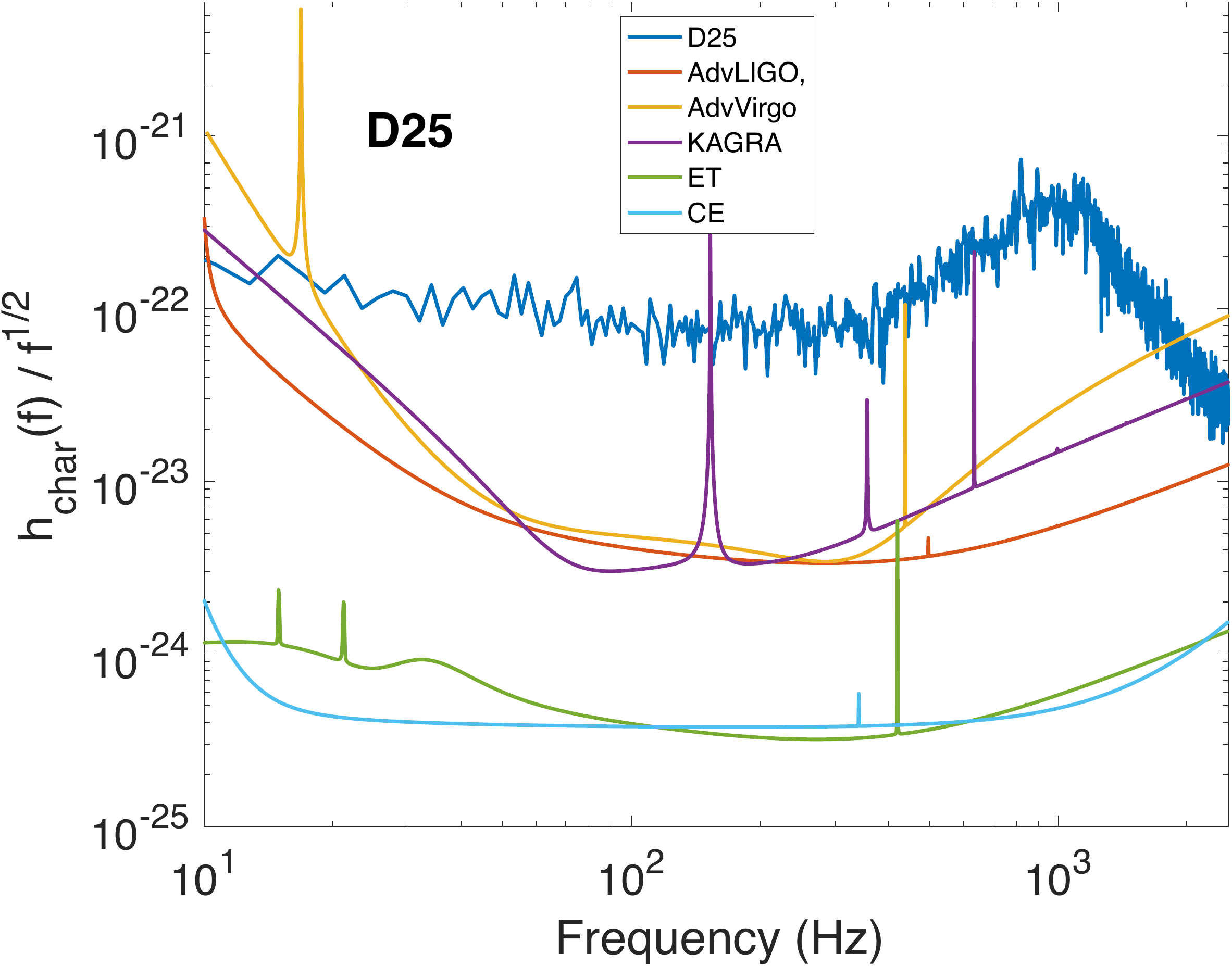}
\caption{Characteristic strains for the D9.6, D15, and D25, respectively.}
\label{fig:characteristicstrains}
\end{figure*}

\subsection{Gravitational Wave Spectra}

Moving now to the spectral classification of the results from our models, in Figure \ref{fig:characteristicstrains} we plot the 
characteristic strains as a function of frequency for all three models. In D9.6, the spectrum of gravitational radiation emission has several features between $\sim$10 Hz and $\sim$2000 Hz worth pointing 
out. There is a gradual increase in gravitational wave emission as we proceed to lower frequency from $\sim$50 Hz. There are 
several ``peaks'' in the spectrum: one between 100 and 200 Hz, another between 400 and 500 Hz, and a third between 1000 and 
2000 Hz. In D15, there is a peak in the spectrum at around 40--50 Hz, after which there is a rise to a distinct peak frequency 
just above 1000 Hz, followed by a rapid fall off. The characteristic strain reaches a magnitude in the range $3-4\times 10^{-22}$ 
in this case. For D25, the peak frequency occurs at a somewhat lower frequency, at $\sim$800 Hz, 
but with a higher amplitude, of $\sim 8 \times 10^{-22}$ . In this model, there is also a gradual increase in gravitational wave emission 
as we move to lower frequencies from $\sim$100 Hz, not present in D15. Also shown are the characteristic 
strains for both existing and planned detectors. Based on our results in D9.6, a detection at the 
distance assumed will be possible, but barely, and only over a reduced frequency range around the maximum sensitivity these detectors 
can provide. On the other hand, for D15 and D25, at the distance assumed, existing detectors are expected to be sensitive 
to most of the emission spectra. Planned detectors -- the Einstein Telescope (ET) and the Cosmic Explorer (CE) -- will be sensitive to the 
full emission spectrum in all three of our models.

\begin{figure*}
\includegraphics[width=\columnwidth,clip]{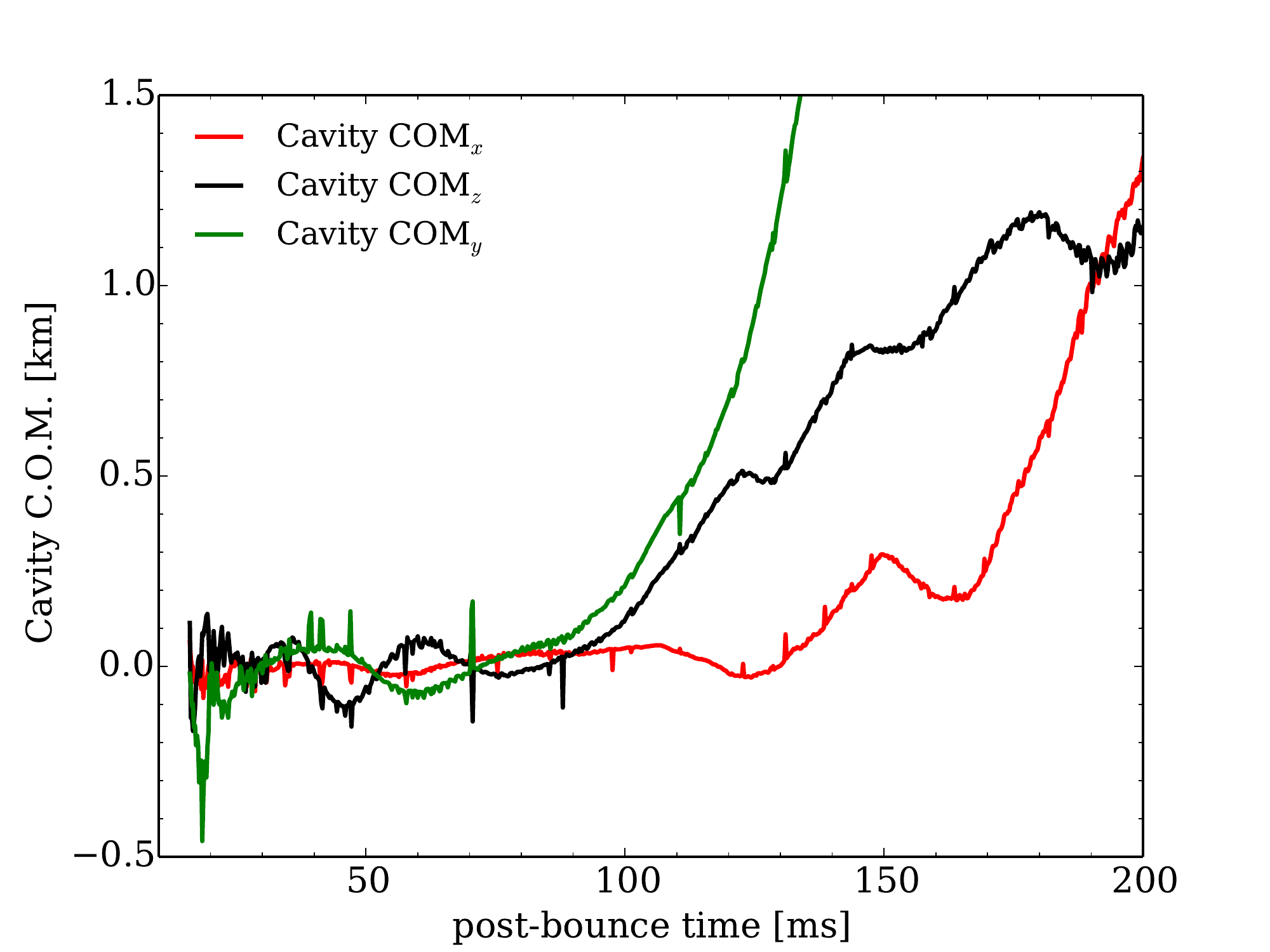}
\includegraphics[width=\columnwidth,clip]{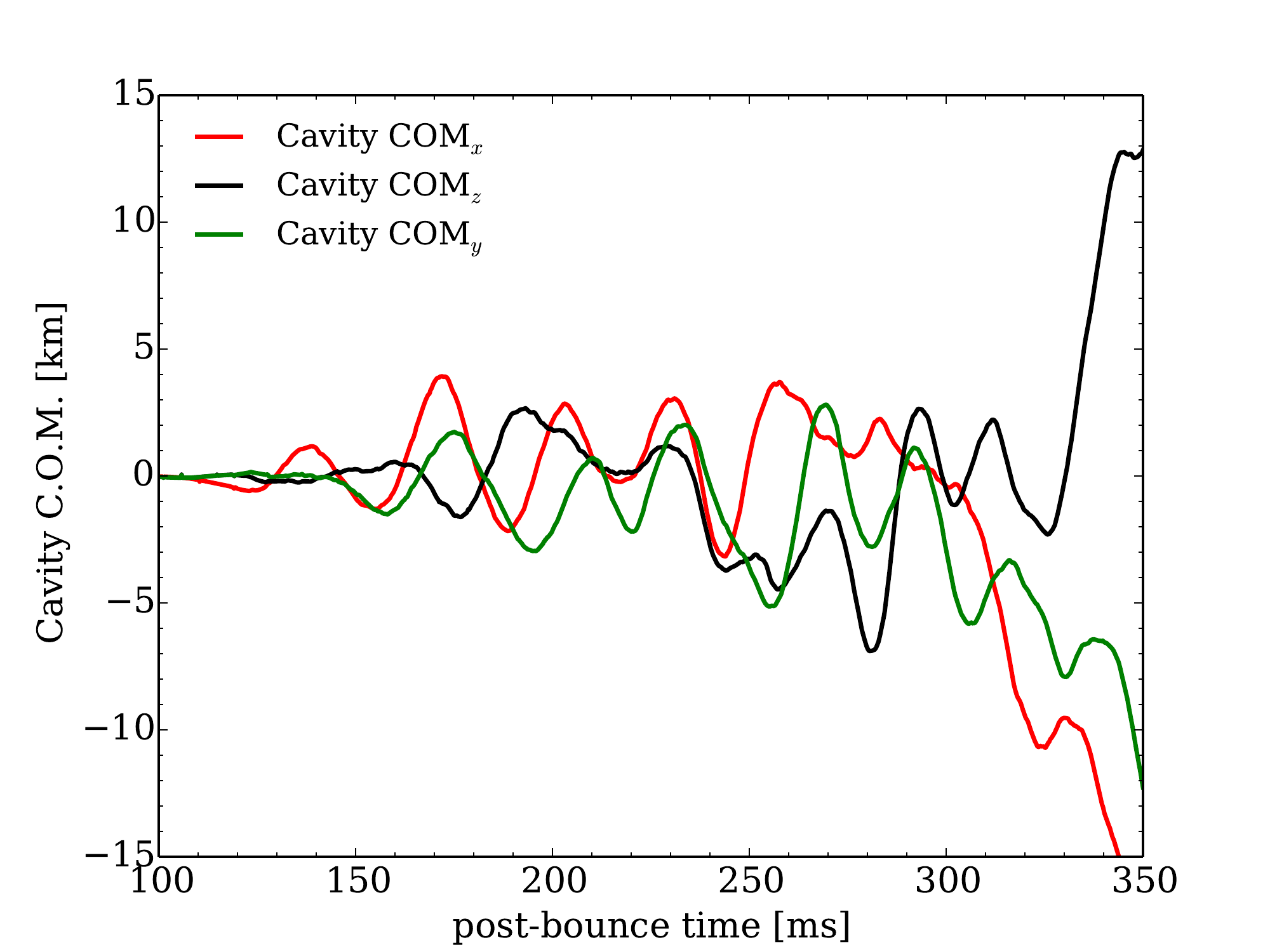}
\includegraphics[width=\columnwidth,clip]{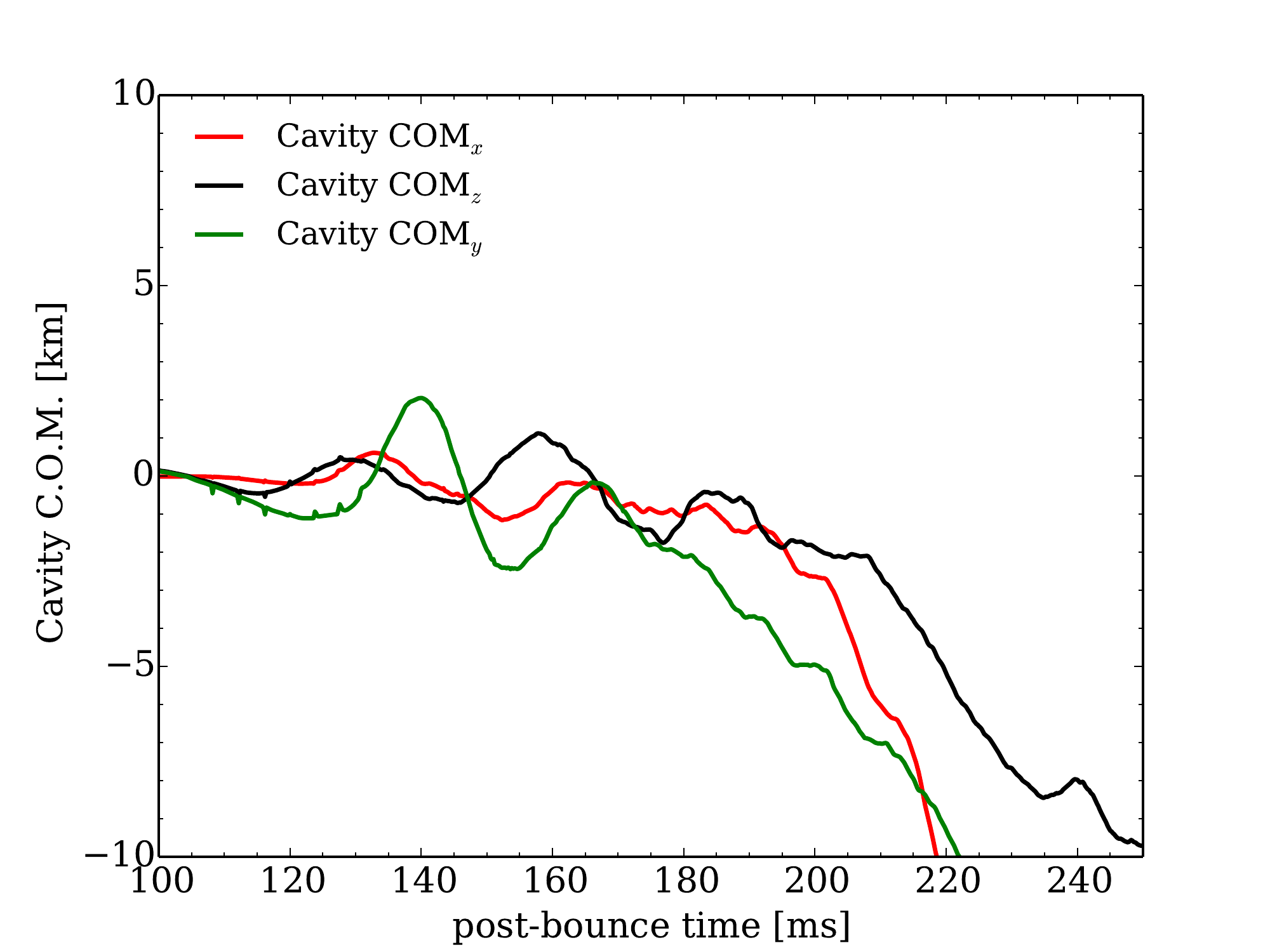}
\caption{Time evolution of the center of mass of the gain region in each of our three models.}
\label{fig:com}
\vspace{-10pt}
\end{figure*}

In Figure \ref{fig:com}, for all three of our models, we plot the evolution of the gain region center-of-mass in each of the three dimensions. 
For each model, the motion of the center-of-mass reflects the global evolution of the fluid flow in the post-shock region and, 
consequently, is an indicator of the presence of the SASI. From the figure, it is evident from the oscillatory behavior of the center-of-mass 
that the SASI is in fact present in all three of our models, with characteristic periods of $\sim$24, $\sim$29, and $\sim$29 ms for 
D9.6, D15, and D25, respectively. These, in turn, correspond to frequencies of $\sim$42, $\sim$34, and $\sim$34 Hz, respectively. Because 
one period of the SASI corresponds to two quadrupolar deformations of the post-shock cavity, the low-frequency gravitational wave 
spectrum should show SASI-associated peaks at twice these frequencies, at $\sim$84, $\sim$68, and $\sim$68 Hz, respectively.

\begin{figure*}
\includegraphics[width=\columnwidth,clip]{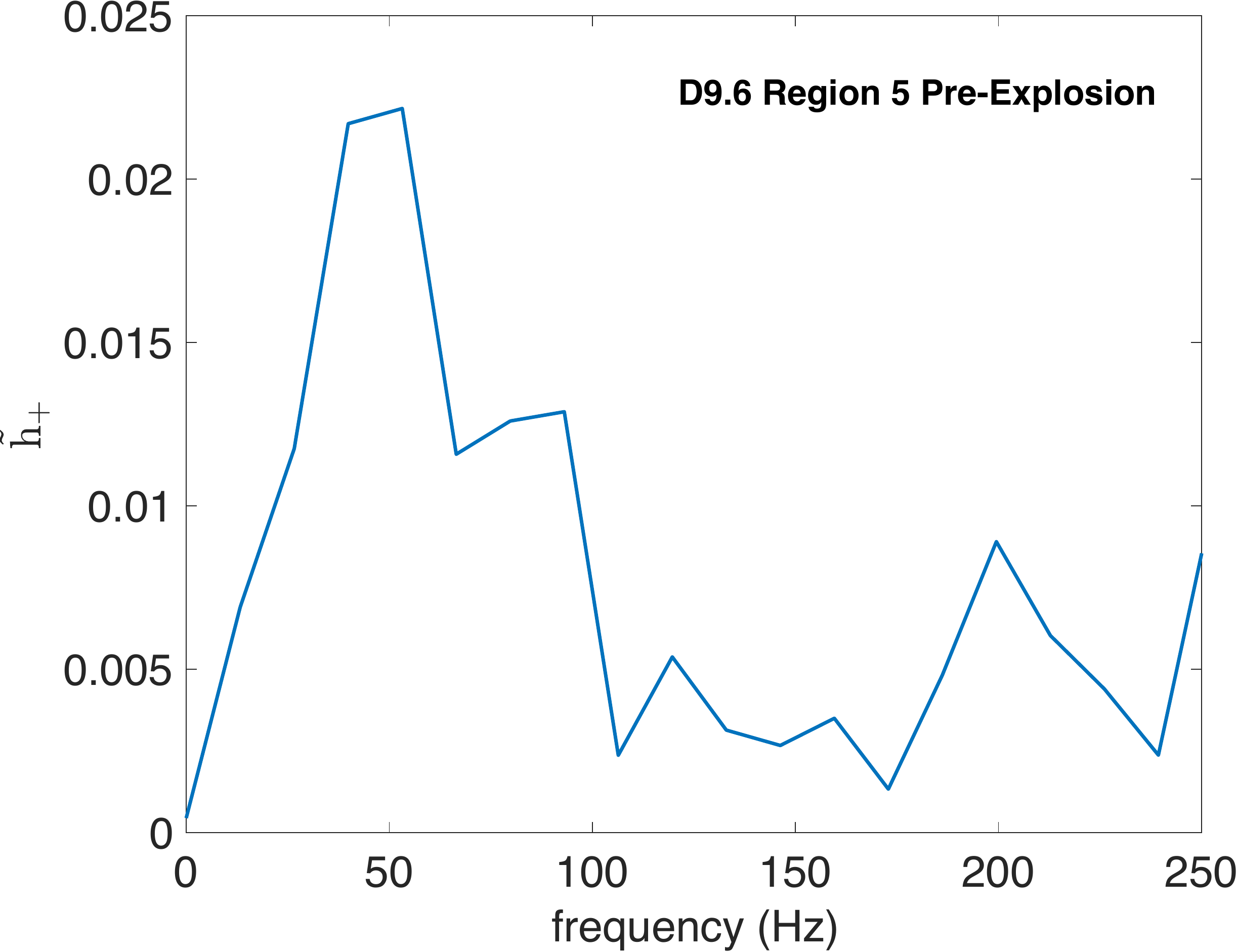}
\includegraphics[width=\columnwidth,clip]{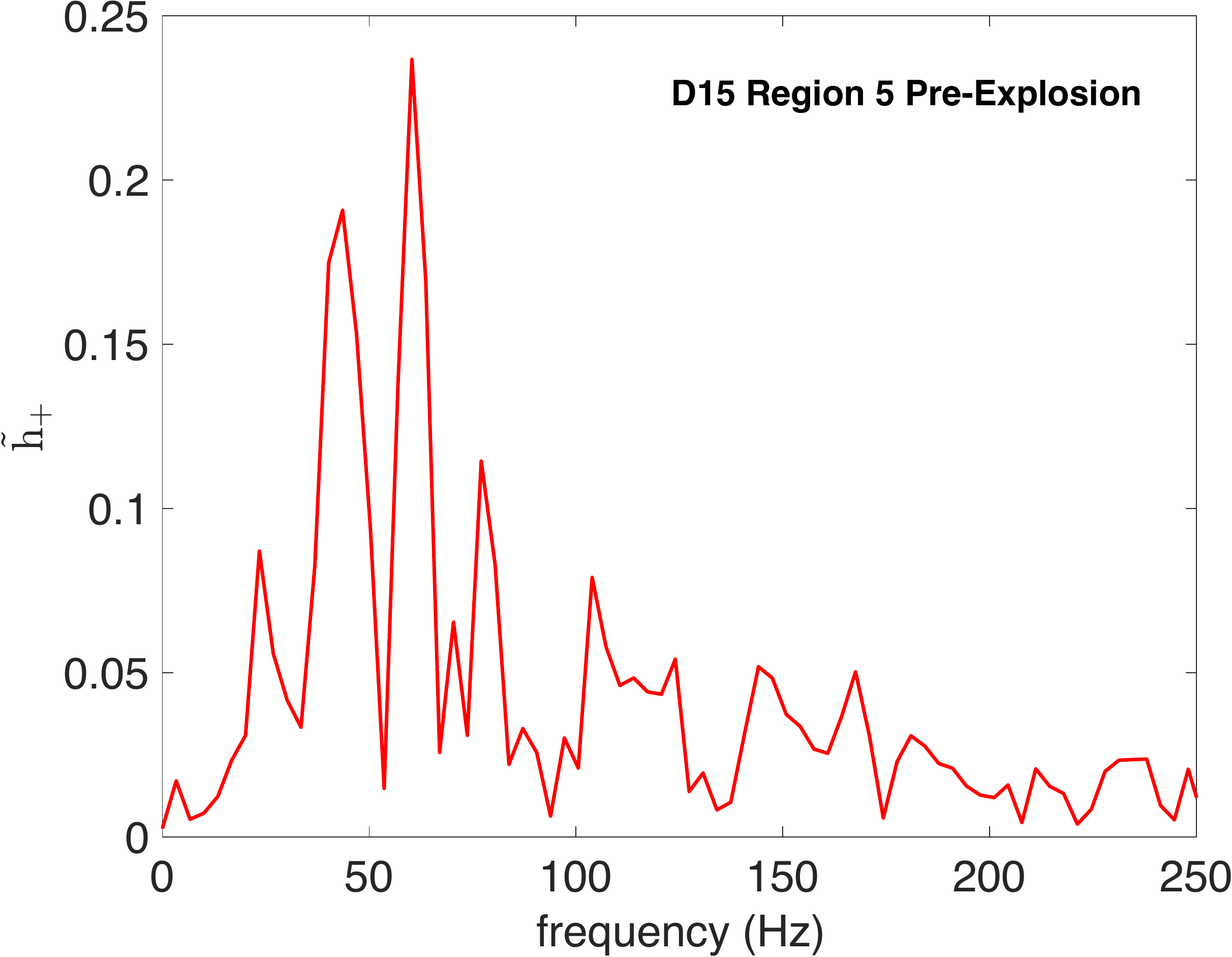}
\includegraphics[width=\columnwidth,clip]{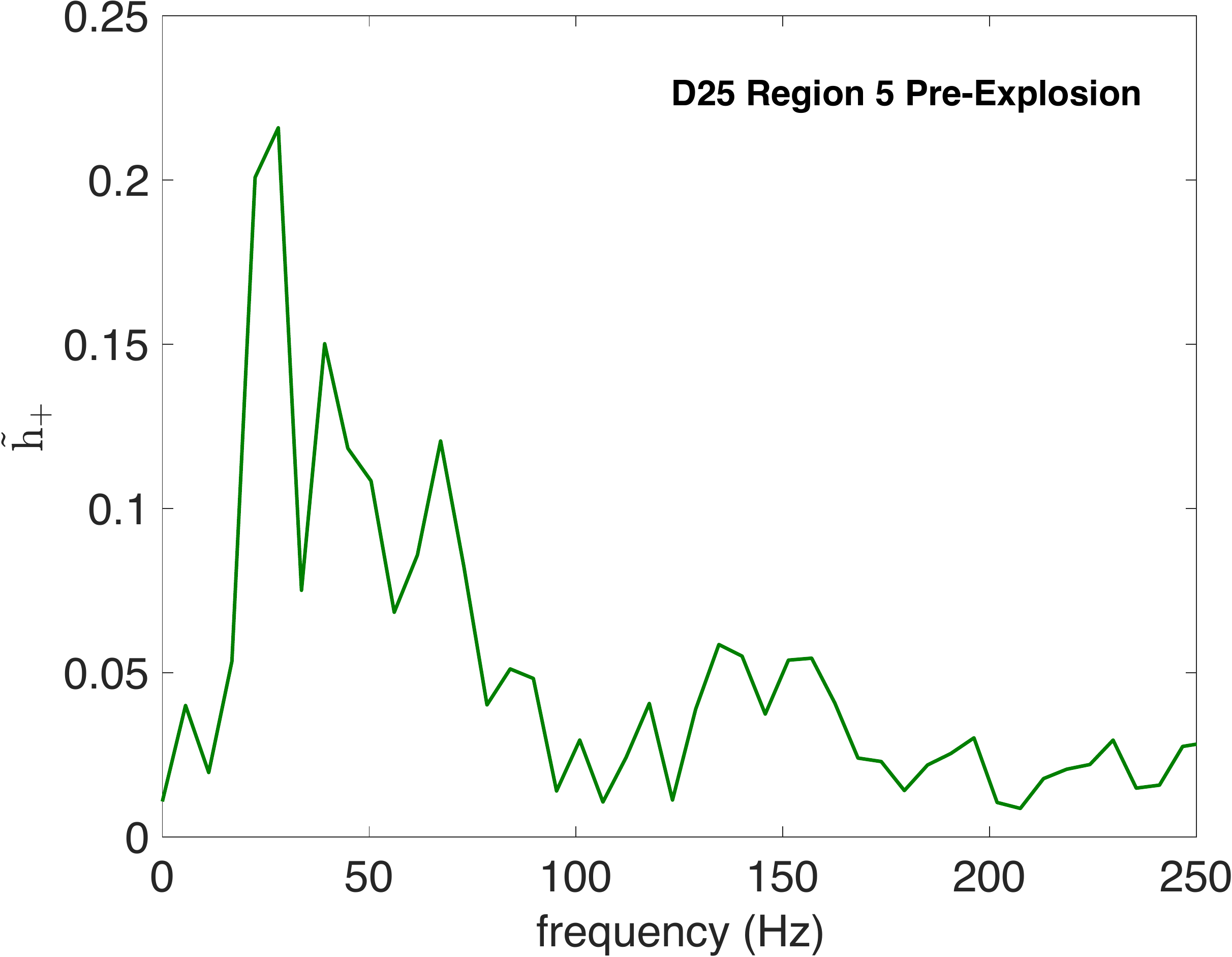}
\caption{Region 5 gravitational wave spectra prior to explosion.}
\label{fig:R5SpectraPreE}
\end{figure*}

\begin{figure*}
\includegraphics[width=\columnwidth,clip]{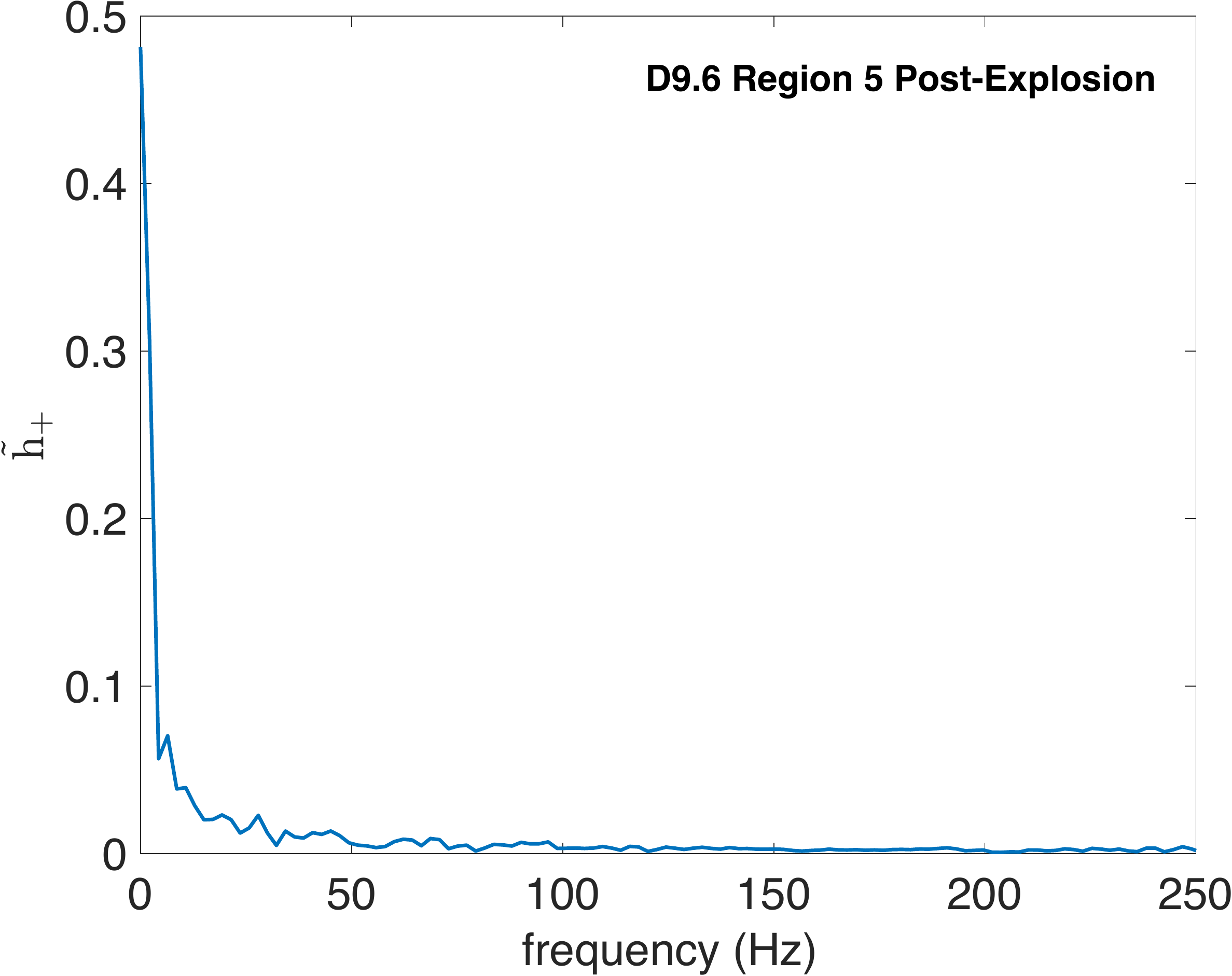}
\includegraphics[width=\columnwidth,clip]{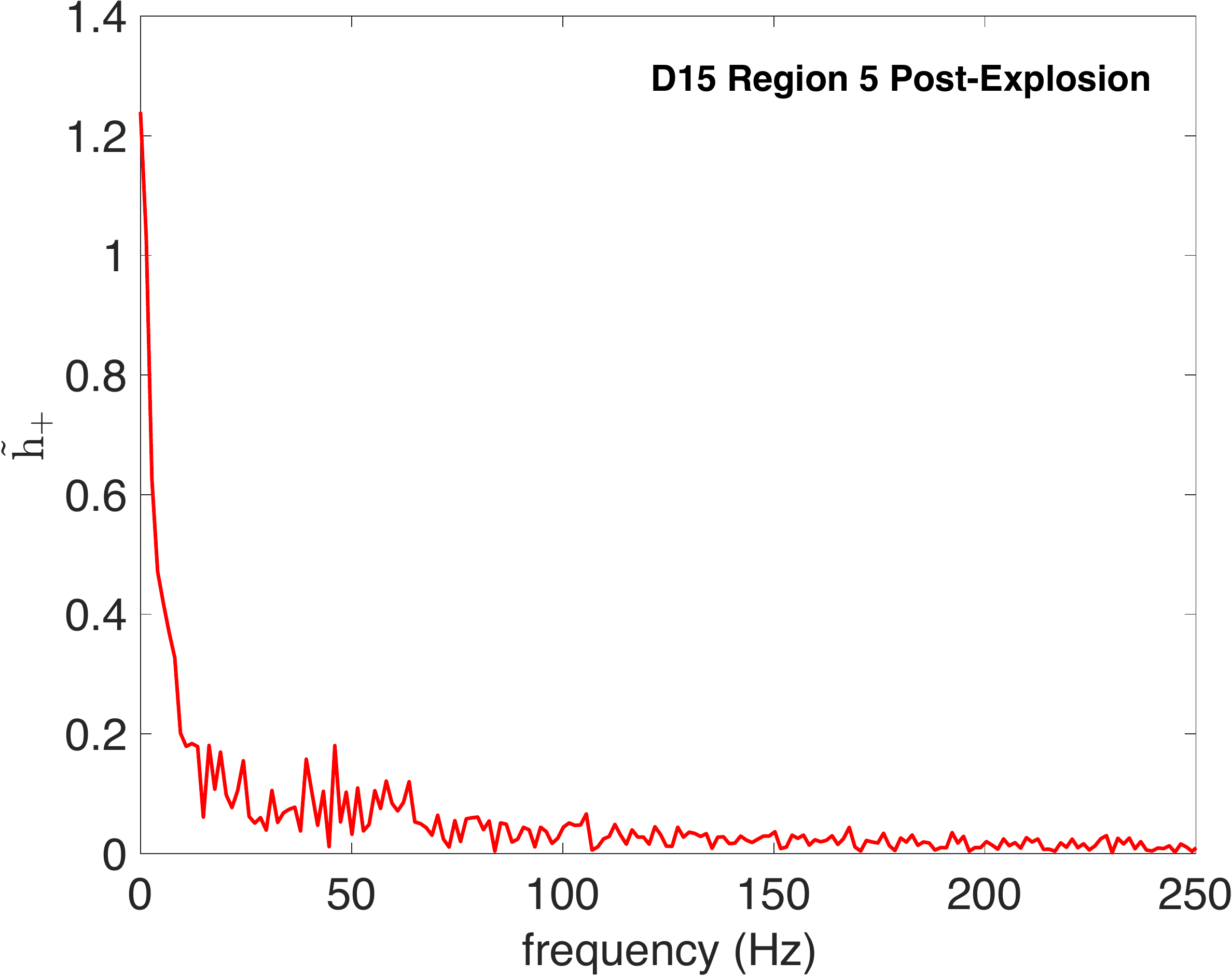}
\includegraphics[width=\columnwidth,clip]{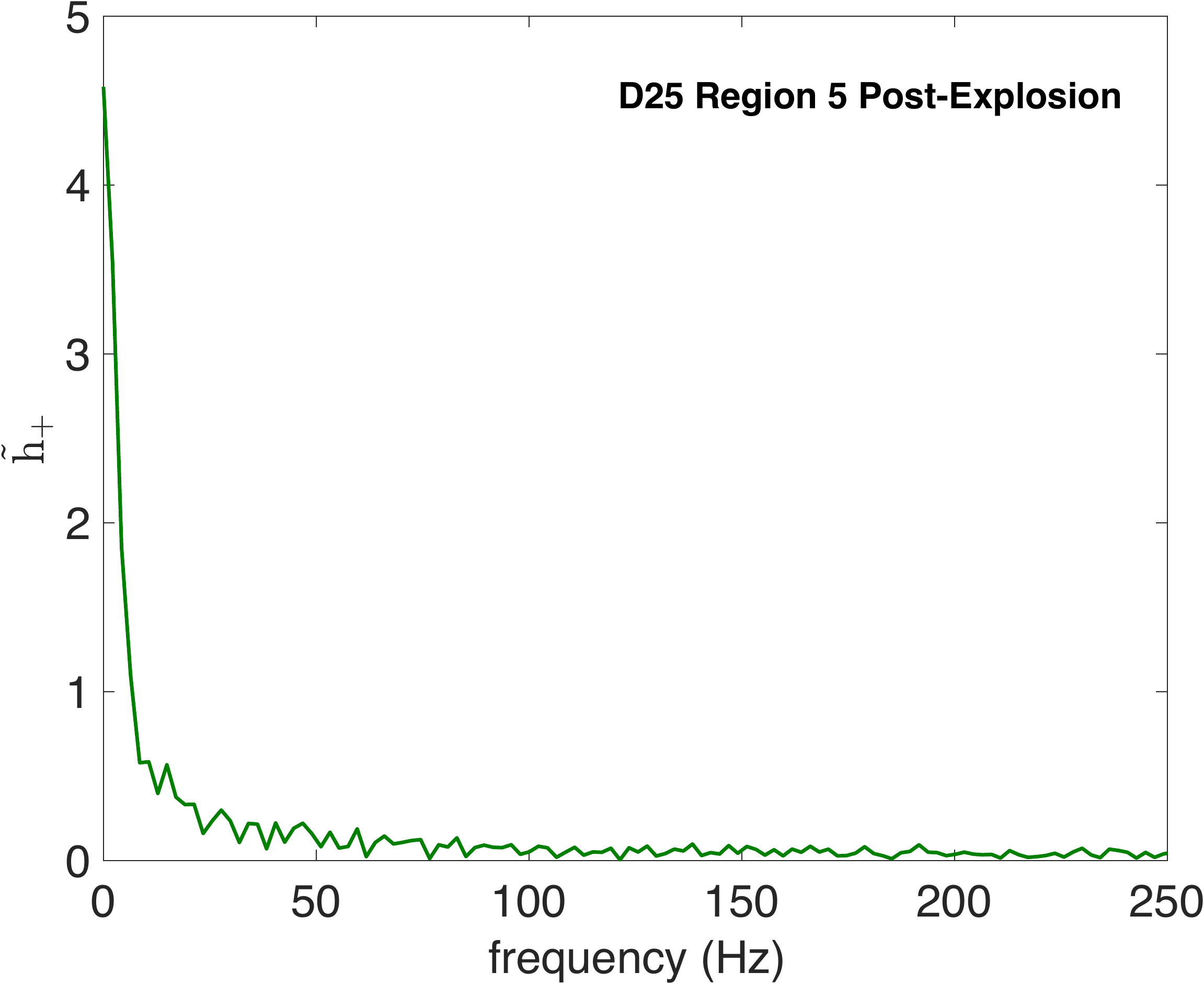}
\caption{Region 5 gravitational wave spectra after explosion.}
\label{fig:R5SpectraPostE}
\end{figure*}

Given the information in Figure \ref{fig:com}, we can now better interpret and, in turn, derive more information from the spectral decomposition 
of the gravitational wave strains from Region 5 in each of our models. In Figures \ref{fig:R5SpectraPreE} and \ref{fig:R5SpectraPostE} we 
plot the Fourier transforms of the plus polarization gravitational wave strains computed using the strain data from (1) before explosion, 
defined by the onset of the near-zero-frequency offset of the gravitational wave strains in each model and (2) the entire run for each 
model. For D9.6, D15, and D25, the ``pre-explosion'' spectra are computed using the temporal strain data up to 75, 300, 
and 180 ms after bounce, respectively. 
Looking first at the pre-explosion spectra, 
our estimates of the gravitational wave frequency for emission associated with the $\ell=1$ SASI modes are consistent with the peaks in 
the spectra at $\sim$60 and $\sim$70 Hz for D15 and D25, respectively, and with the relatively significant emission at a 
frequency $\sim$80 Hz in the spectrum for D9.6. The phase differences between the center of mass motion in 
this region projected along the $x$, $y$, and $z$ axes, shown in Figure \ref{fig:com}, is evidence of the spiral SASI mode, which contributes 
to the gravitational emission at the same frequencies. Thus, we attribute these ``peaks'' to both the sloshing and spiral SASI modes. Peaks 
at approximately half these frequencies in all three cases are consistent with the expected emission from the $\ell=2$ SASI mode. We attribute 
the remaining peaks -- e.g., between 20 and 30 Hz in all three of our models -- and the broadband emission between 0 and 250 Hz to stochastic 
emission from neutrino-driven turbulent convection. Looking now at the same plots but for the ``post-explosion'' case, shown in Figure {\ref{fig:R5SpectraPostE}, 
the low-frequency spectra in all three models are now dwarfed by the emission at near-zero frequencies, corresponding to the near-zero-frequency 
offset below $\sim$10 Hz in the gravitational wave strain that results when explosion occurs in each of these models.

\begin{figure*}
\includegraphics[width=\columnwidth,clip]{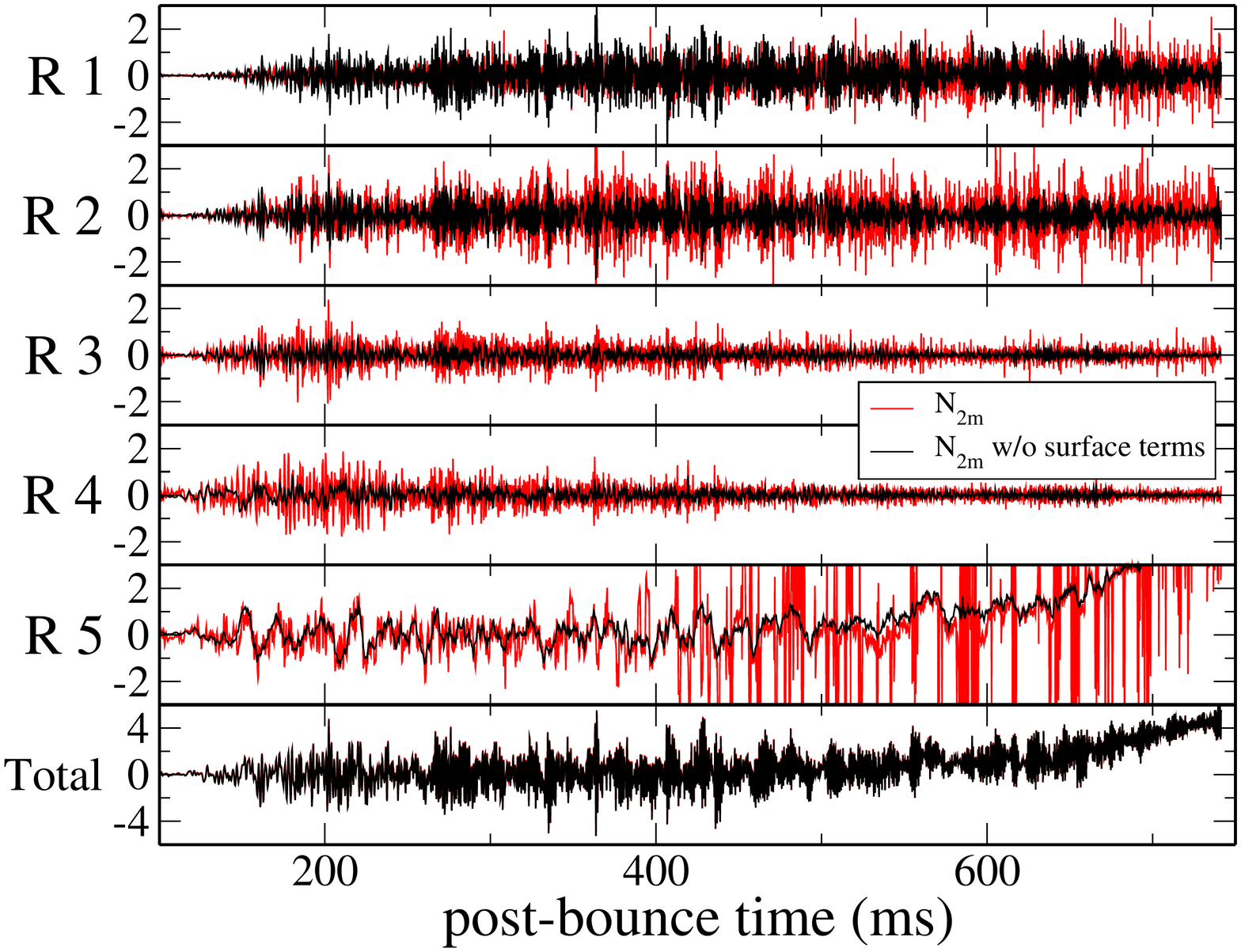}
\includegraphics[width=\columnwidth,clip]{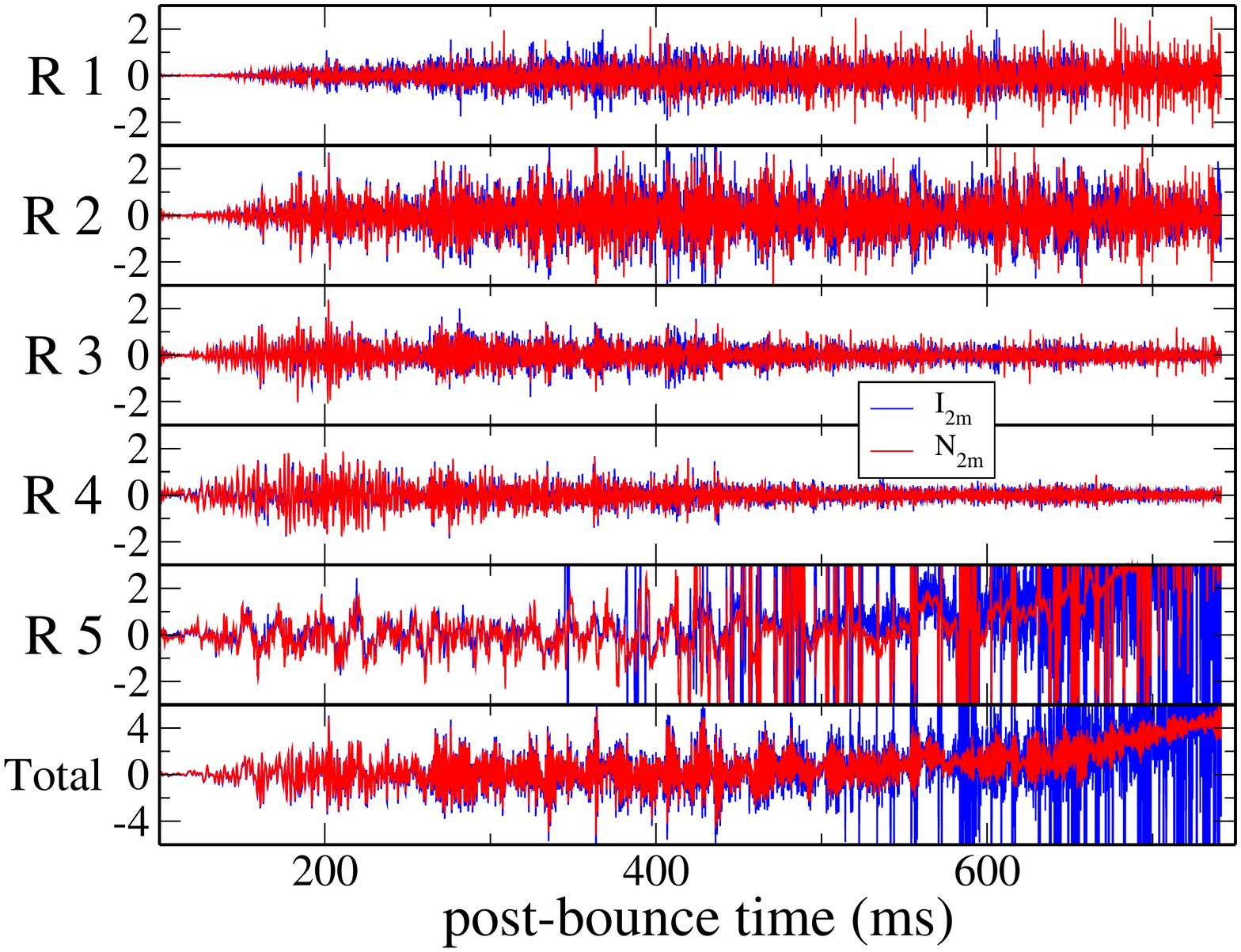}
\caption{Comparisons of the gravitational wave strains by region for D15, computed using the three different 
methods for determining $A_{2m}$, outlined in the text.}
\label{fig:A2m}
\end{figure*}

\begin{figure*}
\includegraphics[width=\columnwidth,clip]{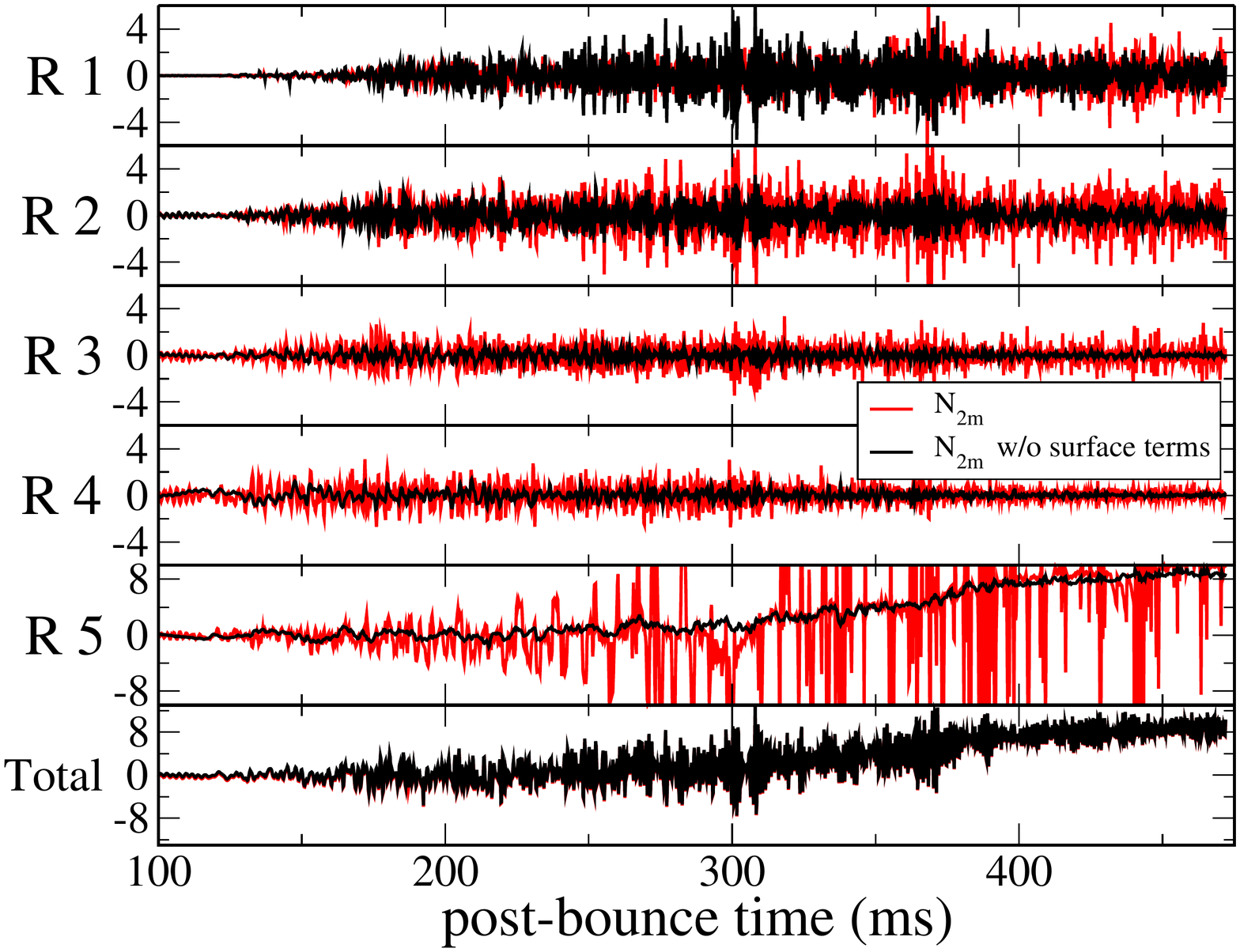}
\includegraphics[width=\columnwidth,clip]{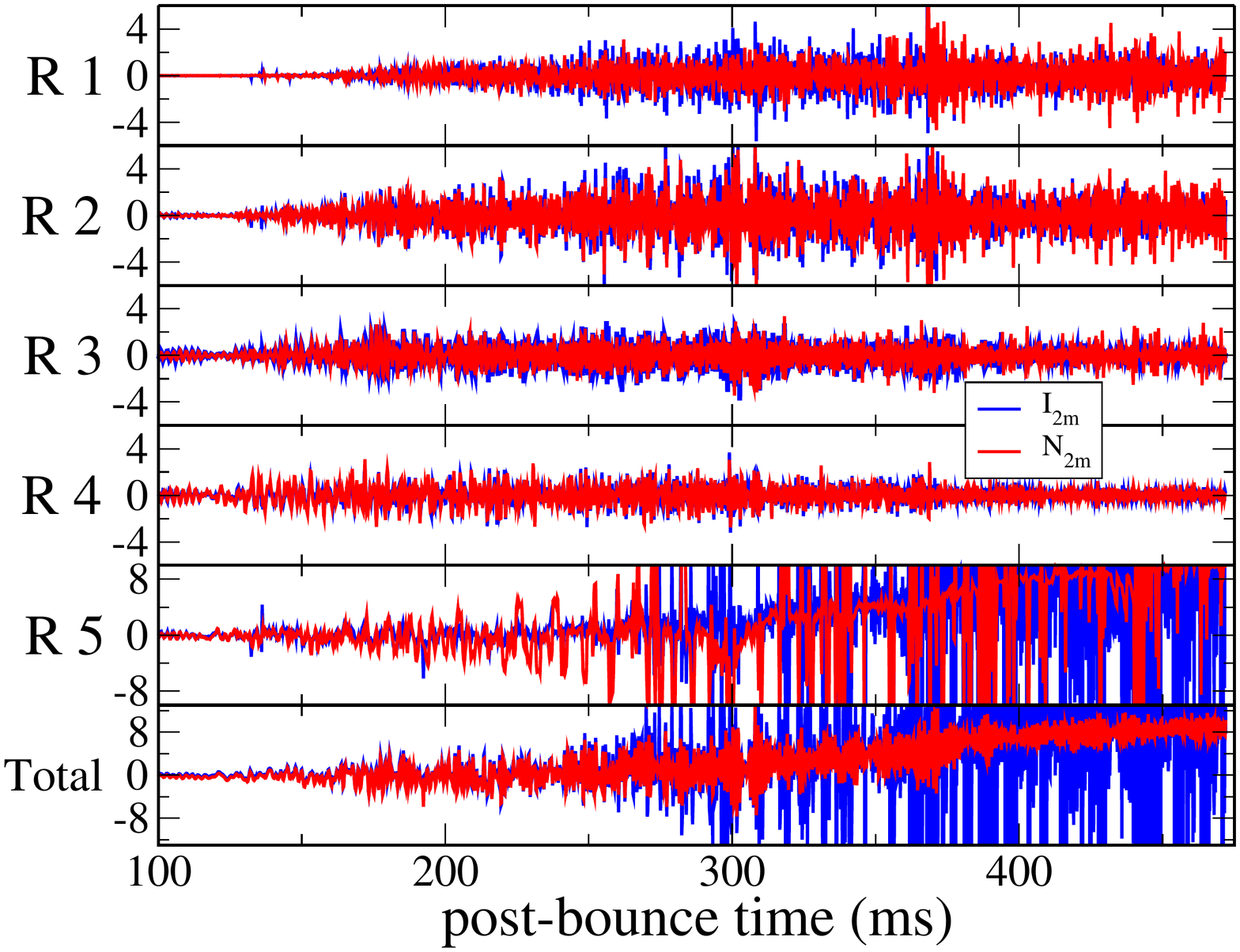}
\caption{Comparisons of the gravitational wave strains by region for D25, computed using the three different 
methods for determining $A_{2m}$, outlined in the text.}
\label{fig:A2m25}
\end{figure*}

\begin{figure*}
\includegraphics[width=\columnwidth,clip]{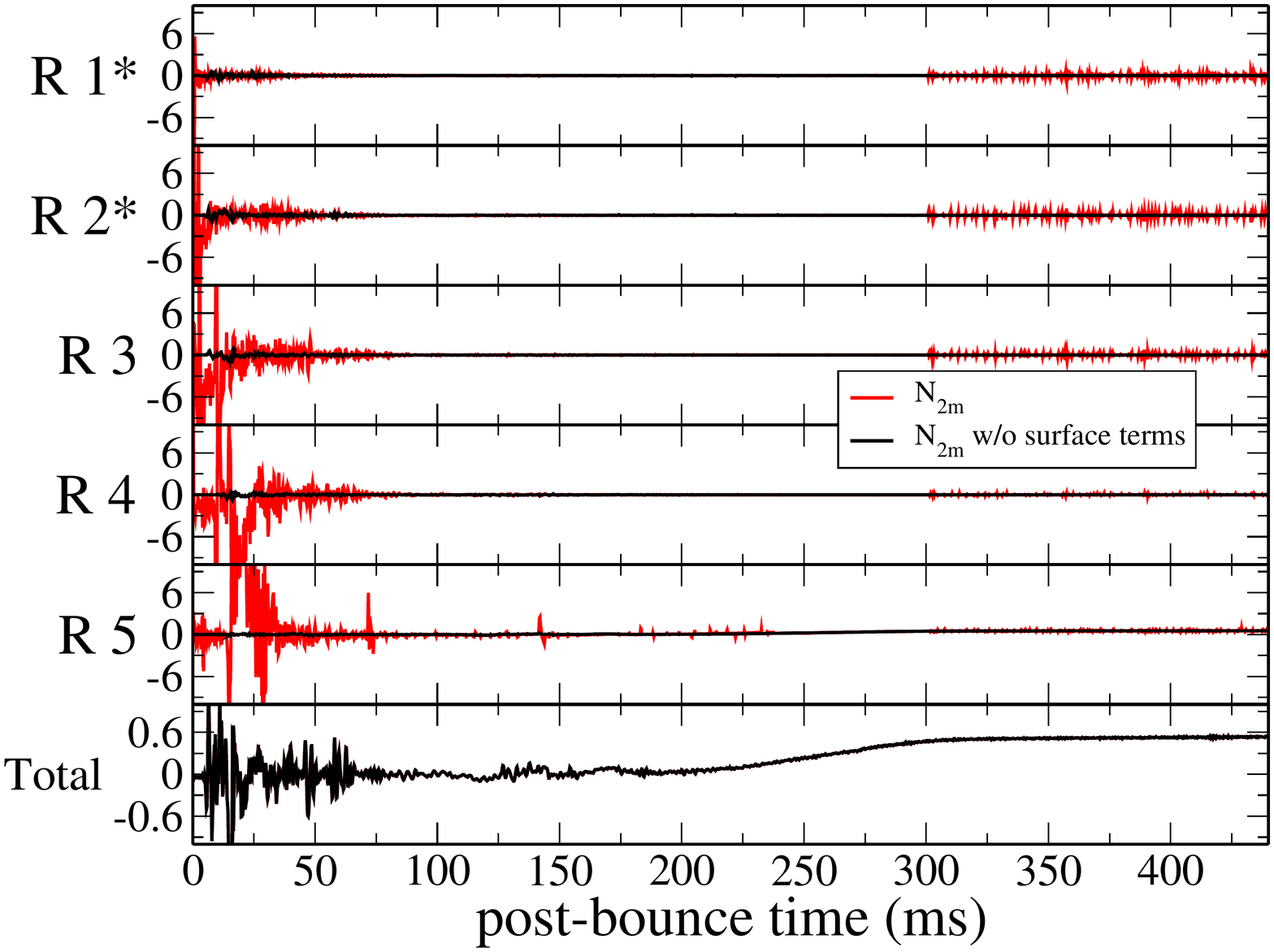}
\includegraphics[width=\columnwidth,clip]{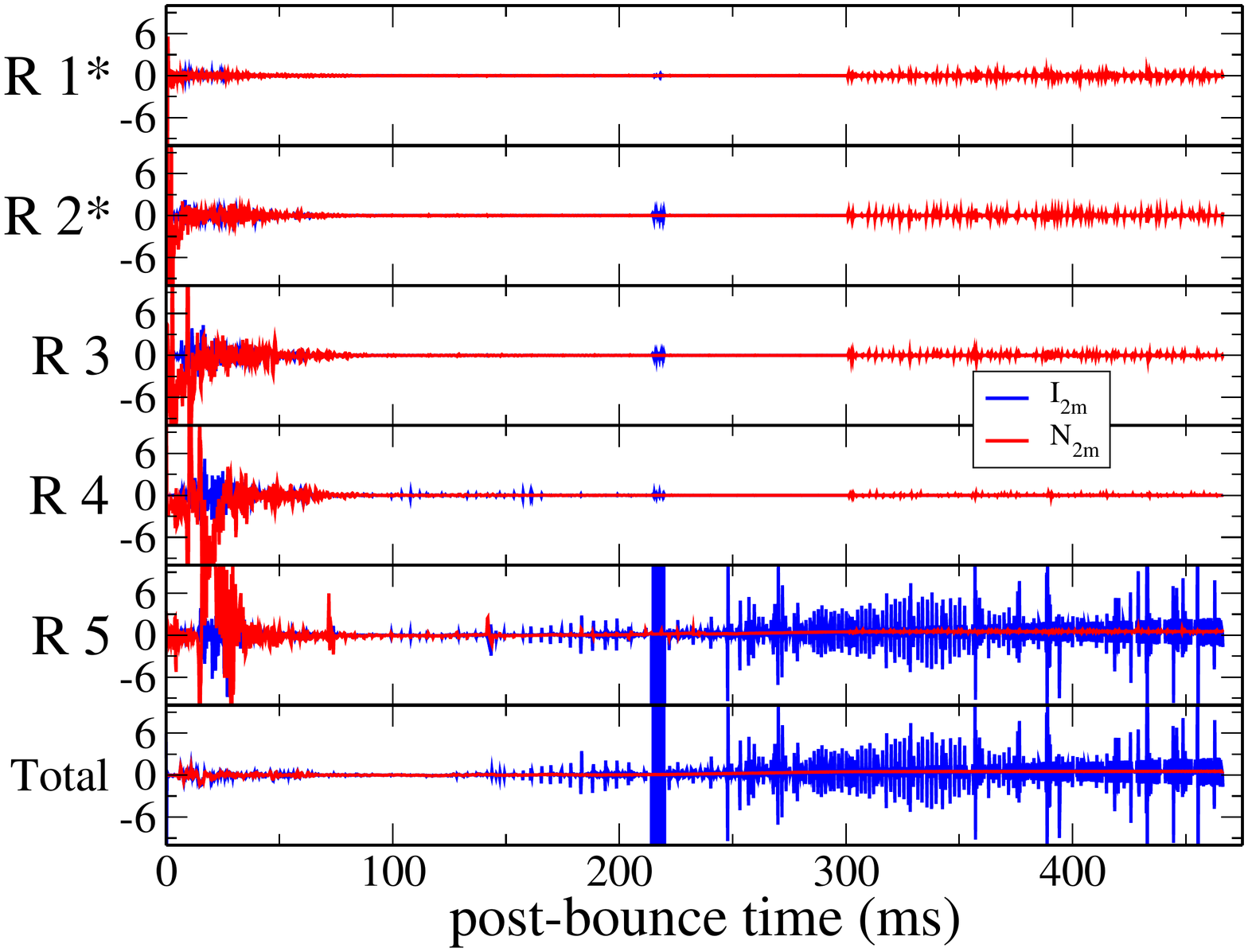}
\caption{Comparisons of the gravitational wave strains by region for D9.6, computed using the three different 
methods for determining $A_{2m}$, outlined in the text.}
\label{fig:A2m96}
\end{figure*}

\subsection{Gravitational Wave Source Identification: Challenges} 

Here we compare the results obtained using the three different approaches to calculating $A_{2m}$. In Figure \ref{fig:A2m}, we compare $Dh_{+}$ by 
region based on determining $A_{2m}$ using (1) the first time derivative of both $N_{2m}$ and $\bar{N}_{2m}$ (left panel) and (2) the first time 
derivative of $N_{2m}$ and the second time derivative of $I_{2m}$ (right panel). In both cases, the results are from D15. In Figures \ref{fig:A2m96} 
and \ref{fig:A2m25}, we repeat the analysis for D9.6 and D25.

First, a note on method: To compute the boundary contributions in $N_{2m}$, we require the boundary values of the quantities entering the boundary terms 
in Equation (\ref{eq:n2m-integration}). To obtain these values, we simply average over the values of these quantities in the zones just below and just above 
the boundary. As we will see, this approach works reasonably well provided that the spherical boundaries used to delineate the regions encapsulate the 
fluid motions giving rise to gravitational wave emission. Once this is no longer true, the boundary terms lead to large, stochastic, and ultimately unphysical 
regional gravitational wave strains, though these contributions cancel when computing the total strain. This was emphasized by Andresen et al. \cite{Andresen2017}
in their attempts to spatially decompose the high-frequency gravitational wave emission in their study, as well.

Given the results of all three methods, for our 15 and 25 M$_\odot$ cases the following conclusions can be drawn: 
(1) All three methods point to Regions 1 and 2 as the primary sources of high-frequency gravitational wave emission. 
(2) The relative contributions of the two regions to the high-frequency signal depends on the method used, with $\bar{N}_{2m}$ overestimating the contribution 
from Region 1 relative to Region 2, and with $N_{2m}$ and $I_{2m}$ yielding similar conclusions that Region 1's contribution is the largest. 
(3) Speaking now to the {\em excitation mechanism} of the high-frequency gravitational wave emission from Regions 1 and 2: The persistence of 
the emission in both models, well beyond the initiation of explosion and well beyond the reduction in high-frequency emission from Regions 3 
and 4, point to persistent Ledoux convection in Region 1 and its convective overshoot into Region 2 as the excitation mechanisms.
(4) The high-frequency contributions from Regions 3 and 4 are larger when $N_{2m}$ or $I_{2m}$ are used, relative to what was obtained using $\bar{N}_{2m}$, 
though both regions remain subdominant for the production of high-frequency gravitational wave production. 
(5) The evolution of the strains in Region 5 show that the results from $N_{2m}$ and $I_{2m}$ become unreliable around the time when explosion is developing in 
the models. In the $N_{2m}$ case, the problem stems from the limitations of our approach discussed above, when our spherical boundaries no longer encapsulate 
the gravitational-wave-generating fluid flows. In the $I_{2m}$ case, the numerical noise associated with computing the second time derivative of $A_{2m}$ swamps 
the signal. By construction, the boundary contributions in the $N_{2m}$ case cancel, and the total strains for the $\bar{N}_{2m}$ and $N_{2m}$ cases agree. This is 
obviously not the case when comparing the $\bar{N}_{2m}$ strain against the $I_{2m}$ strain. 
(6) When viewed in context -- i.e., when viewing the $\bar{N}_{2m}$ results together with the $N_{2m}$ and $I_{2m}$ results  -- it is clear that the $\bar{N}_{2m}$ 
results, when considered alone, provide a great deal of {\em qualitative} information regarding the spatial breakdown and origins of the gravitational wave emission,  and 
are not plagued by issues associated with boundary definitions and numerical noise.

For D9.6, only $\bar{N}_{2m}$ is useful in attempting to understand the breakdown of gravitational wave emission. As we discussed, defining 
the physical boundaries for our five regions is difficult in this case, which renders the use of $N_{2m}$ problematic, and, as before, $I_{2m}$ is dominated by 
numerical noise, especially after the onset of explosion.

\begin{figure*}
\includegraphics[width=\columnwidth,clip]{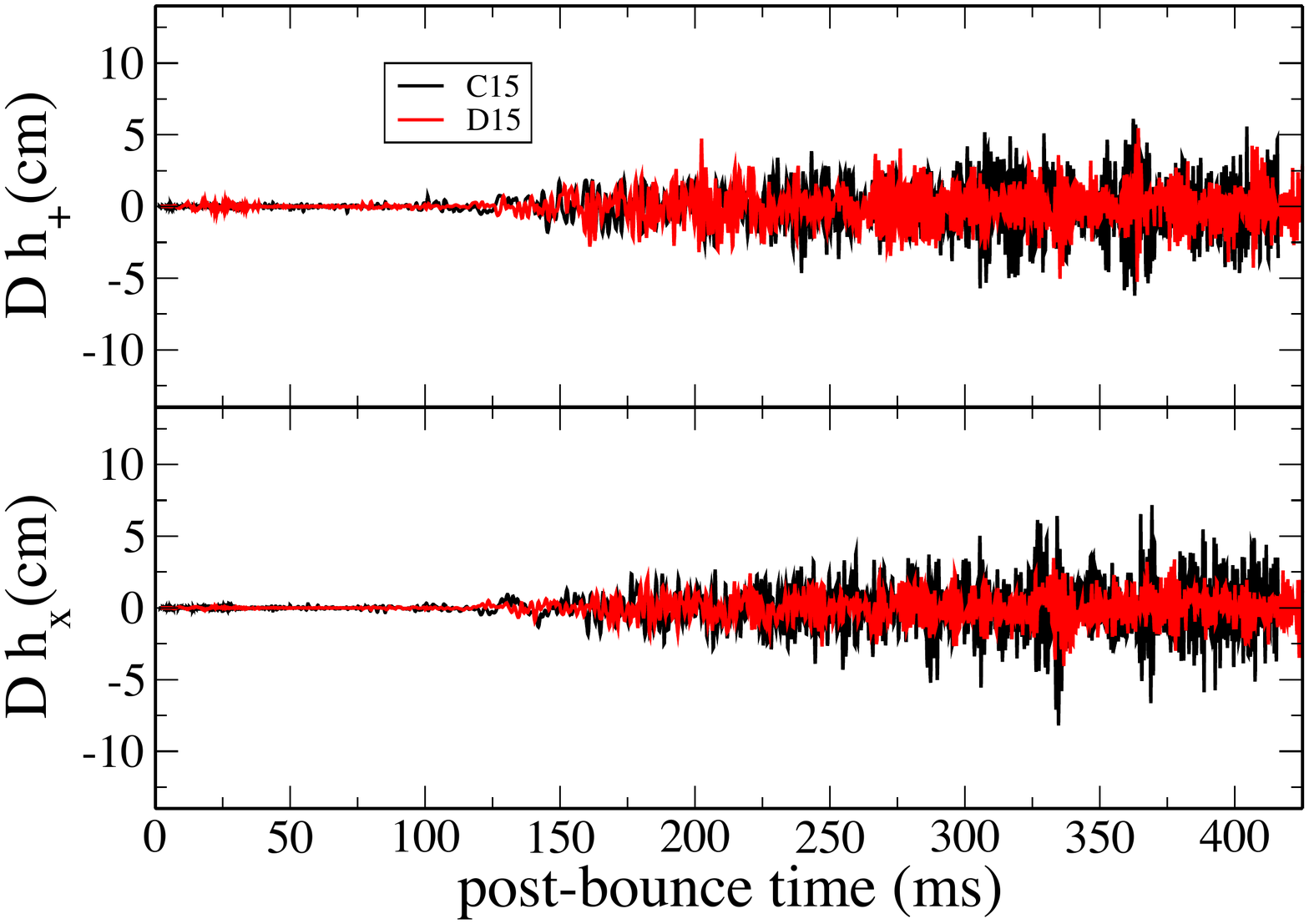}
\caption{Comparison of the gravitational wave strains from our C15-3D and D15-3D models.}
\label{fig:C15vsD15}
\end{figure*}

\subsection{Resolution Dependence of Gravitational Wave Emission Prediction}

Finally, a comparison of the results obtained for the plus- and cross-polarization gravitational wave strains for D15, 
with those obtained for the same progenitor in our first analysis (using model C15; Mezzacappa et al. \cite{MeMaLa20}), is shown in Figure \ref{fig:C15vsD15}. Our new 
strain amplitudes are as much as a factor of 2 smaller. We attribute these differences primarily to the high resolution used in the model presented here 
versus the lower (in some regions significantly lower) resolution afforded by the constant-$\mu$ grid used in our first analysis. This comparison demonstrates 
the need for sufficient numerical resolution for the accurate determination of the gravitational wave emission characteristics.

\section{Summary, Discussion, and Conclusions}

We have presented our predictions for the gravitational wave emission in each of three core collapse supernova 
models, corresponding to progenitors of 9.6, 15, and 25 \msun, all without rotation, with both zero (D9.6) and Solar (D15 and D25)
metallicity.  We have detailed the temporal and spectral characteristics of the emission and have correlated 
the different components of the emission with their physical origin. 

Considering their temporal evolution first, the particular set of models presented here splits into two groups: one 
model (D9.6) is characterized by early post-bounce gravitational wave emission, whereas 
the other two models (D15 and D25) are characterized by late-time emission. In the first 
case, the high-frequency emission is largely confined to a period less than $\sim$75 ms after bounce. (This is 
consistent with the results reported by M\"{u}ller, Janka, and Marek \cite{MuJaMa13} for the same progenitor in the 
context of their two-dimensional simulations.)  In the latter two cases, the high-frequency emission begins after $\sim$125 ms 
and persists for the duration of the run. Thus, given the disparate nature of the progenitors used, we obtain disparate -- 
specifically, complementary -- gravitational wave emission evolution. Our results suggest that across the 
entire range of progenitor masses from which we can expect core collapse supernovae the gravitational 
wave emission from such supernovae will differ significantly, although across a subcategory of progenitors, 
we might expect common emission characteristics.

However, the {\em origin} of the high-frequency emission in all three cases does not exhibit such complementarity. It 
originates from episodes of Ledoux and Schwarzschild convection in the proto-neutron star, in some cases short-lived 
and in others sustained.

The amplitudes of the strains for D9.6 are approximately one fifth of the amplitudes of
the strains seen in D15 and approximately one tenth of the amplitudes of the strains seen 
in D25. Thus, while the former model exhibits the same sources and characteristics of 
high- and low-frequency gravitational wave emission, it is a much weaker source. 

Finally, considering the evolution of the plus polarization strain in all three models, particularly at late times when explosion has set in, 
the explosion appears prolate when viewed along the $z$-axis and oblate when viewed along the $x$-axis.

D9.6 exhibits very high frequency gravitational wave emission, up to $\sim 2000$ Hz. The stochastic nature of the emission is the result of perturbations 
in the core induced by nuclear burning during core collapse. While this is unique to D9.6 in this study, we can expect such high-frequency emission from 
any core collapse supernova model developed using a (non-spherical) perturbed progenitor. Indeed, similar to what we observe here, O'Connor and Couch \cite{OcCo18} 
observed higher-frequency gravitational wave emission in the context of a model using a 20 \msun progenitor with imposed precollapse velocity perturbations.

The very high frequency emission we observe in D9.6 demonstrates a need for a sampling rate that defines a Nyquist frequency 
above the frequencies observed here: 2000 Hz. Adding to this demonstration of need, D15 and D25 clearly show evolving high-frequency 
emission reaching 1500 Hz and above. A Nyquist frequency below $\sim$2000 Hz will not capture the full temporal and, in turn, spectral 
characteristics of gravitational wave emission over the range of progenitor masses we must consider.

The low-frequency gravitational wave emission, below $\sim$250 Hz, in all three of our models is the result of three 
contributing phenomena: (1) neutrino-driven convection, (2) the SASI, and (3) explosion. Neutrino-driven convection 
seeds local, stochastic post-shock flows that contribute to gravitational wave emission over a broad range of frequencies. 
The SASI gives rise to global, organized, sustained features of the post-shock flow that are more easily identifiable in the 
gravitational wave spectra. Explosion gives rise to a near-zero-frequency offset of the gravitational wave strain, which 
is the easiest feature to identify in the spectra. Based on our heat maps and gravitational wave spectra for Region 5 
in each of our models, as well as the time evolution of the Region 5 center of mass in each model, we are able to 
identify with some confidence the contributions to the low-frequency gravitational wave emission from convection, 
the SASI, and explosion. For the case of SASI emission, estimates of the peak frequencies associated with its various 
contributing modes can be made and are consistent with the results we obtained. In all three of our models, the low-frequency 
emission's peak frequency is time dependent, as the models evolve through pre- and post-explosion epochs -- e.g., SASI- versus explosion-dominated.

Using the M\"{u}ller et al. \cite{MuJaMa13} formula relating the evolving peak frequency of high-frequency gravitational wave 
emission to the mass and radius in the proto-neutron star from which such emission largely emanates, we find that the emission 
is best modeled using values of the mass and radius at densities an order of magnitude higher than the density used to define 
the proto-neutron star surface in our models. Thus, attempts to cull the proto-neutron star mass and radius from a detection will, at the very least, 
require further analysis.

Here, we continued our efforts to identify the primary causes of the different components of the gravitational wave emissions 
in our three models (e.g., the high-frequency versus the low-frequency emission). As in Mezzacappa et al. \cite{MeMaLa20}, 
for D15 and D25, we decomposed our gravitational wave strains into their contributions from five 
regions below the shock. Moving inward in radius, they are the gain region (Region 5), the cooling region (Region 4), the proto-neutron star 
surface layer (Region 3), the Ledoux convection overshoot layer (Region 2), and the Ledoux convection layer (Region 1). For 
D9.6, we instead replaced the innermost two regions with regions defined by density contours at, moving 
inward, $10^{13}$ \gcc, and $10^{14}$ \gcc. 

In D15 and D25, we find that the high-frequency (above $\sim$500 Hz) gravitational wave emission stems largely from 
Regions 1 and 2, as the result of sustained Ledoux convection from the continued deleptonization of the proto-neutron star
in Region 1 and convective overshoot into Region 2. This emission is in sync temporally with the development of Ledoux convection 
in the proto-neutron star, which begins after $\sim$125 ms post bounce in both models. 
However, on closer inspection, looking for example at the heat map from Region 2 of D15, there is 
evidence of excitation of gravitational wave emission by accretion onto the proto-neutron star from the turbulent gain region above 
it. At $\sim$500 ms, when explosion develops in this model and accretion is reduced, the high-frequency gravitational wave emission 
from this region decreases. Nonetheless, it is clear that in D15 and D25 the primary causation is from within the 
proto-neutron star.

We evaluated our regional strains using three different methods for calculating $A_{2m}$, all three of which have their drawbacks.
Nonetheless, across all three methods, for D15 and D25 one conclusion remained the same: The convective and 
convective overshoot regions deep within the proto-neutron star, above densities $\sim\rho=10^{12}$ \gcc, dominate the production 
of high-frequency gravitational wave emission. This conclusion is in sync with the conclusions drawn in Andresen et al. \cite{Andresen2017} 
and Mezzacappa et al. \cite{MeMaLa20}. In the former case, Region 2 was identified as the primary source. In the latter case, Region 1 
was identified as supplying more of the gravitational wave emission given that $\bar{N}_{2m}$ was used to compute the strains. Here 
we see that the relative contributions of Regions 1 and 2 to the high-frequency strains depends on the method used, with $N_{2m}$ and 
$I_{2,m}$ favoring Region 2. 

For D9.6, only one of the methods ($\bar{N}_{2m}$) yielded meaningful results given the difficulties associated with 
defining physically meaningful zone boundaries that encapsulate the flows leading to gravitational wave emission. Nonetheless, the results 
from our other two models suggests that meaningful qualitative information regarding the sources of gravitational wave emission can be
obtained through $\bar{N}_{2m}$, which does not suffer from these difficulties. In this model, too, the early high-frequency gravitational wave 
emission is seeded by two episodes of convection in Regions 1 and 2 and then subsequently Region 3 -- i.e., beginning deep within the 
proto-neutron star and then extending up to its surface layers. The interaction of convection with perturbations introduced in the core by 
nuclear burning during collapse lead to very-high-frequency emission up to frequencies $\sim 2000$ Hz. While the largest strain amplitudes 
are confined to the first 75 ms after bounce, evidence of persistent Ledoux convection deep within the proto-neutron star, largely from Region 
1, is evident in the total heat map, where we see the characteristic linear rise in peak frequency over the course of the run. Very low-amplitude 
gravitational wave generation from Regions 4 and 5 is evident as the explosion powers up, but at this time, relative to what is observed in the 
first 75 ms of the run after bounce, the model is gravitational wave quiet. Therefore, in this case it is even easier to discern that the high-frequency 
gravitational wave emission stems from excitation mechanisms within the proto-neutron star.

From plots of our characteristic strains, we see that the gravitational wave emission in all three of our models
is in principle detectable by current generation gravitational wave detectors for the distance to the core collapse 
supernova assumed here (10 kpc). D9.6 is a much weaker source of gravitational 
waves and thus can be observed over a much more restricted frequency range around the detectors' maximum 
sensitivity. Unfortunately, the very high frequency signal obtained in D9.6 is not 
entirely observable at present but would in principle be observable by third-generation detectors.

The primary limitations of our study are the lack of inclusion of rotation and magnetic fields, both of which have been shown to 
affect, even fundamentally alter, the gravitational wave emission in core collapse supernovae \cite{HaKuNa16,Andresen2019,PoMu20,ShKuKo20,PaWaCo21,PaLiCo21,ShKuKo21,VaMuSc22}.

The primary limitations of our \chimera models are the use of ray-by-ray neutrino transport and an effective 
 gravitational potential to capture some aspects of general relativistic gravity. Studies have been performed that demonstrate that no systematic trends were identified, 
 albeit in a limited set of models, when comparing gravitational wave emission predictions from models using 
 ray-by-ray transport with predictions from models using three-dimensional transport \cite{AnGlJa21}. Moreover, 
 the differences that were seen when comparing the two cases were reduced when the resolution in the models 
 was increased. The models we present here are highly resolved in all three spatial dimensions -- e.g., with the 
 Yin--Yang equivalent of one degree resolution in both $\theta$ and $\phi$. Studies that compare emission predictions 
 in models that are fully general relativistic versus models that implement an effective potential have been completed
 in the context of two-dimensional simulations \cite{MuJaMa13}. The authors conclude that the use of an effective 
 potential leads to an overestimate of $\sim$20\% of ``typical'' gravitational wave frequencies. Given that this comparison 
 was made in the context of two-dimensional models, where the primary mode of high-frequency gravitational wave emission 
 is the excitation of the proto-neutron star surface layer by accretion funnels from the gain region, it is not clear 
 what the impact will be when comparing models using general relativistic gravity versus an effective potential 
 when the primary emission occurs deeper in the proto-neutron star, in its convective layer and in the overshoot 
 layer above it. A comparison in the context of three-dimensional models is needed.
  
 Regarding the microphysics employed -- the nuclear equation of state and the neutrino weak interactions: 
 Models that include muons (not included here) and the additional neutrino interactions associated with them can 
 exhibit qualitatively different outcomes \cite{BoJaLo17}. In turn, we should expect quantitative, and 
 perhaps qualitative, differences in the gravitational wave emission characteristics of models that include muons. 
 Moreover, studies have been performed that illuminate the equation of state dependencies of gravitational 
 wave emission predictions \cite{KuKoTa16,KuKoHa17,RiOtAb17,PaLiCo18,EggenbergerAndersen21}. 
 We look forward to reporting on the outcomes of these models in the future.

This research was supported by the National Science Foundation Gravitational Physics Theory Program (PHY 1806692 and 2110177) and by the U.S. Department of Energy Offices of Nuclear Physics and Advanced Scientific Computing Research. 
This research was also supported by an award of computer time provided by the Innovative and Novel Computational Impact on Theory and Experiment (INCITE) Program at the Oak Ridge Leadership Computing Facility (OLCF) and at the Argonne Leadership Computing Facility (ALCF), which are DOE Office of Science User Facilities supported under contracts DE-AC05-00OR22725 and DE-AC02-06CH11357, respectively.
P. M. is supported by the National Science Foundation through its employee IR/D program. The opinions and conclusions expressed herein are those of the authors and do not represent the National Science Foundation.

\end{document}